\newif\ifemulate
\newif\ifastroph

\emulatetrue

\astrophtrue

\ifemulate
	\documentclass[iop,numberedappendix]{emulateapj}
\else
	\documentclass[manuscript]{aastex}
\fi

\newcommand{\myemail}{mulders@lpl.arizona.edu}

\slugcomment{ApJ accepted, 30/10/14}

\shorttitle{A stellar-mass-dependent drop in planet occurrence rates}
\shortauthors{Mulders et al.}

\usepackage{natbib,graphicx,xspace,amsmath,multirow,rotating}
\newcommand{\Teff}{\ensuremath{T_{\rm eff}}\xspace}

\newcommand{\cdpp}{\ensuremath{\sigma_{*}}\xspace}
\newcommand{\cdppnorm}{\ensuremath{\sigma_{\rm LC}}\xspace}
\newcommand{\tlc}{\ensuremath{t_{\rm LC}}\xspace}

\newcommand{\cdppindex}{\ensuremath{\rm cdpp_{index}}\xspace}

\newcommand{\tdur}{\ensuremath{t_{\rm dur}}\xspace}
\newcommand{\snr}{\ensuremath{ {\rm SNR} }\xspace}

\newcommand{\snrobs}{\ensuremath{ {\rm SNR_{tab}}}\xspace}

\newcommand{\fgeo}{\ensuremath{f_{\rm geo}}\xspace}
\newcommand{\feff}{\ensuremath{f_{\rm eff}}\xspace}
\newcommand{\feffi}{\ensuremath{f_{{\rm eff},i}}\xspace}
\newcommand{\focc}{\ensuremath{f_{\rm occ}}\xspace}

\begin{document}

\title{A stellar-mass-dependent drop in planet occurrence rates}

\author{Gijs D. Mulders, Ilaria Pascucci, and D\'aniel Apai\altaffilmark{1}}
\affil{Lunar and Planetary Laboratory, The University of Arizona, Tucson, AZ 85721, USA}
\email{\myemail}
\altaffiltext{1}{Department of Astronomy, The University of Arizona, Tucson, AZ 85721, USA}

\begin{abstract}
The \textit{Kepler Spacecraft} has discovered a large number of planets up to one-year periods and down to terrestrial sizes. 
While the majority of the target stars are main-sequence dwarfs of spectral type F, G, and K, \textit{Kepler} covers stars with effective temperature as low as 2500 K, which corresponds to M stars.
These cooler stars allow characterization of small planets near the habitable zone, yet it is not clear if this population is representative of that around FGK stars.
In this paper, we  calculate the occurrence of planets around stars of different spectral types as a function of planet radius and distance from the star, and show that they are significantly different from each other. We further identify two trends: First, the occurrence of Earth to Neptune-sized planets ($1-4 R_\oplus$) is successively higher toward later spectral types at all orbital periods probed by \textit{Kepler}; Planets around M stars occur twice as frequently as around G stars, and thrice as frequently as around F stars.
Second, a drop in planet occurrence is evident at all spectral types inward of a $\sim 10$ day orbital period, with a plateau further out. By assigning to each spectral type a median stellar mass, we show that the distance from the star where this drop occurs is stellar mass dependent, and scales with semi-major axis as the cube root of stellar mass. By comparing different mechanisms of planet formation, trapping and destruction, we find that this scaling 
best matches the location of the pre-main-sequence co-rotation radius, indicating efficient trapping of migrating planets or planetary building blocks close to the star.
These results demonstrate the stellar-mass dependence of the planet population, both in terms of occurrence rate and of orbital distribution. The prominent stellar-mass dependence of the inner boundary of the planet population shows that the formation or migration of planets is sensitive to the stellar parameters. 
\end{abstract}

\keywords{planetary systems -- stars: low-mass -- planet-star interactions -- planets and satellites: formation -- Protoplanetary Disks}

\section{Introduction}

The \textit{Kepler Spacecraft} has provided an unprecedented view of close-in planetary systems around other stars. The main goal of the mission is to characterize the occurrence rate of terrestrial planets in the habitable zone, while the largest yield of \textit{Kepler} planets and \textit{planetary candidates} (KOIs) are super-earths and sub-Neptunes with orbital periods shorter than any planet in our solar system \citep[e.g.,][]{Borucki:2011cp, 2013ApJS..204...24B}. Though follow-up and characterization of some planets is possible \citep[e.g.,][]{2013Natur.503..381H,2014ApJ...789..154D}, the \textit{Kepler} mission excels in providing statistics on planetary architectures, with a survey-bias that is relatively well understood in terms of false positives \citep{2013ApJ...766...81F}, completeness \citep[e.g.,][]{2013ApJS..204...24B}, and stellar noise \citep{Christiansen:2012bz}. 

Several studies have inferred the occurrence rates of planets from the entire \textit{Kepler} sample of planetary candidates \citep[e.g.,][]{2012ApJS..201...15H, 2013ApJ...766...81F,2013ApJ...770...69P}, which mainly contains planets around main sequence stars with spectral types F, G, and K. In general, the planet occurrence rate increases with distance from the star up to an orbital period of $\sim10$ days \citep{2012ApJS..201...15H}, and becomes flat further out \citep{Youdin:2011gz,Catanzarite:2011cu,2012ApJ...745...20T,2013ApJ...778...53D,Silburt:2014tf}. Special attention has been given to the subset of cooler, lower-mass stars (roughly corresponding to the M dwarf spectral type) where planets smaller and closer to the habitable zone can be detected \citep{2013ApJ...767...95D, 2014ApJ...791...10M}.

A priori, there is no reason to assume planetary architectures are the same across spectral types. The range of spectral types probed by \textit{Kepler} trace a range of stellar masses roughly from $\sim 0.3 M_\odot$ to $\sim 1.5 M_\odot$. The natal environments of planets -- protoplanetary disks -- show strong scaling with stellar mass for fundamental properties such as disk mass \citep{2013ApJ...773..168M,2013ApJ...771..129A}, mass accretion rate \citep[e.g.,][]{2006ApJ...648..484H,2014A&A...561A...2A}, inner disk radius \citep{2002ApJ...579..694M,2007prpl.conf..539M}, dust evolution \citep{2005Sci...310..834A}, and disk chemistry \citep{2009ApJ...696..143P}. Even though the presence of such scaling laws does not necessarily mean the first steps of planet formation are stellar-mass dependent \citep{2012A&A...539A...9M}, the later stages most likely are \citep[e.g.,][]{2007ApJ...669..606R}. One striking trend with spectral type that has not yet received full appreciation is the increasing occurrence rate of $2 - 2.8$ earth radius planets at $<50$ day orbital periods from low-mass M stars to higher-mass F stars reported by \cite{2012ApJS..201...15H}. In order to understand these differences, one needs to take into account that \textit{Kepler} planets are located relatively close to the star, and hence may bear a strong imprint of planet-star interactions -- either during the phase of formation in a protoplanetary disk, or afterwards through tidal interactions -- which may shape the planet distribution in ways different for stars of different masses. 

For a solar mass star, the \textit{Kepler} candidates probe a region where the inner edge of the protoplanetary disk was located, with the co-rotation radius of the gas being located at $\sim 0.05$ to $0.1$ au (see Section \ref{sec:dis}), and the dust sublimation radius at $\sim 0.05$ au for a passive disk \citep{2008ApJ...673L..63P} and $\sim 0.1$ au for an actively accreting one \citep{2011Icar..212..416M}.
Planet occurrence rates may be reduced inside these radii, either by a lack of solid material for in-situ formation \citep{2013MNRAS.431.3444C, Boley:2013ur} or by trapping migrating planets at or outside the co-rotation or sublimation radius \citep{Lin:1996ey,2002ApJ...574L..87K}. For the less massive stars in the Kepler sample, these radii were located much closer to the star -- even during their bright pre-main sequence stage -- and hence a smaller inner disk radius may explain the high \textit{occurrence} of planets. Alternatively, the current distribution of planets may be shaped \textit{after} dissipation of the disk. Tides raised on the star may lead to a spiral-in of close-in planets on gigayear time scales, providing an explanation for the deficiency of hot-Jupiters within 0.05 AU \citep{Jackson:2009cm}. Secular interactions in multi-planet systems may prevent circularization of orbits or excite their eccentricities \citep[e.g.,][]{Correia:2012ea,Greenberg:2013cq} and lead to tidal destruction of planets from farther out \citep{2014MNRAS.443.1451L}. In addition, planets with gaseous envelopes may be partially evaporated \citep{Owen:2013gs}. 

These different mechanisms shape the distribution of planets in various ways, and do so differently for stars of dissimilar masses. Indeed, \cite{Plavchan:2013hl} have confirmed that a scaling law for the location of close-in planets with stellar mass exists, but the authors can not pinpoint the exact mechanism, while \cite{Boley:2013ur} suggest dust sublimation as the most likely origin of the innermost planets in multi-planet systems. We improve on these results by deriving planet occurrence rates as a function of planet radius, orbital period \textit{and} spectral type. We then calculate a median mass for each spectral type bin and show that there is a simple scaling law between the occurrence rate and stellar mass.

In this paper -- the first in a series of two -- we compare planet occurrence rates as a function of spectral type and distance from the star. 
In Section \ref{sec:bins}, we describe how we subdivide our sample into different spectral type bins and assign a median mass to each spectral type.
In Section \ref{sec:occ}, we describe how we convert the observed planet population into planet occurrence rates, taking into account the known biases in the \textit{Kepler} survey and data reduction pipeline, and implement a few improvements for calculating more reliable occurrence rates. In Section \ref{sec:results}, we confirm that the resulting occurrence rates are \textit{significantly} different for stars of different spectral types in the \textit{Kepler} sample, and show that these differences are present at all orbital periods and planet radii. We interpret these trends in terms of processes shaping the inner disk and star-planet interactions, and discuss how they may point the way to discriminating between different planet formation, migration and destruction mechanisms.

\begin{figure}[ht]
	\ifemulate
		\includegraphics[width=\linewidth]{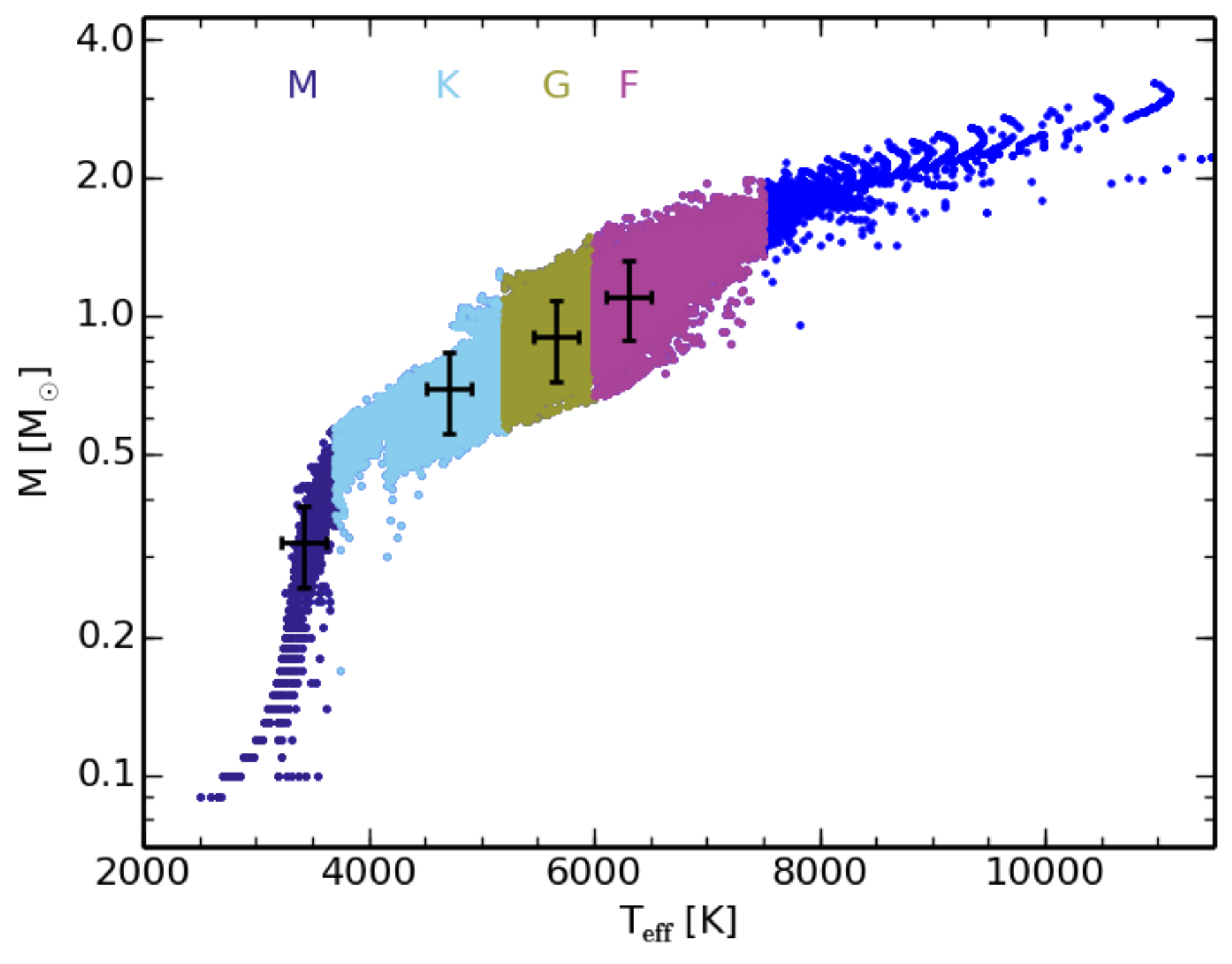}
		\includegraphics[width=\linewidth]{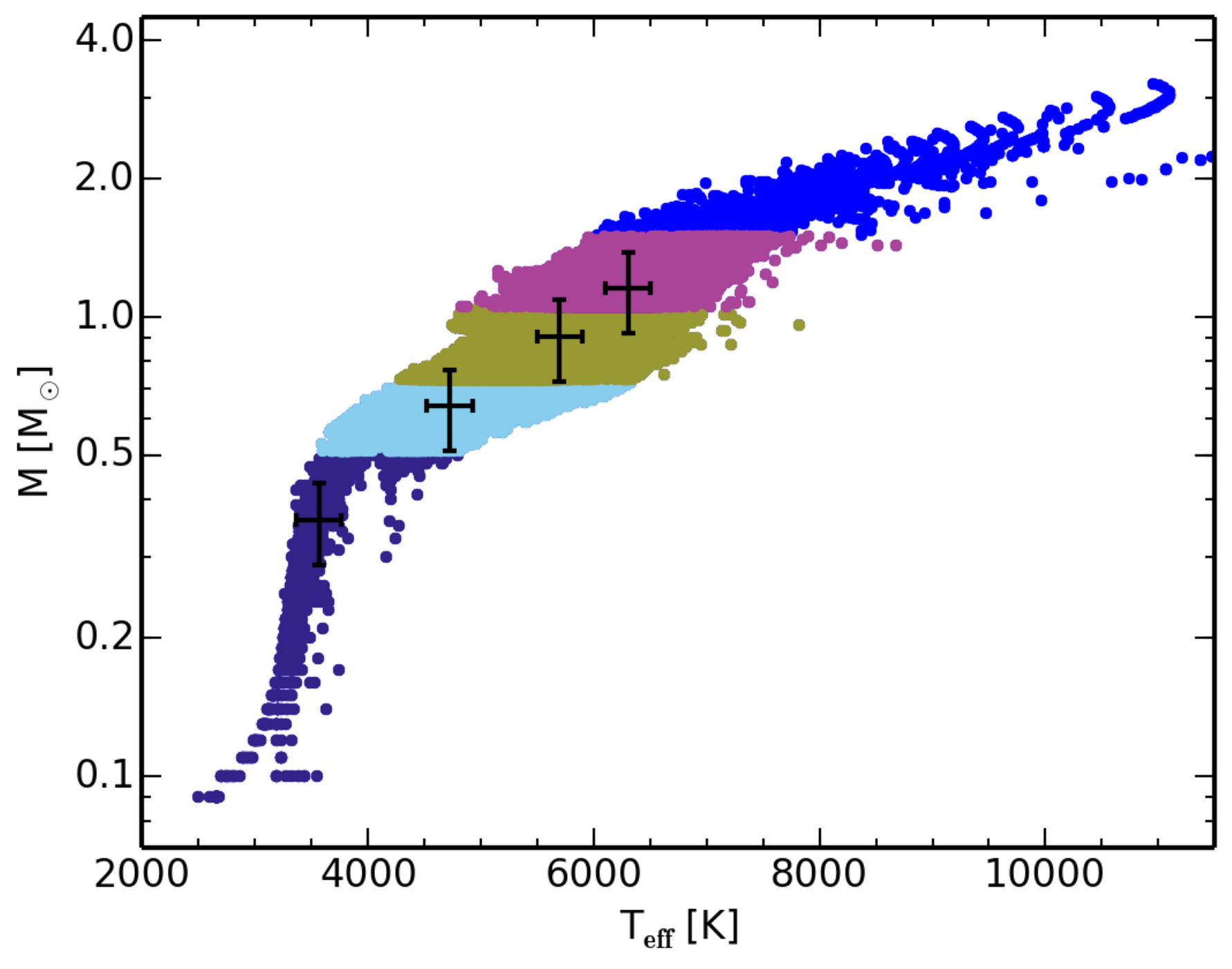}
	\else
		\includegraphics[width=0.7\linewidth]{f1a.pdf}
		\includegraphics[width=0.7\linewidth]{f1b.pdf}
	\fi

		\caption{Mass versus effective temperature of the main sequence stars in the Kepler sample. Different colors represent different bins based on effective temperature (top) and stellar mass (bottom). Black crosses represent typical error bars, and are centered at the median of each bin.
	\label{fig:MTeff}
	}
\end{figure}

\section{The sub-samples of FGKM Kepler stars}\label{sec:bins}

	There are multiple ways of subdividing the Kepler sample, though the most common one is to use effective temperature \citep[e.g.,][]{2012ApJS..201...15H, 2013ApJ...767...95D}. Even if the underlying parameter we are interested in is stellar mass, we will also subdivide the Kepler sample by stellar effective temperature, since, as we will show shortly, the uncertainty on the effective temperature is much smaller than that on stellar mass. 
	
	In general, lower effective temperatures on the main sequence correspond to lower mass stars, though a one-to-one correlation does not exist due to a degeneracy with age and composition. Figure \ref{fig:MTeff} shows the mass and effective temperature for the main sequence stars in the Kepler sample from \cite{Huber:2014dh}, with giants removed according to the prescription in \cite{2011AJ....141..108C}. The \cite{Huber:2014dh} catalogue estimates effective temperature in several ways, but mostly from matching broadband colors to synthetic colors calculated from ATLAS9 model atmospheres \citep{2004astro.ph..5087C}. In this step surface gravity and metallicity are also determined. Typical uncertainties on effective temperature are $\sim3.5\%$ or $200 K$. In a second step, these stellar parameters are used, in combination with the Dartmouth evolutionary models, to estimate stellar masses. Typical uncertainties on stellar masses are as large as $\sim 20\%$.
	
	  Binning in terms of effective temperature (as indicated by colors in top panel of Fig. \ref{fig:MTeff}) rather than stellar mass (bottom panel) results in less stars ending up in adjacent bins due to the size of the error bars. 
	 Hence, \textit{binning in terms of effective temperature creates subsamples that better trace the underlying distribution of stellar masses} than binning in terms of stellar mass that have a larger associated uncertainty.
	
	A natural choice of bin edges is to use spectral type. Table \ref{tab:bins} lists the minimum and maximum effective temperatures that define each bin. The different bins for M, K, G, and F stars have median masses of 0.35, 0.7, 0.9, and 1.1 $M_\odot$, respectively. We have verified that the trends with stellar mass discussed in this paper are also present with bins defined by stellar mass with the same median mass, albeit with lower significance. 
	
\begin{table}
	%\title{}
	\centering
	\begin{tabular}{l l l}\hline\hline
	Spectral type & Effective Temperature [$K$] & Median Mass [$M_\odot$] \\
	\hline
	M & 2400...3700 & 0.35 \\
	K & 3700...5200 & 0.70 \\
	G & 5200...6000 & 0.91 \\
	F & 6000...7500 & 1.08 \\
	\hline\hline\end{tabular}
	\caption{Effective temperatures used to define spectral type bins, and associated stellar masses.}
	\label{tab:bins}
\end{table}

\section{Occurrence rates}\label{sec:occ}
The occurrence rates of planets orbiting other stars can be calculated from the KOIs by taking into account the detection efficiency of \textit{Kepler}. We follow -- conceptually -- the approach of previous studies \citep[e.g.,][]{2012ApJS..201...15H, 2013ApJ...767...95D, 2014ApJ...791...10M}, by simulating the signal of a planet of given size and orbital period \textit{for each star in the \textit{Kepler} field}.
We use the updated stellar dataset from \cite{Huber:2014dh} and the updated planet catalogue from \cite{2014ApJS..210...19B}. Unlike to previous work, we take into account the actual time stars are observed\footnote{For example, about 15\% of the cool stars in 
\cite{2013ApJ...767...95D} have not been observed in the first 6 quarters, underestimating occurrence rates by the same fraction.}, check and correct for mismatches in observed and calculated transit times and signal-to-noise of the transits, implement an additional correction for detection efficiency at long periods, 
calculate upper limits in regions where no planets are detected, and derive planet occurrence rates both as a function of spectral type and for the entire \textit{Kepler} sample. 

This Section is organized as follows:
First, we calculate for each star observed by \textit{Kepler} (\S \ref{sec:stars}) if a planet of given radius and orbital period is detectable (\S \ref{sec:det}) and with what probability (\S \ref{sec:prob}). Comparing these numbers with the observed planetary candidates (\S \ref{sec:planets}) yields the occurrence rates and error bars per period-radius bin, and per spectral type bin (\S \ref{sec:error}) as defined in the previous section. In \S \ref{sec:sma}
we describe how occurrence rates depend on semi-major axis instead of orbital period. 
Table \ref{tab:para} in Appendix \ref{app:para} provides a summary of all the parameters and units used. Appendix \ref{app:occ} provides further details our approach.
The final occurrence rates are shown in Figure \ref{fig:occ}, and in tabular form for different spectral types in Appendix \ref{app:occtable}.

\begin{figure*}[ht]
	\includegraphics[width=\linewidth]{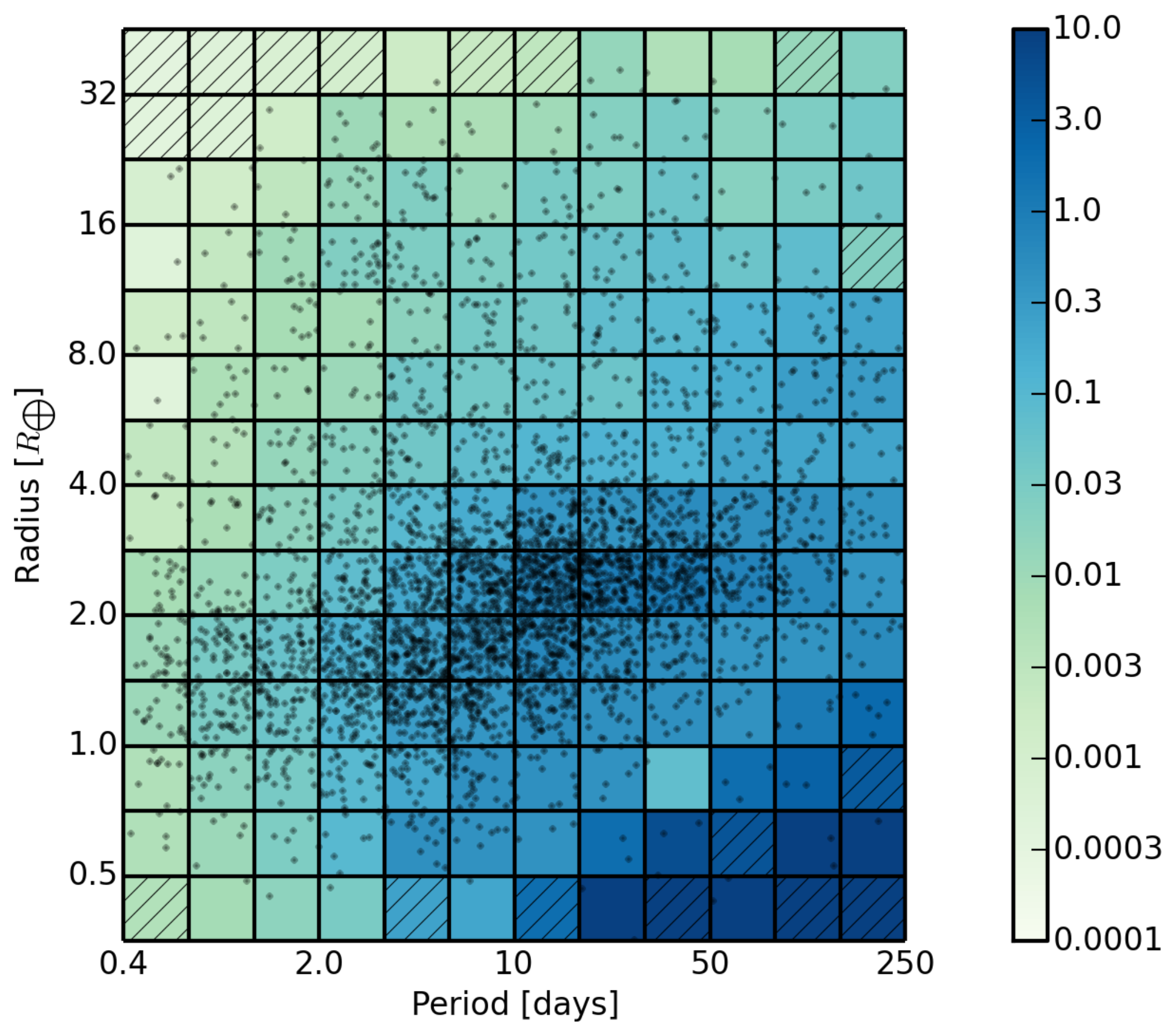}
	\caption{Occurrence rates as a function of orbital period and planet radius. Occurrence rates are calculated by summing the contributions per KOI (dots) per bin, and given per logarithmic area units. Diagonal lines indicate upper limits in areas with no detections. Tabulated occurrence rates are available in appendix \ref{app:occtable}.
	\label{fig:occ}}
\end{figure*}

\subsection{Kepler stars}\label{sec:stars}
We adopt the stellar effective temperature $T_{\rm eff}$,  radius $R_\star$,  mass $M_\star$, and surface gravity $g$ from \cite{Huber:2014dh}, which addressed some of the issues reported in the initial \textit{Kepler Input Catalogue}.
  \cite{Huber:2014dh} also reports in which quarters each star was observed. From this information and the observing time for each quarter (tabulated in the \textit{Kepler} data release notes\footnote{https://archive.stsci.edu/kepler/data\_release.html}) we calculate the total observing time per star $t_{\rm obs}$. We remove giant stars from the sample following the prescription of \cite{2011AJ....141..108C}, which is based on effective temperature and surface gravity. The stellar noise is characterized by the \textit{Combined Differential Photometric Precision} or CDPP \citep{Christiansen:2012bz}, of which we downloaded the December 12th 2013 version from the MAST\footnote{http://archive.stsci.edu/kepler/data\_search/search.php} archive. 

The CDPP is given for 3, 6, and 12 hour periods for each quarter. We describe the stellar noise \cdpp for any transit duration $t$ by fitting a power law with index \cdppindex to the median-combined noise per quarter
\begin{equation}
\label{eq:cdpp}
	\cdpp(t) = \cdppnorm~\left(\frac{t}{\tlc}\right)^{\cdppindex},
\end{equation}	
where we normalize the noise (\cdppnorm) at the shortest period possible period in the long cadence, 1765.5 seconds  (\tlc). Additional details of this approach, in particular the non-Gaussianity of the noise ($\cdppindex \neq -0.5$), are described in Appendix \ref{app:noise}).  After removing stars without a measurement of the CDPP during the first 8 quarters (see below), we are left with a sample of 162,270 stars.

\subsection{Planetary candidates}\label{sec:planets}
We use the list of planetary candidates in \cite{2014ApJS..210...19B}, which presents a set of candidates detected in the first 8 quarters, as well as a re-evaluation of the \cite{Borucki:2011cp} candidates. Transit parameters of the new KOIs were derived using the first \textit{10} quarters of \textit{Kepler} photometry, while KOIs detected in the first 6 quarters have transit parameters determined from 8 quarters \citep{2013ApJS..204...24B}. Though this sample is not meant to be statistically complete, it does currently present the largest uniform sample, since one can assume the KOIs from the first 6 quarters could also have been detected in 8 quarters. Hence, we use only the first 8 quarters for calculating the detection efficiency of each star.

We note that KOIs are available up to Q12 and Q16, but are only partially released to the community. They do not form a uniform sample at the time of submission, which is why we choose to not include them, as it is not clear what the current biases in the Q12 and Q16 lists are. 
After removing single transits and KOIs not matching any stars in our sample, we identify 3,731 planet candidates. This is the list of KOIs we will be using in this paper.

\subsection{Number of stars with a detectable planet}\label{sec:det}
Whether a planet of given radius $R_p$ and orbital period $P$ can be detected around a star depends on the achieved signal-to-noise ratio \snr \textit{for that star}, which is a function of 
transit depth $\delta = (R_p/R_\star)^2$, noise level \cdpp during the transit duration \tdur, and the number of transits $n$.
\begin{equation}
\label{eq:snr}
\snr= \frac{\delta ~ n^{0.5} }{\cdpp (\tdur)}
\end{equation}
Note that, even though the reported noise on 3-12 hour time scales does not appear to be Poissonian for all stars (Appendix \ref{app:noise}), we assume the signal-to-noise still increases with the square root of the number of transits. We demonstrate in Appendix \ref{app:snr} that this  assumption provides the smallest deviations in the signal-to-noise ratios between the calculated values (\snr) and the measured values tabulated in \cite{2014ApJS..210...19B}, \snrobs.
We account for a systematic offset between these two values \citep[see also][]{2014ApJ...791...10M} by multiplying the calculated values from Eq. \ref{eq:snr} by a factor of 1.33 (see Appendix \ref{app:noise} for details).

The transit duration time \tdur is given by:
	\begin{equation}
		\tdur = \frac{PM_\star \sqrt{1-e^2}}{\pi a},
		\label{eq:tdur_used}
	\end{equation}
where $e$ is the orbital eccentricity and $a$ is the semi-major axis given by:
	\begin{equation}
		a= \sqrt[3]{\frac{G M_\star P^2}{4 \pi^2}}
		\label{eq:sma}
	\end{equation}
Using this approach, we have corrected for increased probability of shorter transits in eccentric orbits \citep{2008ApJ...679.1566B}, but not for impact parameter $b$ to account for the \textit{transit duration anomaly} \citep{Plavchan:2012tr}, which we motivate in Appendix \ref{app:tdur}. Unfortunately orbital eccentricities are not available for most planets, but for small eccentricities ($e<0.3$), the deviations in transit duration are less then 5\%. Hence, our results are not influenced by the choice of eccentricity and we assume an average eccentricity of $e=0.1$. This value falls in the range of eccentricities (0.1-0.25) derived by \cite{Moorhead:2011dn}. 

The number of transits $n$ for a given period $P$ is calculated for each individual star from the total observing time: $n= t_{\rm obs}/P$. Since a potential planetary candidate requires three transits to be detected, the detection efficiency $f_n$ increases from zero to one between 2 and 3 potential transits. Following \cite{2013ApJS..204...24B}, we use:

\begin{equation}
\label{eq:fperiod}
\begin{split}
t_{\rm obs} \leq 2 P:\, & f_n= 0 \\
2 P < t_{\rm obs} < 3 P:\, & f_n= (t_{\rm obs}/P -2) \\
t_{\rm obs} \geq 3 P:\, & f_n= 1 \\
\end{split}
\end{equation}

The \textit{Kepler} pipeline considers transits with a signal-to-noise ratio of at least 7.1 a detection. However, the vetting procedure to rule out false positives also removes some bona-fide transits close to the detection limit \citep{2013ApJS..206...25S}, leading to a modified detection efficiency \feff given by a linear ramp between a signal-to-noise of 6 and 12 \citep{2013ApJ...766...81F, 2014ApJ...791...10M}:

	\begin{equation}
	\begin{split}
		\label{eq:feff}
		\snr \leq 6:\, & \feff =0 \\
		6 < \snr \leq 12:\, & \feff = \frac{\snr - 6}{6}  \\
		\snr > 12:\, & \feff = 1 \\
	\end{split}
	\end{equation}
	
The number of stars $N_\star$ in a given spectral type bin $\{T_{\rm eff}\}$ as defined in Section \ref{sec:bins}, around which a planet with given radius and orbital period $\{R_p,P\}$ can be detected based on its signal-to-noise, is then
	\begin{equation}\label{eq:Nstars}
		N_\star(\{T_{\rm eff}\},R_p,P)= \Sigma_{i=0}^{N_\star(\{T_{\rm eff}\})} ~ (\feffi \cdot f_{n,i}),
	\end{equation}
where we round $N_\star$ to the nearest integer.
	
\subsection{Transit probability}\label{sec:prob}
	When calculating occurrence rates one also needs to take into account the geometric transit probability. This probability is given by:
	\begin{equation}
		\label{eq:fgeo}
		\fgeo = \frac{R_p+R_\star}{a(1-e^2)},
	\end{equation}
	where $(1-e^2)$ is a correction factor to take into account the increased transit probability of a planet on an eccentric orbit \citep{2008ApJ...679.1566B}.

\subsection{Occurrence rates and error bars}\label{sec:error}

	The occurrence rate for a planet of given radius and orbital period $\{R_p,P\}$ is 
		\begin{equation}\label{eq:focc}
			\focc(\{T_{\rm eff}\},R_p,P) = \frac{1}{\fgeo~N_\star(\{T_{\rm eff}\},R_p,P)}.
		\end{equation}

	Figure \ref{fig:occ} shows the occurrence rate of the entire sample in the style of \cite{2012ApJS..201...15H} -- but on an extended grid. Occurrence rates per grid cell are calculated by adding the occurrence rate contributions from individual KOIs in that cell, and scaled to a logarithmic area unit. The 1-$\sigma$ errors per grid cell are given by the 16th and 84th percentile of the binomial probability distribution for drawing the detected number of planets out of $N_\star$, where the detection efficiency and probabilities are calculated for a planet at the logarithmic center of the cell, analogous to \cite{2013ApJ...767...95D}. For bins without planets, 1-$\sigma$ upper limits are given by the occurrence rate of a (hypothetical) planet at the logarithmic center of the cell.
	
	The occurrence rates we derive per grid cell are -- within errors -- consistent with those in \cite{2012ApJS..201...15H} for planet radii larger than 2.8 $R_\oplus$. On average, our rates are slightly higher, reflecting the increased detection efficiency of KOIs in \cite{2014ApJS..210...19B}. The deviation at small planet radii arises mainly because \cite{2012ApJS..201...15H} assume their sample was complete at ${\rm SNR} > 10$, while in reality incompleteness is an issue up to much higher SNR \citep{2013ApJ...766...81F}. We are able to reproduce their occurrence rates down to $2 R_\oplus$ if we use a step function rather than the linear ramp in detection efficiency. We have also benchmarked our occurrence rate calculation against the results in \cite{2013ApJ...767...95D}. We are able to reproduce the numbers in their Figure 15 when using their list of stars and KOIs, and by making the same assumptions on detection efficiency.
	
	\subsection{Occurrence rate as a function of semi-major axis}\label{sec:sma}
	When calculating occurrence rates as a function of a parameter other than orbital period, such as semi-major axis, one has to take into account that the orbital period used in calculating the occurrence rate from equations \ref{eq:cdpp} to \ref{eq:focc} is different for each star.  When calculating the occurrence rate as a function of semi-major-axis $a$, the period carries a dependence on stellar mass: 
	\begin{equation}\label{eq:period}
	P= \sqrt{\frac{4\pi^2 a^3}{GM_\star}}
	\end{equation}
	replacing equation \ref{eq:sma}, and $P$ has to be replaced by $a$ in equations \ref{eq:Nstars} and \ref{eq:focc}.

\section{results}\label{sec:results}

\begin{figure*}[ht]
	\includegraphics[width=\linewidth]{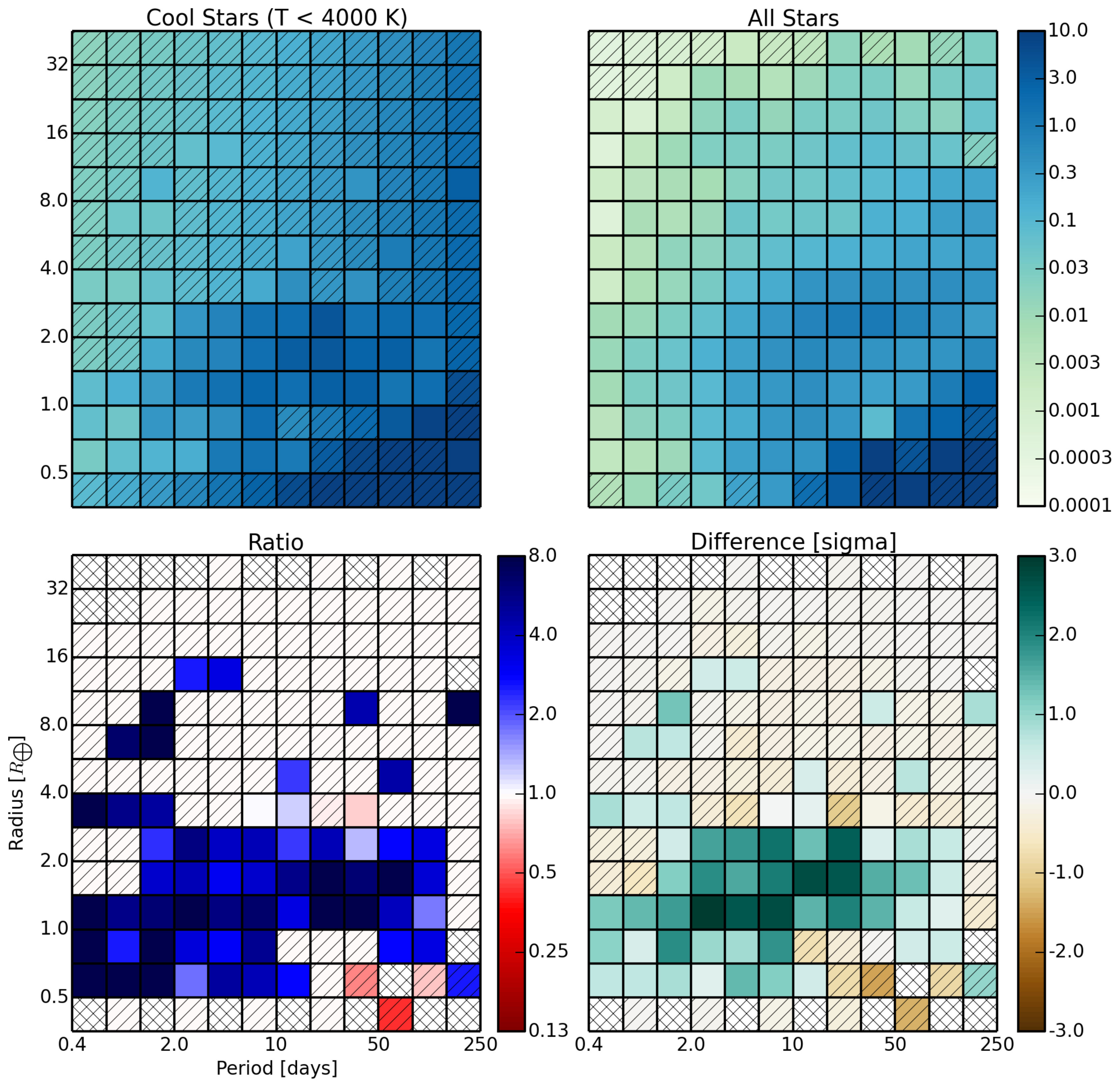}
	\caption{Differences between occurrence rates of the cool star subsample as defined in \cite{2013ApJ...767...95D} ($\Teff <4000 K$) and the entire \textit{Kepler sample}. Top panels: Occurrence rates for cool stars (left, hereafter $f_1$) and all stars (right, hereafter $f_2$) on the same scale as figure \ref{fig:occ}. Bottom left panel: ratio between the top panels, defined as $f_1 / f_2$. Bottom right panel: difference between top panels expressed in significance of the result: $(f_1-f_2)/\sqrt{\sigma_1^2 + \sigma_2^2}$. 
	Hatched diagonals indicate one upper limit on the occurrence rate was used in the comparison, hatched crosses indicate no comparison could be made between two upper limits.
	\label{fig:diag}}
\end{figure*}

\begin{figure}[ht]
	\ifemulate
		\includegraphics[width=0.99\linewidth]{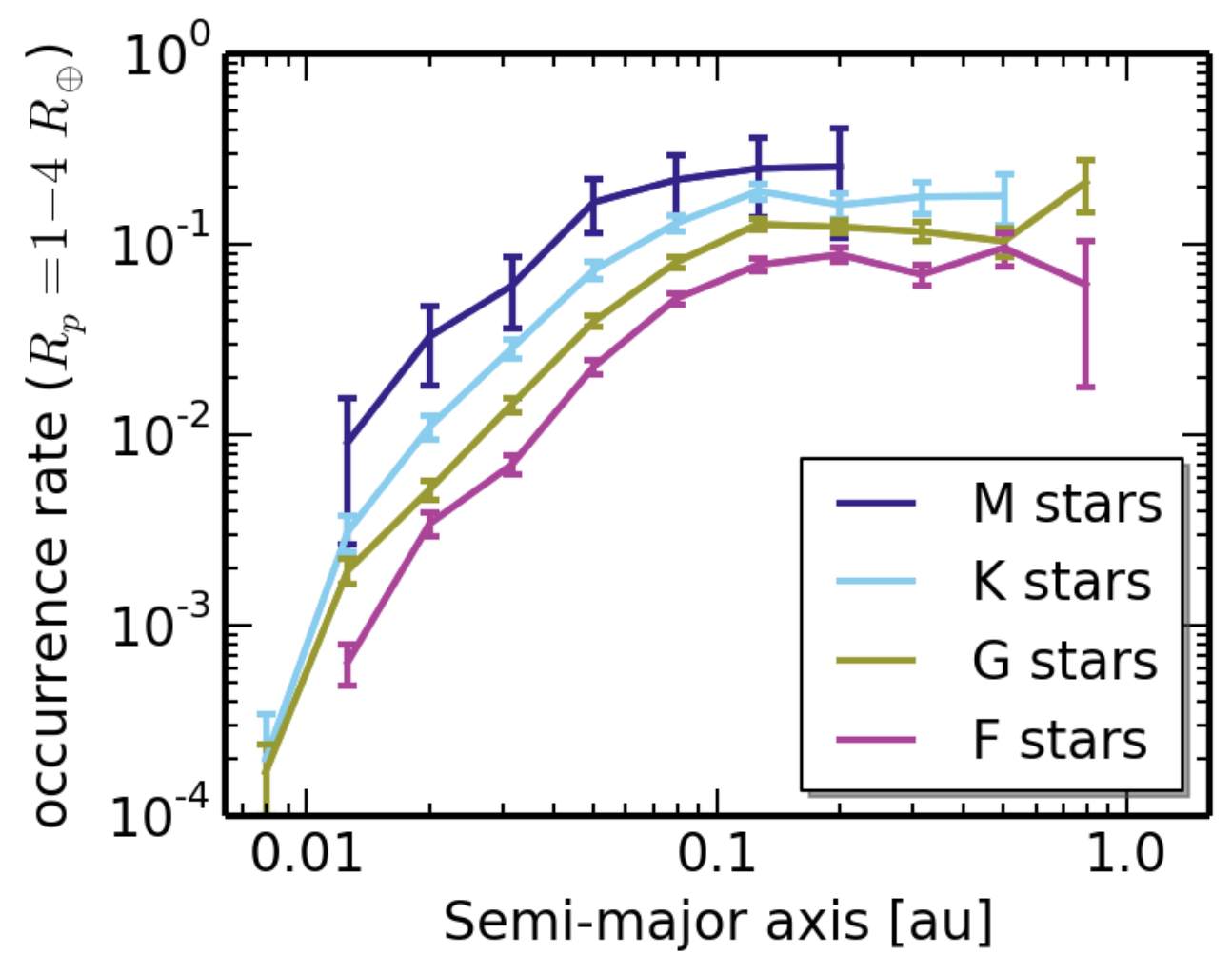}\\
		\includegraphics[width=0.99\linewidth]{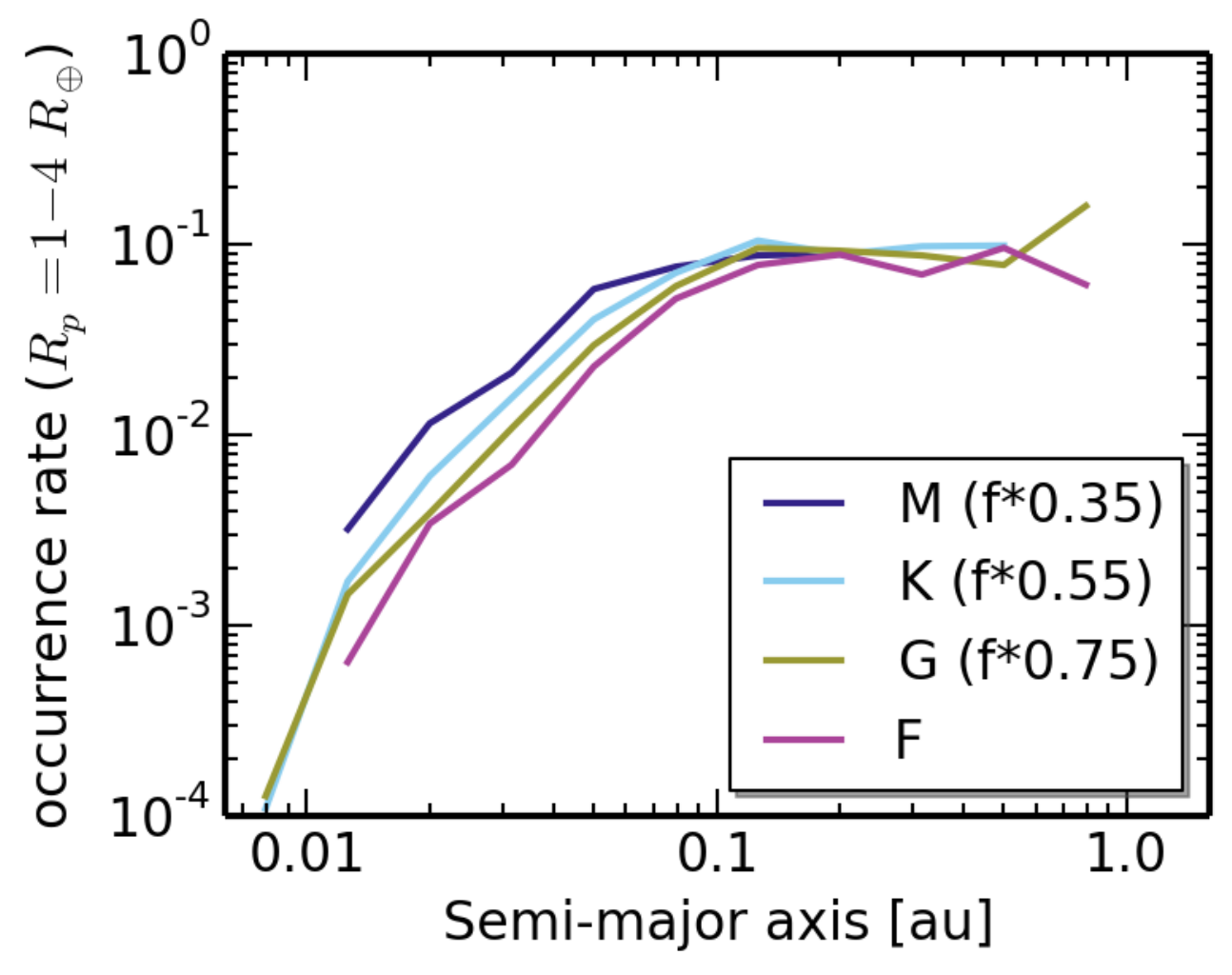}\\
		\includegraphics[width=0.99\linewidth]{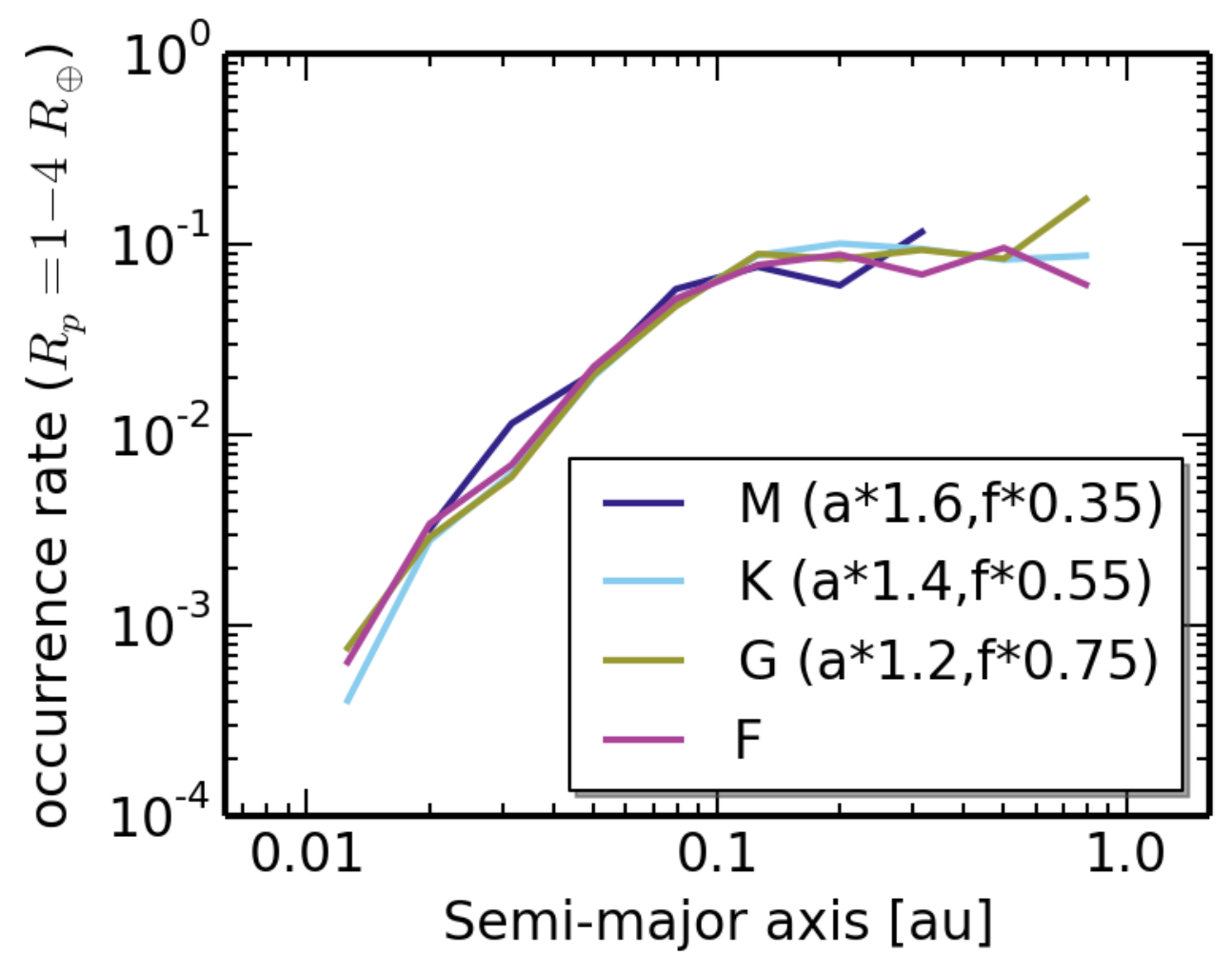}
	\else
		\includegraphics[width=0.49\linewidth]{f4a.pdf}\\
		\includegraphics[width=0.49\linewidth]{f4b.pdf}\\
		\includegraphics[width=0.49\linewidth]{f4c.pdf}
	\fi

	\caption{Occurrence rate as a function of distance from the star, for spectral types M to F. Error bars in the top panel are given by the square root of planets in each bin. Occurrence rates in the middle panel are calculated by multiplying the occurrence rate $f$ \textit{of each planet} by the number in brackets. The bottom panel is calculated by also multiplying the semi-major axis $a$ \textit{of each planet} by the factor in brackets.
	\label{fig:radial-sma}
	}
\end{figure}

	Figure \ref{fig:diag} compares the planet occurrence rate for the cool star subsample as defined in \cite{2013ApJ...767...95D} ($\Teff <4000 K$, median mass $0.47 M_\odot$) with that of the entire Kepler sample (median mass $0.95 M_\odot$). The bottom left panel shows the ratio per grid cell, where blue means a higher occurrence rate for cool stars and red a lower one. The bottom right panel shows these differences are significant even at the level of individual grid cells, especially between one to four Earth radii and one to fifty day periods. To interpret and quantify these differences, we calculate the occurrence rates of planets between one and four earth radii in size as a function of \textit{semi-major axis}\footnote{Note that the orbital period $P$ carries a dependence on semi-major axis, see \S \ref{sec:sma}.} for the spectral type bins defined in \S \ref{sec:bins} (See Fig. \ref{fig:radial-sma}) These rates are different at the 4.2, 9.3, and 11.6 sigma level with G stars for M, K, and F stars respectively. The lower significance toward later spectral types is mainly a result of the smaller sample size.

	The curves appear self-similar (bottom panel), with a plateau in occurrence rates and a steep decrease toward the star. The location of the cutoff $a_{\rm cut}$ falls between $\sim 0.05 \, {\rm au}$ to $\sim 0.1 \, {\rm au}$ for M to F stars, respectively. Inside the cutoff, the occurrence rate is best described by a power law of index $2$.
Different authors have found different slopes for the occurrence rates outside the cutoff:
Earlier results based on fewer KOIs and without incompleteness corrections at low signal-to noise, found a decay \citep{Youdin:2011gz,Catanzarite:2011cu,2012ApJ...745...20T}, whereas more recent results find a flat \citep{2013ApJ...778...53D,Silburt:2014tf}, to slightly rising \citep{2013ApJ...770...69P} slope, though the latter result is consistent within errors with a flat slope.

To match the plateaus to the F stars we apply scaling factors to the overall occurrence rate of 0.35, 0.55 and 0.75 for the M, K, and G stars, respectively. This indicates that the occurrence rate increase for $2$-$2.8 R_\oplus$ planets identified by \cite{2012ApJS..201...15H} extends down to earth-sized planets included in our $1$-$4 R_\oplus$ bin. However, a difference in cutoff location between bins remains present at the $2.4$ (M-G), $5.4$ (K-G) and $5.9$ (F-G) sigma level (middle panel of Figure \ref{fig:radial-sma}). The curves can be matched to that of the F stars by shifting the semi-major axis by a factor 1.6, 1.4, and 1.2 for M, K, and G stars, respectively. These factors are progressively larger for later spectral types -- and hence lower stellar masses -- confirming that the planet population extends closer in to lower mass stars, as already noted by \cite{Plavchan:2012tr}.

\subsection{Truncation mechanisms}\label{sec:mech}

\begin{figure*}
	\centering
	\ifemulate
		\includegraphics[width=0.4\linewidth]{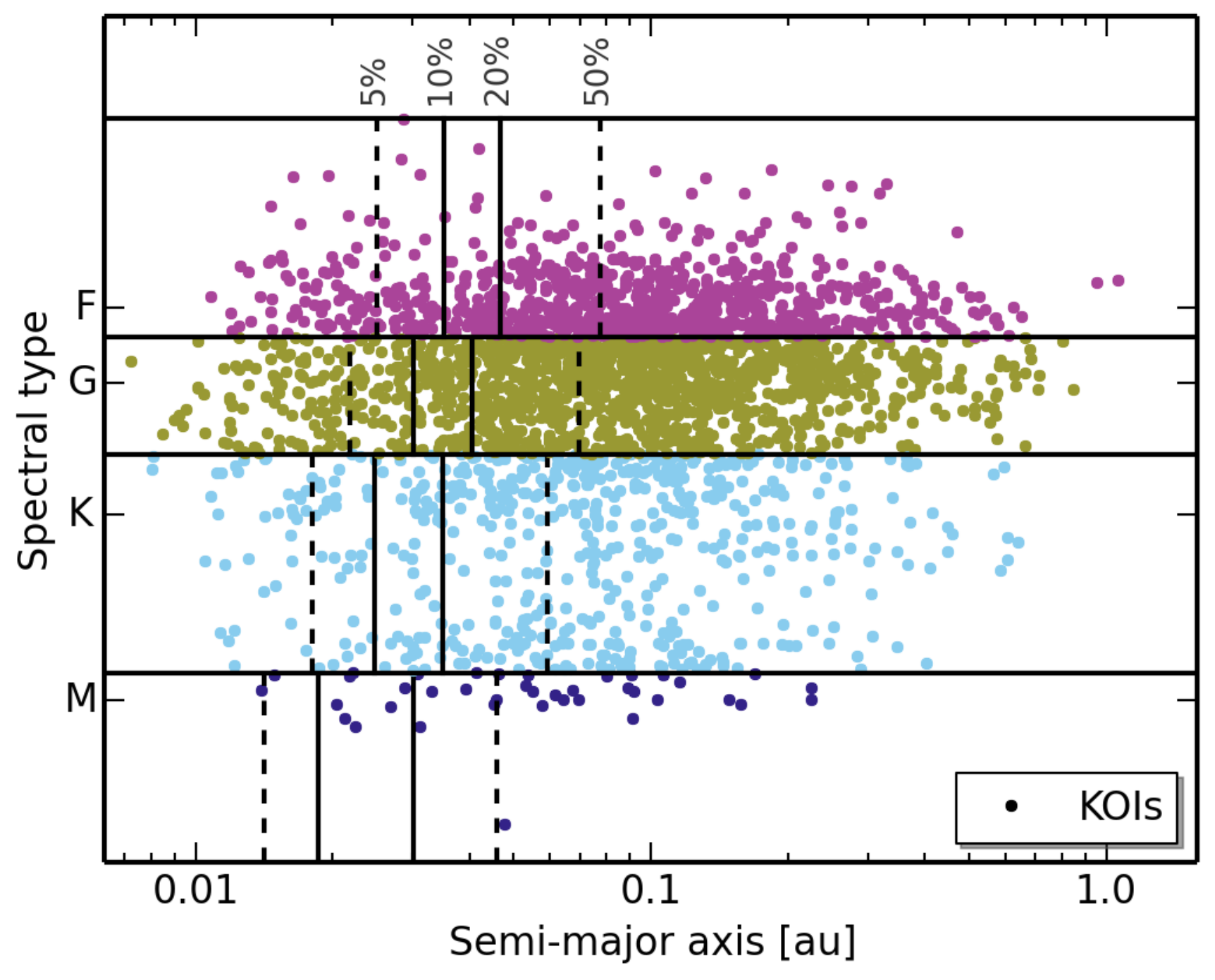}
		\includegraphics[width=0.4\linewidth]{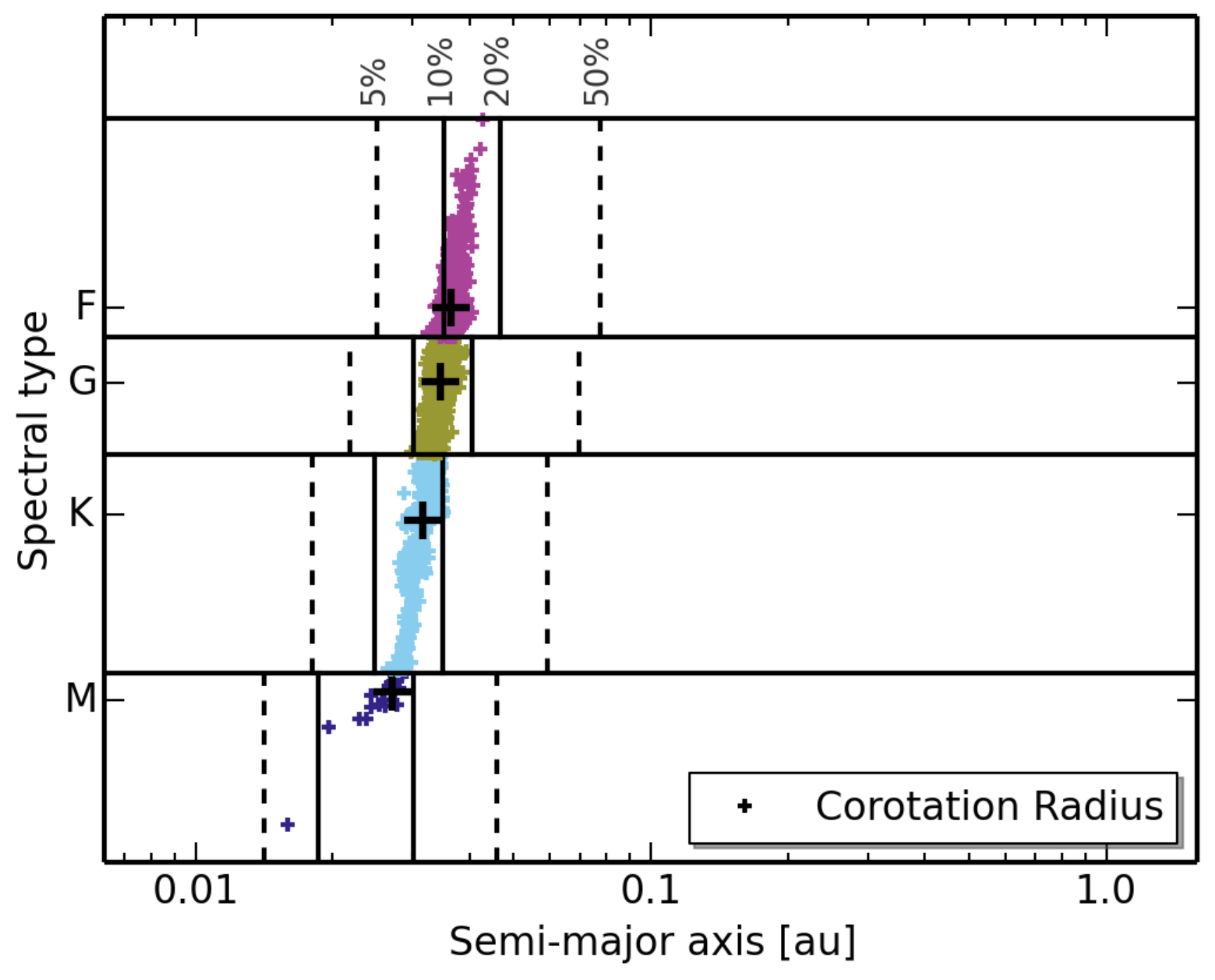}\\
		\includegraphics[width=0.4\linewidth]{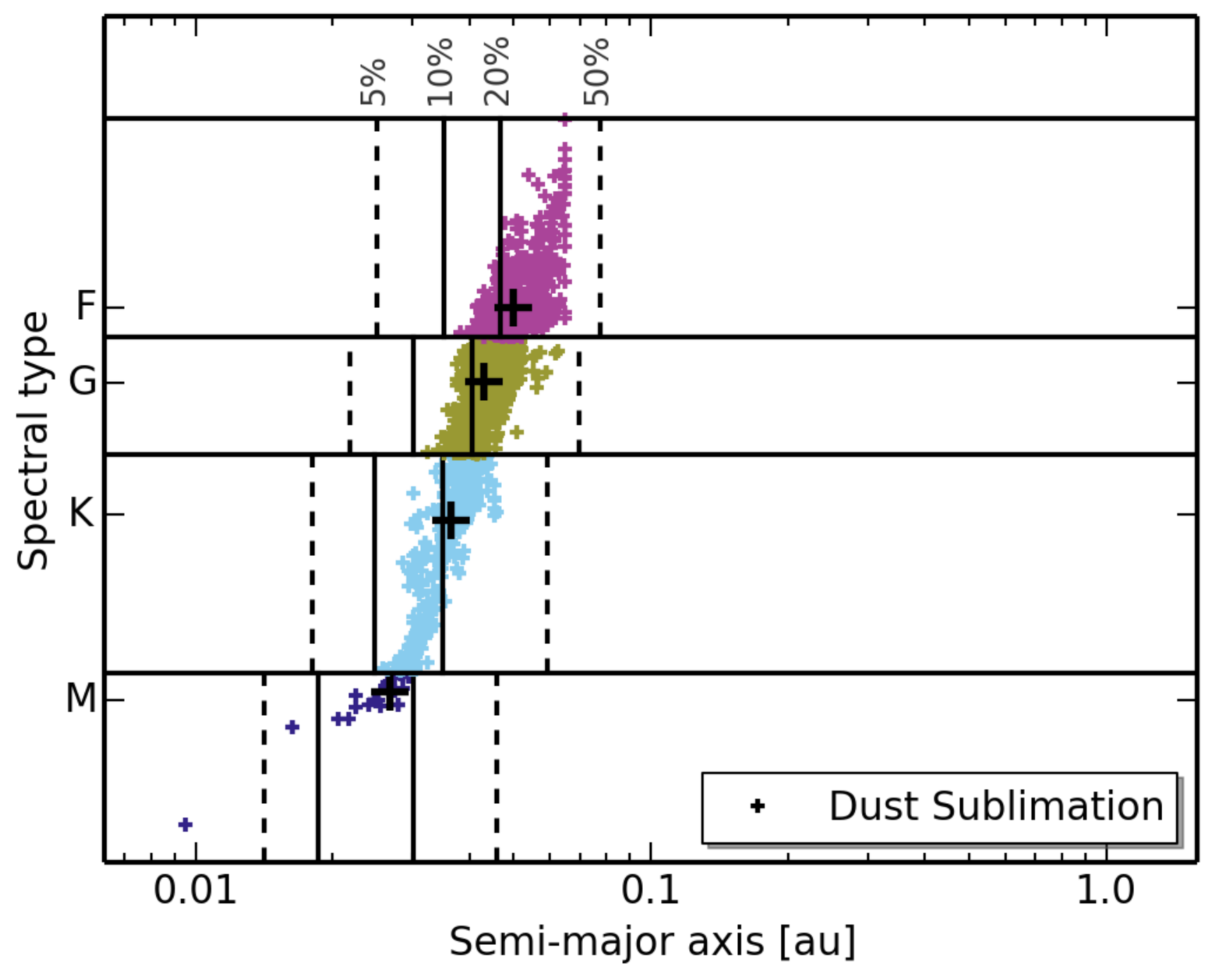}
		\includegraphics[width=0.4\linewidth]{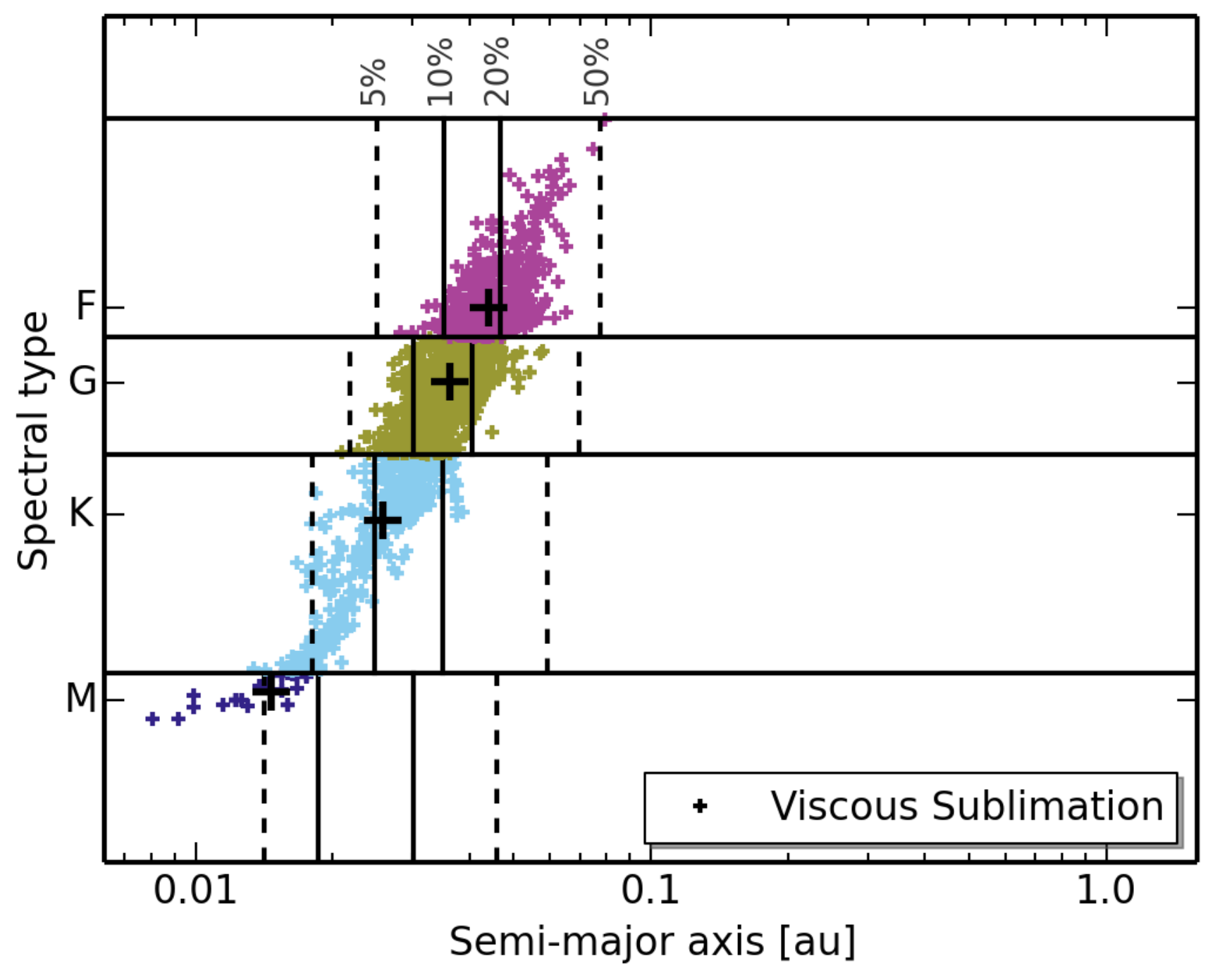}\\
		\includegraphics[width=0.4\linewidth]{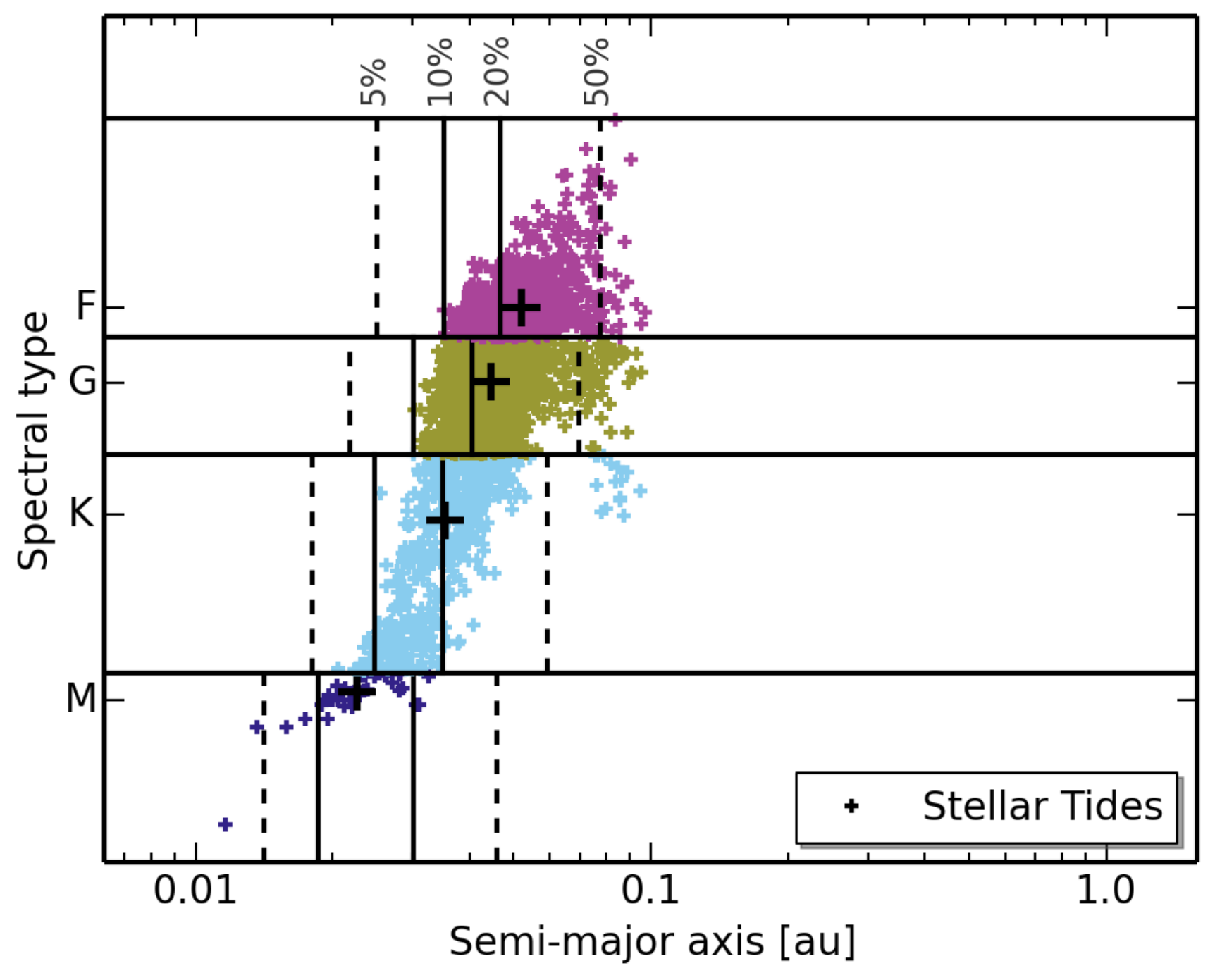}
		\includegraphics[width=0.4\linewidth]{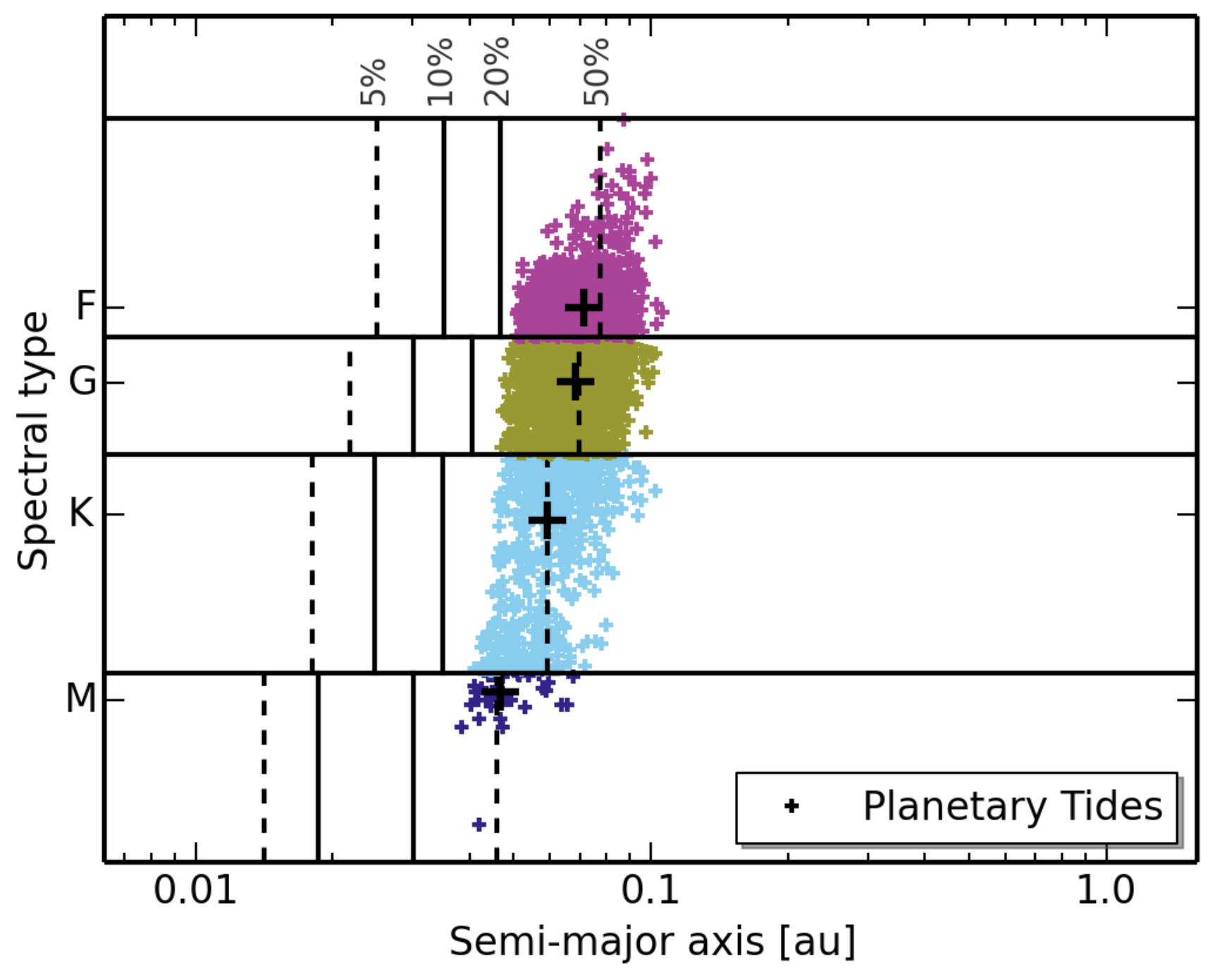}\\
	\else
		\includegraphics[width=0.34\linewidth]{f5a.pdf}
		\includegraphics[width=0.34\linewidth]{f5b.pdf}\\
		\includegraphics[width=0.34\linewidth]{f5c.pdf}
		\includegraphics[width=0.34\linewidth]{f5d.pdf}\\
		\includegraphics[width=0.34\linewidth]{f5e.pdf}
		\includegraphics[width=0.34\linewidth]{f5f.pdf}\\
	\fi
		\caption{The location of a cutoff in planet occurrence rate for different mechanisms in different spectral type bins. Colored crosses indicate this location for each individual KOI, while thick black crosses indicate the median location in each bin. Vertical lines indicate where the observed occurrence rates drop by a factor of 20, 10, 5 and 2 (dashed, solid, solid, dashed, respectively) with respect to the plateau. Horizontal lines indicate transitions between spectral type bins. 
	From left to right, top to bottom: 
	Kepler objects of interest as function of effective temperature, for reference; Pre-main-sequence (PMS) co-rotation radius (Eq. \ref{eq:corot}); PMS dust sublimation radius of a passive disk (Eq. \ref{eq:sub}); PMS dust sublimation radius of an actively accreting disk with $q=2$ (Eq. \ref{eq:subvis}); Planet destruction by stellar tides (Eq. \ref{eq:startides}); Destruction by planetary tides maintained through secular interactions (Eq. \ref{eq:planettides}).
	\label{fig:mech}
	}
\end{figure*}

To identify which mechanism might be responsible for setting the steep decrease in planet occurrence rate toward the star, we estimate the location where a cutoff would occur as a function of stellar mass for a range of processes. These mechanisms, described in the introduction, include: inhibiting planet formation inside a disk inner edge; trapping migrating planets at that edge; or removing planets by tidal interactions. 

In Figure \ref{fig:mech}, we show how these processes match the observed turnover in planet occurrence rate, each colored cross representing the calculated turnover location for each KOI based on the stellar mass, radius and effective temperature in \cite{Huber:2014dh}. 
To guide the eye, we have indicated the location where the occurrence rates are reduced by 5, 10, 20, and 50\% with respect to the plateau for different spectral type bins. The median truncation location per bin is shown by a thick black cross. 
These mechanisms include additional free parameters that can be tuned to match the exact location of the cutoff, and the purpose of this plot is mainly to show how the cutoff radius scales with the median mass per spectral type bin. The power-law index of each scaling law with stellar mass is also given in Table \ref{tab:turnover}. 

Though in principle the index of the power law inside of $a_{\rm cut}$ contains information on the truncation process, we did not attempt a full forward modeling to explain the shape of this curve, but simply note that it may reflect either a spread in initial disk or observed stellar parameters that define the location of the turnover, or may be a result of long-term dynamical evolution of the planet population such as planet-planet scattering. Our approach is very similar to that of \cite{Plavchan:2012tr}, though it is independent of the functional form chosen to describe the radial planet occurrence rates, leading to a different stellar-mass scaling for $a_{\rm cut}$, which is discussed in detail in \S \ref{sec:plavchan}.

\begin{table}
	%\title{}
	\centering
	\begin{tabular}{l l l l}\hline\hline
	Mechanism & Location & Assumption & Eq. \\
	\hline
	co-rotation		& $M_\star^{1/3}$ 	&  					& \ref{eq:corot}	\\
	sublimation 	& $M_\star^{0.7}$ 	& $L_{\rm PMS} \propto M_\star^{1.4} $ 	& \ref{eq:sub} \\
				& $M_\star^{11/9}$ 	& $q=2$ 				& \ref{eq:subvis}	\\
				& $M_\star^{7/9}$ 	& $q=1$ 				& \ref{eq:subvis}	\\
	tides (stellar)	& $M_\star^{9/13}$ 	& $R_\star \propto M_\star $ 		& \ref{eq:startides} \\
		(planetary)	& $M_\star^{3/13}$ 	&  					& \ref{eq:planettides} \\
	\hline\hline\end{tabular}
	\caption{Power-law index of turnover radius as function of stellar mass for the different truncation mechanisms discussed in \S \ref{sec:mech}.
	The third column lists any additional assumption required for turning the listed equation in a pure stellar-mass dependency. The pre-main-sequence luminosity scaling is based on the \cite{1998A&A...337..403B} evolutionary tracks at 1 Myr.}
	\label{tab:turnover}
\end{table} 

Protoplanetary disks are truncated at or around the co-rotation radius, from where material is funneled onto the star. This radius is defined as the location where the stars angular velocity $\Omega$ equals the Keplerian frequency:
\begin{equation}
\label{eq:corot}
a_{\rm corot}= \sqrt[3]{\frac{G M_\star}{\Omega_{\rm PMS}^2}},
\end{equation}
where $\Omega_{\rm PMS}$ is the angular velocity of the star. The angular velocity is independent of stellar mass during the protoplanetary disk lifetime ($\Omega_{\rm PMS} \sim 10 \Omega_\odot$), albeit with large scatter \citep{2007IAUS..243..231B}. This scatter results in a range of co-rotation radii for any given stellar mass, but we will use the average value for the purpose of showing how the cutoff location is different in each spectral type bin, as illustrated in the top right panel of Figure \ref{fig:mech}. We note that the co-rotation radius from the minimum and maximum angular velocity ($\Omega_{\rm PMS} \sim 2-30  \Omega_\odot$) bracket the factor 2 and 20 drop in occurrence rate, respectively.

The dust sublimation radius for a passive, illuminated disk is located at:
\begin{equation}
\label{eq:sub}
a_{\rm sub}= a_0 \sqrt{L_{\rm PMS}/L_\odot}
\end{equation}
where the pre-main-sequence luminosity $L_{\rm PMS}$ is calculated from the \cite{1998A&A...337..403B} evolutionary tracks at an age of 1 Myr, and $a_0$ is a normalization factor containing all uncertainties in determining the dust sublimation radius (including age). More sophisticated formulas for calculating the dust sublimation radius are available \citep[e.g.,][]{2009A&A...506.1199K}, though these do not significantly alter the stellar-mass-dependence. Picking a later age (up to 10 Myr) results in smaller hole sizes, but does not affect the luminosity-mass dependence in the given mass-range, which is of order $L_{\rm PMS} \propto M_\star^{1.5}$.  The exact value of the normalization factor is only instrumental in determining the stellar-mass-dependency, and we take it to be $a_0 = 0.04 ~ {\rm au}$, to be consistent with the dust hole inferred from interferometric observations of disks around low-mass stars \citep{2008ApJ...673L..63P}. 

We also explore dust sublimation by viscous heating \textit{within} the disk, as typical mid-plane temperatures of a low-mass pre-main sequence star within 1 au are not dominated by stellar heating \citep{1998ApJ...500..411D}. Using the analytic expression based on detailed radiative transfer models from \cite{2011Icar..212..416M}, the dust\footnote{The physics of dust and ice sublimation are essentially the same, and we have checked with the original model setup that this expression also applies to dust sublimation.} sublimation radius scales as:
\begin{equation}
\label{eq:subvis}
a_{\rm sub}= a_1 \left(\frac{M_\star}{M_\odot}\right)^{ (3+4q) / 9}
\end{equation}
where $q$ is the power-law index of the scaling between mass-accretion rate and stellar mass  ($\dot{M} \propto M_\star^q$), which is observed to be $q \sim 2$ \citep{2014A&A...561A...2A}, and $a_1$ is a normalization factor that we take be $a_1=a_0$. This dependence is shown in Fig. \ref{fig:mech} second row right panel.

A drop in the number of close-in exoplanets has been noted before Kepler and attributed to tides raised on a planet's host star by \cite{Jackson:2009cm}. The authors assume constant planet formation efficiency and subsequent removal of close-in planets by dissipation of tidal energy over gigayear time scales, which results in a planet distribution as a function of semi-major axis (Fig. 4 in \citealt{Jackson:2009cm}) closely resembling our occurrence rate plots in Figure \ref{fig:radial-sma}. The location of the turnover is given by equation 1 in \cite{Jackson:2009cm}:
\begin{equation}
\label{eq:tides}
\begin{split}
\frac{1}{a} \frac{{\rm d} a}{{\rm d} t} = & -\left[   \frac{63}{2} \sqrt{G M_\star^3} \frac{R_p^5}{Q_p M_p} e^2 + \right. \\ 
 & \left. \frac{9}{2} \sqrt{G / M_\star} \frac{R_\star^5 M_p}{Q_\star} \left( 1+\frac{57}{4} e^2 \right) \vphantom{\frac{57}{4}} \right] a^{-13/2} \\
 = & \, {\rm  const} = - a_2^{-13/2}
\end{split}
\end{equation}
where $Q_\star$ and $Q_p$ are the modified tidal dissipation parameters for the star and planet, respectively, which we take, as in previous work of \cite{Jackson:2009cm}, to be $Q_\star=10^6$ and $Q_p = 10^3$. For tidally circularized orbits ($e \sim 0$), stellar tides dominate, and equation \ref{eq:tides} simplifies to:
	\begin{equation}
	\label{eq:startides}
	\begin{split}
	a_{{\rm tides},*} = &  a_2  \left(  \frac{9}{2} \sqrt{G / M_\star} \frac{R_\star^5 M_p}{Q_\star} \right)^{2/13},
	\end{split}
	\end{equation}
	where $a_2$ is a constant that we take to be  $a_2 = 8 \cdot 10^{-11} s^{2/13}$, such that truncation occurs at 0.05 AU for a G type star, as in \cite{Jackson:2009cm}. 
	
	If eccentricities can be maintained, for example through secular interactions in a multi-planet system, planetary tides -- the first term in equation \ref{eq:tides} -- dominate even for small eccentricities ($e \sim 0.01$), and the tidal truncation radius lies further in, at
	\begin{equation}
	\label{eq:planettides}
	\begin{split}
	a_{{\rm tides},p} =  &a_2  \left( \frac{63}{2} \sqrt{G M_\star^3} \frac{R_p^5}{Q_p M_p} \right)^{2/13}
	\end{split}
	\end{equation}
	There are significant uncertainties in the location of the tidal truncation radius, due to uncertainties in the tidal parameter $Q$ and orbital eccentricities, though these can be mitigated through a different choice of the normalization parameter $a_2$. Hence, the tidal truncation radius shown in Figure \ref{fig:mech} can move significantly inward or outward to match the observed truncation radius by a different choice of parameters, but these uncertainties do not affect the stellar-mass dependency that is of interest here.
	
We do not consider photo-evaporation \citep{Owen:2013gs}, as this process does not reduce the number of planets as long as the planet detection limit is above the core mass, which is always the case for the periods of interest here. 

The pre-main-sequence co-rotation radius best matches the location of the stellar-mass dependent turnover, closely followed by planetary tides and the dust sublimation radius of a passive disk. Stellar tides, as well as dust sublimation in a viscous disk, provide a poor fit to the observations. Although this list of processes is not comprehensive, it should be kept in mind that any model invoked to explain the population of \textit{Kepler} planets must be able to reproduce this trend, which is close to $a_{\rm cut} \propto M_\star^{1/3} \propto P(M_\star)$. The power-law index of each scaling law with stellar mass is also given in Table \ref{tab:turnover}.

	\section{Discussion}\label{sec:dis}
	We find that the two truncation mechanisms that best match the observed cutoff are disk truncation at the pre-main-sequence co-rotation radius and planet destruction by planetary tides. Discriminating between these two is hard, since the indices of these processes lie close together (Table \ref{tab:turnover}), and modifications to the tidal theory -- for example if the tidal dissipation parameter $Q_\star$ depends on stellar mass -- may predict a distribution not considered here.
	However, the tidal hypothesis does make two strong predictions that can be tested in the near future:
	\begin{itemize}
	\item Planets on low-eccentricity orbits can only migrate over significant distances if they maintain their orbital eccentricity through secular interactions with other planets in the system.
	Though undetected planets -- either due to long periods, low signal-to-noise or high mutual inclination -- may complicate such an analysis from the \textit{Kepler} data alone, the magnitude of the drop in occurrence rate ($\sim100$) implies that, to first order, \textit{all} short-period exoplanets ($P<10$ days, $R<1-4R_\oplus$) have companions at larger separations that maintain the non-zero eccentricity of their orbits.
	\item Tidal interactions take place over gigayear time scales, and hence the truncation radius is expected to move out over time. Ages of stars in the \textit{Kepler} sample are not well determined, though gyrochronology has provided estimates for a subsample of them \citep{2013ApJ...775L..11M,2014ApJS..211...24M}. The K2 mission provides an excellent opportunity to test the tidal hypothesis, as its mission strategy (observing multiple fields for short times) is well-suited for studying close-in planets around stars of different ages, where the tidal truncation radius should lie closer in for younger stars.
	\end{itemize}
	
		\subsection{Higher occurrence rates for cooler stars}
	Using updated stellar parameters and additional KOIs, we confirm the trend identified by \cite{2012ApJS..201...15H} that planet occurrence rates increase toward cooler stars. Thanks to the larger sample of KOIs we can also show that this trend is present up to 150-day periods and extend it down to earth-sized planets (Figure \ref{fig:diag}). The scaling factor we derive follows the same linear dependence on effective temperature described by \cite{2012ApJS..201...15H} equation 9, with $f_0=0.17$ and $k_T=-0.07$, consistent within their errors. Surprisingly, there is no simple linear or power-law correlation with stellar \textit{mass}.
	
	One possibility is that the higher occurrence rate for cooler stars is due to observational biases.
	\textit{Kepler} is a magnitude-limited survey, and different spectral types probe different populations of stars. On the main sequence, earlier (later) spectral types are brighter (fainter), and hence are located further away (closer by) for the same magnitude. Hence, the earlier spectral types in the \textit{Kepler} sample probe a population that is \textit{on average} further away. As \textit{Kepler} looks out of the galactic plane, these more distant stars are also located higher above the galactic plane, where stars are on average older, have a lower metallicity, and have a higher probability of belonging to the galactic thick disk which is also enriched in $\alpha$-elements. In the next two paragraphs we will discuss if the differences in stellar composition between the different spectral types in the \textit{Kepler} sample can explain the increasing planet occurrence toward later spectral types.
	
	The occurrence of giant planets correlates with metallicity \citep[e.g.,][]{2000A&A...363..228S,Fischer:2005gi}, though lower-mass planets are not preferentially found around metal-rich stars \citep[e.g.,][]{2008A&A...487..373S,Buchhave:2014dm}. Metallicities for stars in the Kepler field are not determined with high enough precision to test a possible metallicity-occurrence relation. \cite{2012ApJS..201...15H} explored the metallicity gradient as a function of spectral type in the \textit{Kepler} sample using the Besancon galactic model \citep{2003A&A...409..523R}. The authors find that the metallicity gradient is not large enough to explain the increased occurrence toward later spectral types even in the optimistic case when planet occurrence scales with metallicity as it does for giant planets. In addition, they find an upturn in metallicity for F stars due to an age cutoff, which contradicts the persistently lower occurrence toward earlier spectral types.
	
	In the low-metallicity regime, a high abundance of $\alpha$ elements correlates with planet occurrence \citep{2012A&A...547A..36A}. These elements, such as Mg, O and Si, make up the bulk mass of the planet-forming solids.
	The abundances of $\alpha$ elements for the Kepler stars have not been measured, yet inferences for the entire population can be made based on a similar distance argument as for metallicity. Earlier spectral types are located higher above the galactic plane, and hence contain a larger fraction of stars belonging to the galactic thick disk -- which is enriched in $\alpha$ elements -- than to the galactic thin disk. Hence they contain on average a higher abundance of $\alpha$ elements, predicting an \textit{increased} occurrence of planets toward \textit{earlier} spectral types, contrary to what is observed. 
	
	Hence, it is unlikely that the trend of increased planet occurrence toward later spectral types in the Kepler sample is caused by differences in stellar populations. It may be an imprint of a different formation efficiency of planets around low-mass stars, for example, the formation of more numerous, but smaller planets \citep[e.g.,][]{2006ApJ...642.1131K,2007ApJ...669..606R}. We will leave this analysis for future work, as it requires a thorough investigation of star-disk-scaling laws, planet mass-radius relations and a comparison of different regions in the disk.
	
	Another potential explanation for this trend is the presence of binary companions that may inhibit planet formation. The binary fraction is a strong function of spectral type, rising from $\sim 35\%$ in M stars to $\sim 60\%$ in F stars \citep[e.g.,][]{2010ApJS..190....1R}. However, only close binaries are able to dynamically disrupt planets within 1 AU. \cite{Parker:2013jv} estimate that 65-90\% of G type stars can host a planet within 1 AU, and this number is likely higher for the shorter separations considered in this work. Hence, we consider it unlikely that binaries are the sole explanation for the decreased occurrence by a factor of $\sim 2$ between M and G stars, unless they inhibit planet formation in a way other than through dynamical instability.
	 	 	 
	\subsection{Differences from Plavchan \& Bilinski (2013)}\label{sec:plavchan}
	We derive a smaller power-law index for the planet occurrence-stellar mass scaling law (0.33) than \cite{Plavchan:2013hl} for the \textit{Kepler} KOIs (0.38-0.9). We attribute this difference mainly to the functional form chosen by the authors to compare with the observed planet distribution within 0.1 AU. This functional form (Gaussian) is a good approximation for hot-Jupiters, with an observed pile-up at 0.03 AU. However, we see no evidence for such a pile-up in the smaller \textit{Kepler} planets ($1 R_\oplus$ to $4 R_\oplus$) considered in this work, as the occurrence rate does not show a steep drop outwards of the cutoff (Figure \ref{fig:radial-sma}). A different functional form, for example a double power law, will likely result in a scaling law more consistent with our result.
	
	The different distributions for Hot-Jupiters (pile-up) versus sub-Neptunes and super-Earths (plateau with cutoff) likely indicate a different mechanism is at work. Hence, the equations for tidal interactions presented here are different from those in \cite{Plavchan:2013hl}, as they describe different processes. The equations in the latter paper are mainly based on a single planet scattered on an extremely eccentric orbit and being tidally circularized. However, it is not clear how such a model could account for the flat occurrence rates well outside the cutoff, where tides do not play a role in circularizing the orbits. Instead, we rely on \textit{removal of planets} interior to a certain cutoff radius, either by in-spiral of planets on a circular orbit due to tides raised on the star, or by tides raised on planets with (low-)eccentricity orbits that are maintained due to secular interactions. However, we note that the power-law index for tidal circularization after Kozai interactions with a binary companion \citep[$3/13$, ][]{Wu:2007gp}, is the same as that for secular interaction in multi-planet systems (Table \ref{tab:turnover}). 

	\subsection{Migration versus in-situ formation}
	In this section we explore the implication that the stellar mass scaling of the cutoff location occurs early on and reflects the disk co-rotation radius. The occurrence rates in Figure \ref{fig:radial-sma} show a planet formation rate or stalling mechanism that is independent of distance from the star until it is truncated at the co-rotation radius. 
	
	In an in-situ formation context, this cutoff would lie at the dust sublimation radius, inward of which no solid building blocks can condense from the gas phase. The dust sublimation radius in protoplanetary disks  is typically observed to be outside that of the gas disk \citep[e.g.,][]{Eisner:2009jg,Eisner:2010ck,2011ApJ...743..112S}, and hence inconsistent with the derived stellar-mass scaling of the cutoff. Only a hybrid scenario \citep[e.g.,][]{2012ApJ...751..158H,2014A&A...569A..56C}, in which building blocks (that are sufficiently large to resist sublimation) migrate inward and get trapped at the co-rotation and further out, can explain the truncation at the co-rotation radius.
	
	In a planet-migration context, planets would form further out, migrate inward through the gas disk, and get trapped at the co-rotation radius instead of falling onto the star \citep[e.g.,][]{2013ApJ...764..105S}. 
	The lack of a pronounced peak in occurrence rate at the cutoff requires additional planets to get trapped behind the first planet to create the plateau in occurrence rate outwards of a 10 day period. Trapping migrating planets in mean-motion resonances and subsequent destabilization has also been proposed to explain the presence of \textit{Kepler} multi-planet systems \citep{Baruteau:2013gf,2014AJ....147...32G}.
	Typical planet formation times inward of the snow line and for a minimum-mass solar nebula disk are much longer than gas disk lifetimes
Hence, these planets need to form outside the snow line \citep[e.g.,][]{2008ApJ...682.1264K}, or form in a disk with a higher surface density inside the snow line \citep{2013MNRAS.431.3444C}.

\section{Conclusion}\label{sec:con}

We have derived planet occurrence rates as a function of spectral type, orbital period, and planet radius for the first 8 quarters of \textit{Kepler} data. We find that:

\begin{itemize}
\item Planet occurrence rates are successively higher toward cooler stars, at all orbital periods and planet radii, most significantly between 1-4 $R_\oplus$ and up to 50-day periods. Planets around M stars occur twice as frequently as around G stars, and thrice as frequently as around F stars.
\item The occurrence rates of Earth to Neptune-sized planets (1-4 $R_\oplus$) as a function of distance from the central star are self-similar for spectral types M, K, G and F. These rates are characterized by an increase up to ten-day orbital periods and a plateau farther out.
\item By assigning to each spectral type bin a median stellar mass, we show that the semi-major axis $a$ of the cutoff in planet occurrence scales with stellar mass $M_\star$ as $a \propto M_\star^{1/3}$. The stellar-mass-dependence of this location is consistent with the location of the pre-main-sequence co-rotation radius, and -- to a lesser degree -- destruction by planetary tides in multi-planet systems, but inconsistent with truncation by stellar tides or the location of the dust sublimation radius.
\item We confirm the linear scaling of occurrence rate with stellar effective temperature identified by \cite{2012ApJS..201...15H}, and show that it is persistent at all orbital periods up to 150 days and extends down to earth-sized planets.
\end{itemize}

Overall, our results show that the planet formation or migration process is strongly stellar-mass dependent.

\ifemulate
	\vspace{1em}{\it Acknowledgments:}\\
\else
	\acknowledgments
\fi
This paper includes data collected by the Kepler mission. Funding for the Kepler mission is provided by the NASA Science Mission directorate. DA acknowledges support from the NASA Origins of Solar Systems grant NNX11AG57G. We would like to thank Richard Greenberg for helpful discussion on tidal destruction of exoplanets.

{\it Facilities:} \facility{Kepler}

\ifemulate
	\ifastroph		
		\bibliography{kepler-accepted.bbl}
	\else	
		\bibliography{/Users/mulders/Dropbox/papers/papers3}
	\fi
\else	
	\bibliography{kepler-accepted.bbl}

\begin{thebibliography}{}
\expandafter\ifx\csname natexlab\endcsname\relax\def\natexlab#1{#1}\fi

\bibitem[{Adibekyan {et~al.}(2012)Adibekyan, Delgado~Mena, Sousa, Santos,
  Israelian, Gonz{\'a}lez~Hern{\'a}ndez, Mayor, \&
  Hakobyan}]{2012A&A...547A..36A}
Adibekyan, V.~Z., Delgado~Mena, E., Sousa, S.~G., {et~al.} 2012, Astronomy {\&}
  Astrophysics, 547, A36

\bibitem[{Alcal{\'a} {et~al.}(2014)Alcal{\'a}, Natta, Manara, Spezzi, Stelzer,
  Frasca, Biazzo, Covino, Randich, Rigliaco, Testi, Comeron, Cupani, \&
  D'Elia}]{2014A&A...561A...2A}
Alcal{\'a}, J., Natta, A., Manara, C.~F., {et~al.} 2014, Astronomy {\&}
  Astrophysics, 561, A2

\bibitem[{Andrews {et~al.}(2013)Andrews, Rosenfeld, Kraus, \&
  Wilner}]{2013ApJ...771..129A}
Andrews, S., Rosenfeld, K.~A., Kraus, A.~L., \& Wilner, D.~J. 2013, The
  Astrophysical Journal, 771, 129

\bibitem[{Apai {et~al.}(2005)Apai, Pascucci, Bouwman, Natta, Henning, \&
  Dullemond}]{2005Sci...310..834A}
Apai, D., Pascucci, I., Bouwman, J., {et~al.} 2005, Science, 310, 834

\bibitem[{Baraffe {et~al.}(1998)Baraffe, Chabrier, Allard, \&
  Hauschildt}]{1998A&A...337..403B}
Baraffe, I., Chabrier, G., Allard, F., \& Hauschildt, P.~H. 1998, Astronomy
  {\&} Astrophysics, 337, 403

\bibitem[{Baruteau \& Papaloizou(2013)}]{Baruteau:2013gf}
Baruteau, C., \& Papaloizou, J. 2013, Astrophysical Journal, 778, 7

\bibitem[{Batalha {et~al.}(2013)Batalha, Rowe, Bryson, Barclay, Burke,
  Caldwell, Christiansen, Mullally, Thompson, Brown, Dupree, Fabrycky, Ford,
  Fortney, Gilliland, Isaacson, Latham, Marcy, Quinn, Ragozzine, Shporer,
  Borucki, Ciardi, Gautier, Haas, Jenkins, Koch, Lissauer, Rapin, Basri, Boss,
  Buchhave, Carter, Charbonneau, Christensen-Dalsgaard, Clarke, Cochran,
  Demory, D{\'e}sert, DeVore, Doyle, Esquerdo, Everett, Fressin, Geary,
  Girouard, Gould, Hall, Holman, Howard, Howell, Ibrahim, Kinemuchi, Kjeldsen,
  Klaus, Li, Lucas, Meibom, Morris, Prsa, Quintana, Sanderfer, Sasselov,
  Seader, Smith, Steffen, Still, Stumpe, Tarter, Tenenbaum, Torres, Twicken,
  Uddin, Van~Cleve, Walkowicz, \& Welsh}]{2013ApJS..204...24B}
Batalha, N., Rowe, J.~F., Bryson, S.~T., {et~al.} 2013, The Astrophysical
  Journal Supplement, 204, 24

\bibitem[{Boley \& Ford(2013)}]{Boley:2013ur}
Boley, A.~C., \& Ford, E.~B. 2013, eprint arXiv:1306.0566, 1306.0566

\bibitem[{Borucki {et~al.}(2011)Borucki, Koch, Basri, Batalha, Boss, Brown,
  Caldwell, Christensen-Dalsgaard, Cochran, DeVore, Dunham, Dupree, Gautier,
  Geary, Gilliland, Gould, Howell, Jenkins, Kjeldsen, Latham, Lissauer, Marcy,
  Monet, Sasselov, Tarter, Charbonneau, Doyle, Ford, Fortney, Holman, Seager,
  Steffen, Welsh, Allen, Bryson, Buchhave, Chandrasekaran, Christiansen,
  Ciardi, Clarke, Dotson, Endl, Fischer, Fressin, Haas, Horch, Howard,
  Isaacson, Kolodziejczak, Li, MacQueen, Meibom, Prsa, Quintana, Rowe, Sherry,
  Tenenbaum, Torres, Twicken, Van~Cleve, Walkowicz, \& Wu}]{Borucki:2011cp}
Borucki, W.~J., Koch, D.~G., Basri, G., {et~al.} 2011, Astrophysical Journal,
  728, 117

\bibitem[{Bouvier(2007)}]{2007IAUS..243..231B}
Bouvier, J. 2007, Star-Disk Interaction in Young Stars, 243, 231

\bibitem[{Buchhave {et~al.}(2014)Buchhave, Bizzarro, Latham, Sasselov, Cochran,
  Endl, Isaacson, Juncher, \& Marcy}]{Buchhave:2014dm}
Buchhave, L.~A., Bizzarro, M., Latham, D.~W., {et~al.} 2014, Nature, 509, 593

\bibitem[{Burke(2008)}]{2008ApJ...679.1566B}
Burke, C.~J. 2008, The Astrophysical Journal, 679, 1566

\bibitem[{Burke {et~al.}(2014)Burke, Bryson, Mullally, Rowe, Christiansen,
  Thompson, Coughlin, Haas, Batalha, Caldwell, Jenkins, Still, Barclay,
  Borucki, Chaplin, Ciardi, Clarke, Cochran, Demory, Esquerdo, Gautier,
  Gilliland, Girouard, Havel, Henze, Howell, Huber, Latham, Li, Morehead,
  Morton, Pepper, Quintana, Ragozzine, Seader, Shah, Shporer, Tenenbaum,
  Twicken, \& Wolfgang}]{2014ApJS..210...19B}
Burke, C.~J., Bryson, S.~T., Mullally, F., {et~al.} 2014, The Astrophysical
  Journal Supplement, 210, 19

\bibitem[{Castelli \& Kurucz(2004)}]{2004astro.ph..5087C}
Castelli, F., \& Kurucz, R.~L. 2004, arXiv.org, 5087

\bibitem[{Catanzarite \& Shao(2011)}]{Catanzarite:2011cu}
Catanzarite, J., \& Shao, M. 2011, Astrophysical Journal, 738, 151

\bibitem[{Chiang \& Laughlin(2013)}]{2013MNRAS.431.3444C}
Chiang, E., \& Laughlin, G. 2013, Monthly Notices of the Royal Astronomical
  Society, 431, 3444

\bibitem[{Christiansen {et~al.}(2012)Christiansen, Jenkins, Caldwell, Burke,
  Tenenbaum, Seader, Thompson, Barclay, Clarke, Li, Smith, Stumpe, Twicken, \&
  Van~Cleve}]{Christiansen:2012bz}
Christiansen, J.~L., Jenkins, J.~M., Caldwell, D.~A., {et~al.} 2012,
  Publications of the Astronomical Society of the Pacific, 124, 1279

\bibitem[{Ciardi {et~al.}(2011)Ciardi, von Braun, Bryden, van Eyken, Howell,
  Kane, Plavchan, Ram{\'\i}rez, \& Stauffer}]{2011AJ....141..108C}
Ciardi, D.~R., von Braun, K., Bryden, G., {et~al.} 2011, The Astronomical
  Journal, 141, 108

\bibitem[{Correia {et~al.}(2012)Correia, Bou{\'e}, \& Laskar}]{Correia:2012ea}
Correia, A. C.~M., Bou{\'e}, G., \& Laskar, J. 2012, Astrophysical Journal,
  744, L23

\bibitem[{Cossou {et~al.}(2014)Cossou, Raymond, Hersant, \&
  Pierens}]{2014A&A...569A..56C}
Cossou, C., Raymond, S.~N., Hersant, F., \& Pierens, A. 2014, Astronomy {\&}
  Astrophysics, 569, 56

\bibitem[{D'Alessio {et~al.}(1998)D'Alessio, Canto, Calvet, \&
  Lizano}]{1998ApJ...500..411D}
D'Alessio, P., Canto, J., Calvet, N., \& Lizano, S. 1998, Astrophysical Journal
  v.500, 500, 411

\bibitem[{Dong \& Zhu(2013)}]{2013ApJ...778...53D}
Dong, S., \& Zhu, Z. 2013, The Astrophysical Journal, 778, 53

\bibitem[{Dressing \& Charbonneau(2013)}]{2013ApJ...767...95D}
Dressing, C., \& Charbonneau, D. 2013, The Astrophysical Journal, 767, 95

\bibitem[{Dumusque {et~al.}(2014)Dumusque, Bonomo, Haywood, Malavolta,
  Segransan, Buchhave, Collier~Cameron, Latham, Molinari, Pepe, Udry,
  Charbonneau, Cosentino, Dressing, Figueira, Fiorenzano, Gettel, Harutyunyan,
  Horne, Lopez-Morales, Lovis, Mayor, Micela, Motalebi, Nascimbeni, Phillips,
  Piotto, Pollacco, Queloz, Rice, Sasselov, Sozzetti, Szentgyorgyi, \&
  Watson}]{2014ApJ...789..154D}
Dumusque, X., Bonomo, A.~S., Haywood, R.~D., {et~al.} 2014, The Astrophysical
  Journal, 789, 154

\bibitem[{Eisner {et~al.}(2009)Eisner, Graham, Akeson, \&
  Najita}]{Eisner:2009jg}
Eisner, J.~A., Graham, J.~R., Akeson, R.~L., \& Najita, J. 2009, Astrophysical
  Journal, 692, 309

\bibitem[{Eisner {et~al.}(2010)Eisner, Monnier, Woillez, Akeson, Millan-Gabet,
  Graham, Hillenbrand, Pott, Ragland, \& Wizinowich}]{Eisner:2010ck}
Eisner, J.~A., Monnier, J.~D., Woillez, J., {et~al.} 2010, Astrophysical
  Journal, 718, 774

\bibitem[{Fischer \& Valenti(2005)}]{Fischer:2005gi}
Fischer, D.~A., \& Valenti, J. 2005, Astrophysical Journal, 622, 1102

\bibitem[{Fressin {et~al.}(2013)Fressin, Torres, Charbonneau, Bryson,
  Christiansen, Dressing, Jenkins, Walkowicz, \& Batalha}]{2013ApJ...766...81F}
Fressin, F., Torres, G., Charbonneau, D., {et~al.} 2013, The Astrophysical
  Journal, 766, 81

\bibitem[{Goldreich \& Schlichting(2014)}]{2014AJ....147...32G}
Goldreich, P., \& Schlichting, H.~E. 2014, The Astronomical Journal, 147, 32

\bibitem[{Greenberg {et~al.}(2013)Greenberg, Van~Laerhoven, \&
  Barnes}]{Greenberg:2013cq}
Greenberg, R., Van~Laerhoven, C., \& Barnes, R. 2013, Celestial Mechanics and
  Dynamical Astronomy, 117, 331

\bibitem[{Hansen \& Murray(2012)}]{2012ApJ...751..158H}
Hansen, B. M.~S., \& Murray, N. 2012, The Astrophysical Journal, 751, 158

\bibitem[{Hartmann {et~al.}(2006)Hartmann, D'Alessio, Calvet, \&
  Muzerolle}]{2006ApJ...648..484H}
Hartmann, L.~W., D'Alessio, P., Calvet, N., \& Muzerolle, J. 2006, The
  Astrophysical Journal, 648, 484

\bibitem[{Howard {et~al.}(2012)Howard, Marcy, Bryson, Jenkins, Rowe, Batalha,
  Borucki, Koch, Dunham, Gautier, Van~Cleve, Cochran, Latham, Lissauer, Torres,
  Brown, Gilliland, Buchhave, Caldwell, Christensen-Dalsgaard, Ciardi, Fressin,
  Haas, Howell, Kjeldsen, Seager, Rogers, Sasselov, Steffen, Basri,
  Charbonneau, Christiansen, Clarke, Dupree, Fabrycky, Fischer, Ford, Fortney,
  Tarter, Girouard, Holman, Johnson, Klaus, Machalek, Moorhead, Morehead,
  Ragozzine, Tenenbaum, Twicken, Quinn, Isaacson, Shporer, Lucas, Walkowicz,
  Welsh, Boss, DeVore, Gould, Smith, Morris, Prsa, Morton, Still, Thompson,
  Mullally, Endl, \& MacQueen}]{2012ApJS..201...15H}
Howard, A.~W., Marcy, G.~W., Bryson, S.~T., {et~al.} 2012, The Astrophysical
  Journal Supplement, 201, 15

\bibitem[{Howard {et~al.}(2013)Howard, Sanchis-Ojeda, Marcy, Johnson, Winn,
  Isaacson, Fischer, Fulton, Sinukoff, \& Fortney}]{2013Natur.503..381H}
Howard, A.~W., Sanchis-Ojeda, R., Marcy, G.~W., {et~al.} 2013, Nature, 503, 381

\bibitem[{Huber {et~al.}(2014)Huber, Silva~Aguirre, Matthews, Pinsonneault,
  Gaidos, Garc{\'\i}a, Hekker, Mathur, Mosser, Torres, Bastien, Basu, Bedding,
  Chaplin, Demory, Fleming, Guo, Mann, Rowe, Serenelli, Smith, \&
  Stello}]{Huber:2014dh}
Huber, D., Silva~Aguirre, V., Matthews, J.~M., {et~al.} 2014, The Astrophysical
  Journal Supplement Series, 211, 2

\bibitem[{Jackson {et~al.}(2009)Jackson, Barnes, \& Greenberg}]{Jackson:2009cm}
Jackson, B., Barnes, R., \& Greenberg, R. 2009, Astrophysical Journal, 698,
  1357

\bibitem[{Kama {et~al.}(2009)Kama, Min, \& Dominik}]{2009A&A...506.1199K}
Kama, M., Min, M., \& Dominik, C. 2009, Astronomy {\&} Astrophysics, 506, 1199

\bibitem[{Kennedy \& Kenyon(2008)}]{2008ApJ...682.1264K}
Kennedy, G., \& Kenyon, S.~J. 2008, The Astrophysical Journal, 682, 1264

\bibitem[{Kokubo {et~al.}(2006)Kokubo, Kominami, \& Ida}]{2006ApJ...642.1131K}
Kokubo, E., Kominami, J., \& Ida, S. 2006, The Astrophysical Journal, 642, 1131

\bibitem[{Kuchner \& Lecar(2002)}]{2002ApJ...574L..87K}
Kuchner, M., \& Lecar, M. 2002, The Astrophysical Journal, 574, L87

\bibitem[{Lanza \& Shkolnik(2014)}]{2014MNRAS.443.1451L}
Lanza, A.~F., \& Shkolnik, E.~L. 2014, Monthly Notices of the Royal
  Astronomical Society, 443, 1451

\bibitem[{Lin {et~al.}(1996)Lin, Bodenheimer, \& Richardson}]{Lin:1996ey}
Lin, D. N.~C., Bodenheimer, P., \& Richardson, D.~C. 1996, Nature, 380, 606

\bibitem[{McQuillan {et~al.}(2013)McQuillan, Mazeh, \&
  Aigrain}]{2013ApJ...775L..11M}
McQuillan, A., Mazeh, T., \& Aigrain, S. 2013, The Astrophysical Journal
  Letters, 775, L11

\bibitem[{McQuillan {et~al.}(2014)McQuillan, Mazeh, \&
  Aigrain}]{2014ApJS..211...24M}
McQuillan, A., Mazeh, T., \& Aigrain, S.  2014, The Astrophysical Journal Supplement, 211, 24

\bibitem[{Millan-Gabet {et~al.}(2007)Millan-Gabet, Malbet, Akeson, Leinert,
  Monnier, \& Waters}]{2007prpl.conf..539M}
Millan-Gabet, R., Malbet, F., Akeson, R., {et~al.} 2007, Protostars and Planets
  V, 539

\bibitem[{Min {et~al.}(2011)Min, Dullemond, Kama, \&
  Dominik}]{2011Icar..212..416M}
Min, M., Dullemond, C.~P., Kama, M., \& Dominik, C. 2011, Icarus, 212, 416

\bibitem[{Mohanty {et~al.}(2013)Mohanty, Greaves, Mortlock, Pascucci, Scholz,
  Thompson, Apai, Lodato, \& Looper}]{2013ApJ...773..168M}
Mohanty, S., Greaves, J., Mortlock, D., {et~al.} 2013, The Astrophysical
  Journal, 773, 168

\bibitem[{Monnier \& Millan-Gabet(2002)}]{2002ApJ...579..694M}
Monnier, J.~D., \& Millan-Gabet, R. 2002, The Astrophysical Journal, 579, 694

\bibitem[{Moorhead {et~al.}(2011)Moorhead, Ford, Morehead, Rowe, Borucki,
  Batalha, Bryson, Caldwell, Fabrycky, Gautier, Koch, Holman, Jenkins, Li,
  Lissauer, Lucas, Marcy, Quinn, Quintana, Ragozzine, Shporer, Still, \&
  Torres}]{Moorhead:2011dn}
Moorhead, A.~V., Ford, E.~B., Morehead, R.~C., {et~al.} 2011, The Astrophysical
  Journal Supplement Series, 197, 1

\bibitem[{Morton \& Swift(2014)}]{2014ApJ...791...10M}
Morton, T.~D., \& Swift, J. 2014, The Astrophysical Journal, 791, 10

\bibitem[{Mulders \& Dominik(2012)}]{2012A&A...539A...9M}
Mulders, G.~D., \& Dominik, C. 2012, Astronomy {\&} Astrophysics, 539, A9

\bibitem[{Owen \& Wu(2013)}]{Owen:2013gs}
Owen, J.~E., \& Wu, Y. 2013, Astrophysical Journal, 775, 105

\bibitem[{Parker \& Quanz(2013)}]{Parker:2013jv}
Parker, R.~J., \& Quanz, S.~P. 2013, Monthly Notices of the Royal Astronomical
  Society, 436, 650

\bibitem[{Pascucci {et~al.}(2009)Pascucci, Apai, Luhman, Henning, Bouwman,
  Meyer, Lahuis, \& Natta}]{2009ApJ...696..143P}
Pascucci, I., Apai, D., Luhman, K., {et~al.} 2009, The Astrophysical Journal,
  696, 143

\bibitem[{Petigura {et~al.}(2013)Petigura, Marcy, \&
  Howard}]{2013ApJ...770...69P}
Petigura, E.~A., Marcy, G.~W., \& Howard, A.~W. 2013, The Astrophysical
  Journal, 770, 69

\bibitem[{Pinte {et~al.}(2008)Pinte, Menard, Berger, Benisty, \&
  Malbet}]{2008ApJ...673L..63P}
Pinte, C., Menard, F., Berger, J.~P., Benisty, M., \& Malbet, F. 2008, The
  Astrophysical Journal, 673, L63

\bibitem[{Plavchan \& Bilinski(2013)}]{Plavchan:2013hl}
Plavchan, P., \& Bilinski, C. 2013, Astrophysical Journal, 769, 86

\bibitem[{Plavchan {et~al.}(2012)Plavchan, Bilinski, \&
  Currie}]{Plavchan:2012tr}
Plavchan, P., Bilinski, C., \& Currie, T. 2012, eprint arXiv:1203.1887,
  1203.1887

\bibitem[{Raghavan {et~al.}(2010)Raghavan, McAlister, Henry, Latham, Marcy,
  Mason, Gies, White, \& ten Brummelaar}]{2010ApJS..190....1R}
Raghavan, D., McAlister, H.~A., Henry, T.~J., {et~al.} 2010, The Astrophysical
  Journal Supplement, 190, 1

\bibitem[{Raymond {et~al.}(2007)Raymond, Scalo, \&
  Meadows}]{2007ApJ...669..606R}
Raymond, S.~N., Scalo, J., \& Meadows, V.~S. 2007, The Astrophysical Journal,
  669, 606

\bibitem[{Robin {et~al.}(2003)Robin, Reyl{\'e}, Derri{\`e}re, \&
  Picaud}]{2003A&A...409..523R}
Robin, A.~C., Reyl{\'e}, C., Derri{\`e}re, S., \& Picaud, S. 2003, Astronomy
  {\&} Astrophysics, 409, 523

\bibitem[{Salyk {et~al.}(2011)Salyk, Blake, Boogert, \&
  Brown}]{2011ApJ...743..112S}
Salyk, C., Blake, G.~A., Boogert, A. C.~A., \& Brown, J.~M. 2011, The
  Astrophysical Journal, 743, 112

\bibitem[{Santos {et~al.}(2000)Santos, Israelian, \&
  Mayor}]{2000A&A...363..228S}
Santos, N.~C., Israelian, G., \& Mayor, M. 2000, Astronomy {\&} Astrophysics,
  363, 228

\bibitem[{Seader {et~al.}(2013)Seader, Tenenbaum, Jenkins, \&
  Burke}]{2013ApJS..206...25S}
Seader, S., Tenenbaum, P., Jenkins, J.~M., \& Burke, C.~J. 2013, The
  Astrophysical Journal Supplement, 206, 25

\bibitem[{Silburt {et~al.}(2014)Silburt, Gaidos, \& Wu}]{Silburt:2014tf}
Silburt, A., Gaidos, E., \& Wu, Y. 2014, arXiv.org, 1406.6048v1

\bibitem[{Sousa {et~al.}(2008)Sousa, Santos, Mayor, Udry, Casagrande,
  Israelian, Pepe, Queloz, \& Monteiro}]{2008A&A...487..373S}
Sousa, S.~G., Santos, N.~C., Mayor, M., {et~al.} 2008, Astronomy {\&}
  Astrophysics, 487, 373

\bibitem[{Swift {et~al.}(2013)Swift, Johnson, Morton, Crepp, Montet, Fabrycky,
  \& Muirhead}]{2013ApJ...764..105S}
Swift, J.~J., Johnson, J.~A., Morton, T.~D., {et~al.} 2013, The Astrophysical
  Journal, 764, 105

\bibitem[{Traub(2012)}]{2012ApJ...745...20T}
Traub, W.~A. 2012, The Astrophysical Journal, 745, 20

\bibitem[{Wu {et~al.}(2007)Wu, Murray, \& Ramsahai}]{Wu:2007gp}
Wu, Y., Murray, N.~W., \& Ramsahai, J.~M. 2007, Astrophysical Journal, 670, 820

\bibitem[{Youdin(2011)}]{Youdin:2011gz}
Youdin, A.~N. 2011, Astrophysical Journal, 742, 38

\end{thebibliography}
\fi

\appendix

\section{Parameters and units}\label{app:para}
	A list of all parameters used in Section \ref{sec:occ} is shown in Table \ref{tab:para}.

	\begin{table}[ht]
		\title{Mathematical symbols used in Section \ref{sec:occ}}
		\centering
		\begin{tabular}{l| l l l l}\hline\hline
		\multicolumn{2}{l}{Parameter} & Description & Unit & Source \\
		\hline
		\multirow{9}{*}{\begin{sideways}Star\end{sideways}} 
			& \Teff		& Stellar effective temperature	& $K$ 		& 1 \\
			& $R_\star$ 		& Stellar Radius				& cm	& 1 \\
			& $g$		& Gravity at stellar surface		& cm$/s^2$	& 1 \\
			& $M_\star$		& Stellar Mass				& $g$		& 1 \\
			& \cdpp 		& Stellar noise	& ppm & 2 \\
			& \tlc 		& Time-resolution of \textit{Kepler} long cadence (1765.5 sec) & $s$	& \\ 
			& \cdppnorm	& Fitted noise at long cadence			& ppm & Eq. \ref{eq:cdpp} \\
			& \cdppindex 	& Fitted noise power-law index				& 	& Eq. \ref{eq:cdpp} \\
			& $t_{\rm obs}$			& Time each star is observed			& s  & 1 \\
		\hline
		\multirow{14}{*}{\begin{sideways}Planet candidate\end{sideways}} 
			& $P$ 		& Orbital period					& $s$		& 3 or Eq. \ref{eq:period} \\
			& $\delta$		& Transit depth					&	 		& 3 \\
			& \tdur		& Calculated transit duration		& $s$		&  Eq. \ref{eq:tdur_used}  \\
			& \snr		& Calculated transit signal to noise	&			& Eq. \ref{eq:snr} \\
			& \snrobs		& Tabulated transit signal to noise	&		& 3 \\
			& b			& Impact parameter [0]				&			& fitted \\
			& e			& Eccentricity	[0.1]					&			& fixed \\
			& $a$		& Semi-major axis of planet orbit	& cm 		& Eq. \ref{eq:sma} \\
			& $R_p$ 		& Planet radius						& cm 		& 3 \\	
			& $N_\star$	& Number of stars where a planet is detectable &	& Eq. \ref{eq:Nstars} \\
			& \feff		& Detection efficiency	from SNR			&			& Eq. \ref{eq:feff} \\
			& $f_n$ 		& Detection efficiency at long $P$ 		& 			& Eq. \ref{eq:fperiod}\\
			& \fgeo		& Geometric transit probability		& 			& Eq. \ref{eq:fgeo} \\
			& \focc 		& Planet occurrence rate 			& 			& Eq. \ref{eq:focc} \\
		
		\hline\hline\end{tabular}
		\caption{References -- (1) \citep{Huber:2014dh} / IPAC; (2) MAST; (3) \citep{2014ApJS..210...19B}}
		\label{tab:para}
	\end{table}

\section{Additional notes on occurrence rates.}\label{app:occ}

\subsection{Transit duration anomaly, and eccentricities}\label{app:tdur}

	The transit duration of a hypothetical planet with period $P$, semi-major axis $a$, eccentricity $e$, pericenter orientation $\omega$ and impact parameter $b$ is
	\begin{equation}
		\tdur = \frac{PM_\star}{\pi a}\sqrt{1-b^2}~f(e,\omega), 
		\label{eq:app:tdur_complex}
	\end{equation}
	where $f(e,\omega)$ is a pre-factor to correct for the orbital eccentricity, given by \cite[][Eq. 7]{2008ApJ...679.1566B}:
	\begin{equation}
		f(e,\omega)= \frac{\sqrt{1-e^2}}{1+e \cos(\omega)}.
	\end{equation}
	Marginalizing over all possible pericenter orientations $\omega$, taking into account increased probability of transit near pericenter, but assuming an edge-on orbit (b=0), gives \cite[][Eq. 18]{2008ApJ...679.1566B}:
		\begin{equation}
		f(e,\omega)= \sqrt{1-e^2}
	\end{equation}
	Taking all this into account, the transit duration simplifies to:
	\begin{equation}
		\tdur = \frac{PM_\star \sqrt{1-e^2}}{\pi a} \sqrt{1-b^2}.
		\label{eq:app:tdur_simple}
	\end{equation}
	
	Marginalizing over a uniform impact parameter distribution between $b=0$ and $b=1$ gives:
	\begin{equation}
	\int_0^1 \sqrt{1-b^2} ~db= \frac{\pi}{4}.
	\end{equation}
		
\begin{figure}
		\centering
		\includegraphics[width=0.7\linewidth]{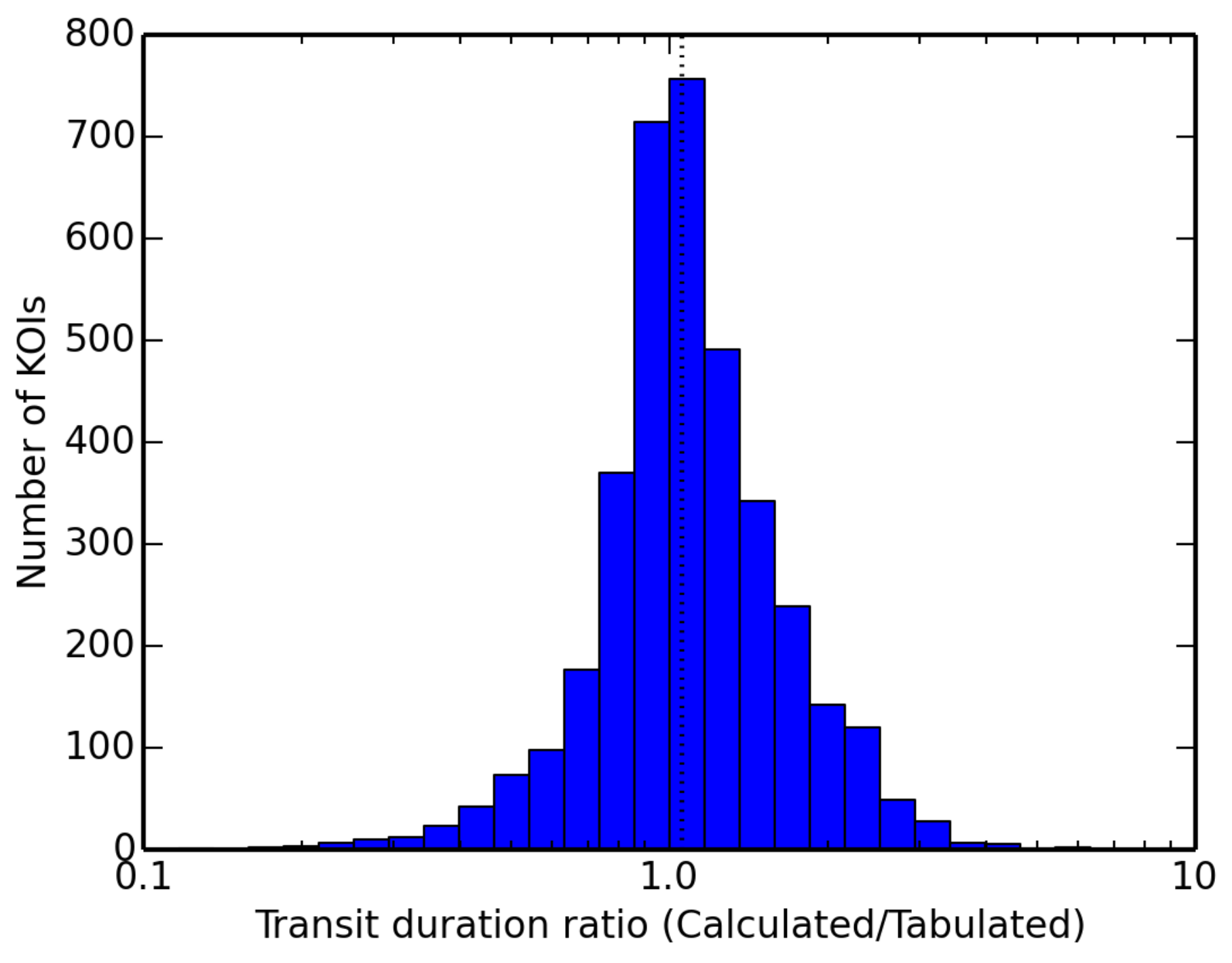}
		\caption{Histogram of calculated versus tabulated transit duration based on equation \ref{eq:app:tdur}. The median transit duration anomaly is given by the dotted line. This ratio is close to 1 for $b=0$ displayed here, and hence we do not use the expected average impact parameter ($b=0.5$) for calculating the transit duration.
		\label{fig:tdurcomp}}
	\end{figure}

	However, ignoring the impact parameter ($b=0$) gives a better fit to the observed transit duration (Figure \ref{fig:tdurcomp})
	\begin{equation}
		\tdur = \frac{PM_\star \sqrt{1-e^2}}{\pi a},
		\label{eq:app:tdur}
	\end{equation}
	which probably indicates stellar radii are systematically underestimated in the Kepler catalogue\citep[e.g.,][]{Plavchan:2012tr}. 
	
\subsection{Non-Gaussianity of stellar noise}\label{app:noise}

\begin{figure}
		
		\includegraphics[width=0.49\linewidth]{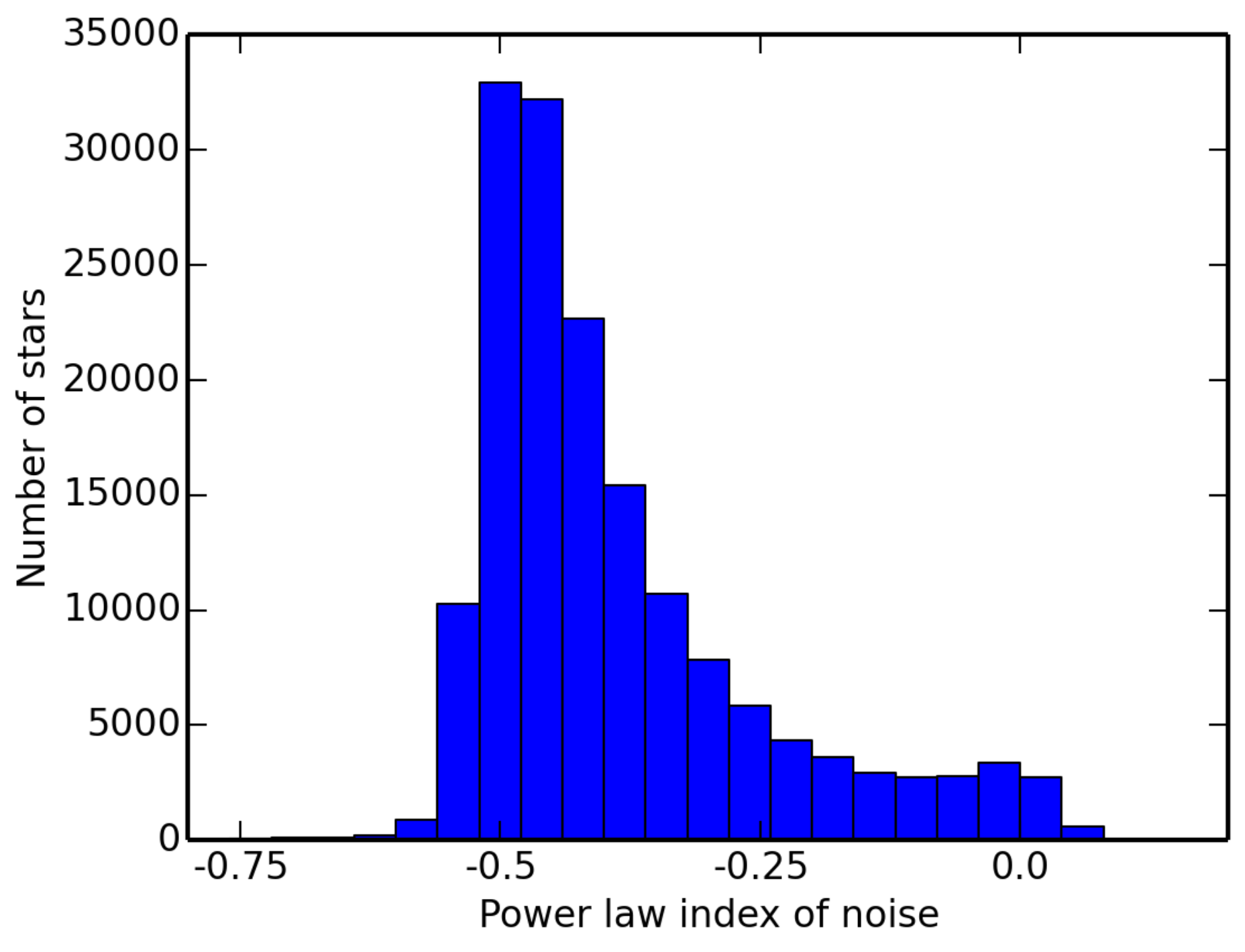}
		\includegraphics[width=0.49\linewidth]{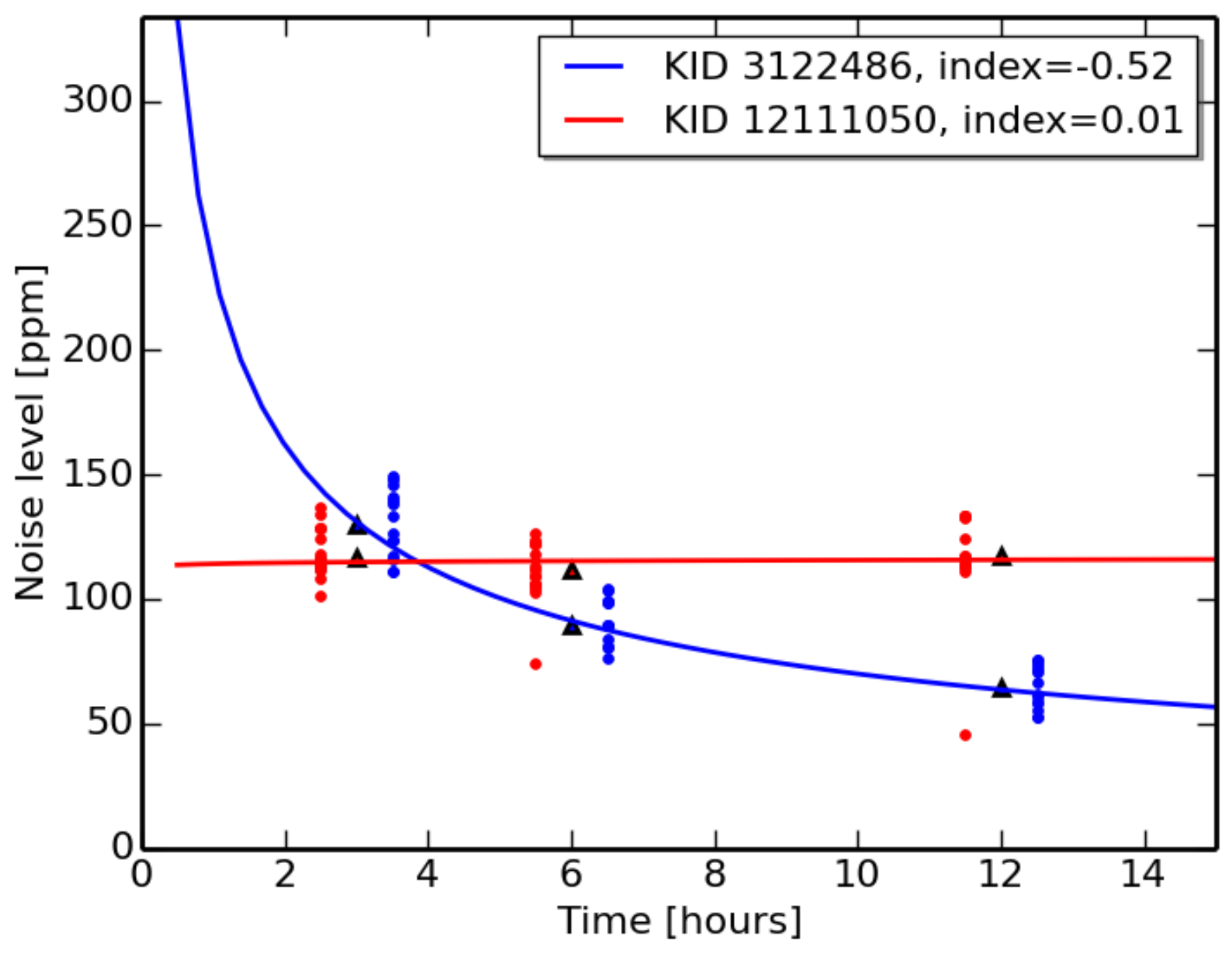}
		\caption{Left panel: Power-law index of the time-behavior of the median noise per quarter (\cdpp, Eq. \ref{eq:cdpp}) in the \textit{Kepler} star sample. 
		For most of the stars the noise is well approximated by Poissonian noise corresponding to an index of $-0.5$. A significant part of the stars deviates from this trend and show a flatter noise profile, with a peak index around $0$.
		Right panel: example of a Poissonian (KIC 3122486, blue line) and flat (KIC 12111050, red line) noise profile. 
		Crosses indicate the individual CDPP measurements per quarter, offset by plus(minus) half an hour for the blue(red) symbols. Flat noise profiles may lead to an underestimation of the noise at large transit durations if Poissonian noise is assumed.
		\label{fig:noiseindex}
		}
	\end{figure}

	The CDPP is reported for three different time scales, which allows for a first-order characterization of the noise profiles as a function of time. We have fitted the noise using a simple power law in time (Eq. \ref{eq:cdpp}). The distribution of power-law indices is shown in Figure \ref{fig:noiseindex}. A significant fraction of the stars deviates from a pure Poissonian noise profile, which corresponds to a power-law index of $-0.5$. There is a distinct tail toward shallower noise profiles, with an additional peak around $\cdppindex=0$ corresponding to noise that does not decrease with time, at least on three to twelve hour time scales. The choice of noise profile impacts how the signal-to-noise increases as a function of both transit time and number of transits, and does so differently for each star. Different studies have used different assumptions. We will test them in the next subsection.
		
\subsection{Signal-to-noise-ratio}\label{app:snr}
	As reported by \cite{2014ApJ...791...10M}, calculating the signal-to-noise of a transit using the CDPP from the MAST database does not always reproduce the signal-to-noise ratios of KOIs reported by the \textit{Kepler} team, which are measured from the light-curve. In addition, a significant fraction of the \textit{Kepler} stars exhibit non-Poissonian noise profiles as a function of time (appendix \ref{app:noise}). In this Section, we explore which assumptions for calculating the total signal-to-noise (\snr) best reproduces the ratios measured by the Kepler team (\snrobs, \citealt{2014ApJS..210...19B}) We will use those assumptions for the calculation of \snr for all stars in the \textit{Kepler} sample in the main text.

	When comparing calculated to tabulated signal-to-noise ratios two things should be kept in mind. First, the tabulated value \snrobs is based on a transit with given impact parameter $b$ and eccentricity $e$ and is measured from the transit light curve, while we use \textit{average} eccentricities and impact parameters to calculate the signal-to-noise for each star (Appendix \ref{app:tdur}). Hence, we expect a significant scatter when comparing calculated to measured values. Second, \snrobs is measured from the light curve fitting -- which is done on ten quarters of data for detection from the first 8 quarters, and 8 quarters of data for detection in the first six quarters. The calculated \snr, however, depends -- among other things -- on how accurately we know the stellar parameters, in which systematic errors are still present \citep[e.g.,][]{Plavchan:2012tr}. Hence, the purpose of this comparison is not to exactly reproduce \snrobs, but to find the best way of correcting for known and unknown biases in the data. We will assume the best solution minimizes the scatter in $\snr/\snrobs$, and apply an overall scaling factor akin to \cite{2014ApJ...791...10M} to correctly calculate the \textit{average} detection efficiency in the entire \textit{Kepler} sample of stars.
	
\begin{figure}
		\centering
		\includegraphics[width=0.49\linewidth]{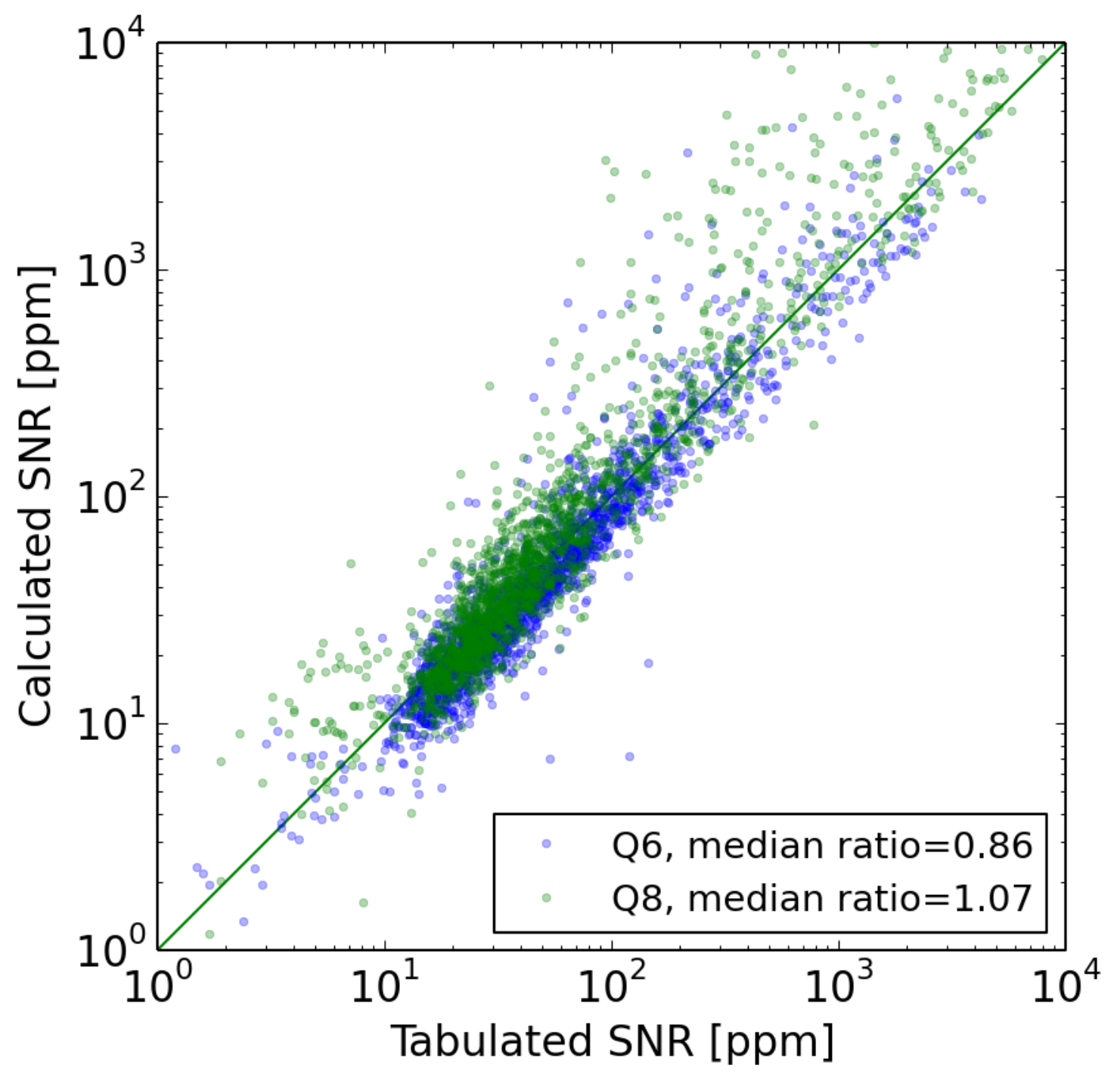}
		\includegraphics[width=0.49\linewidth]{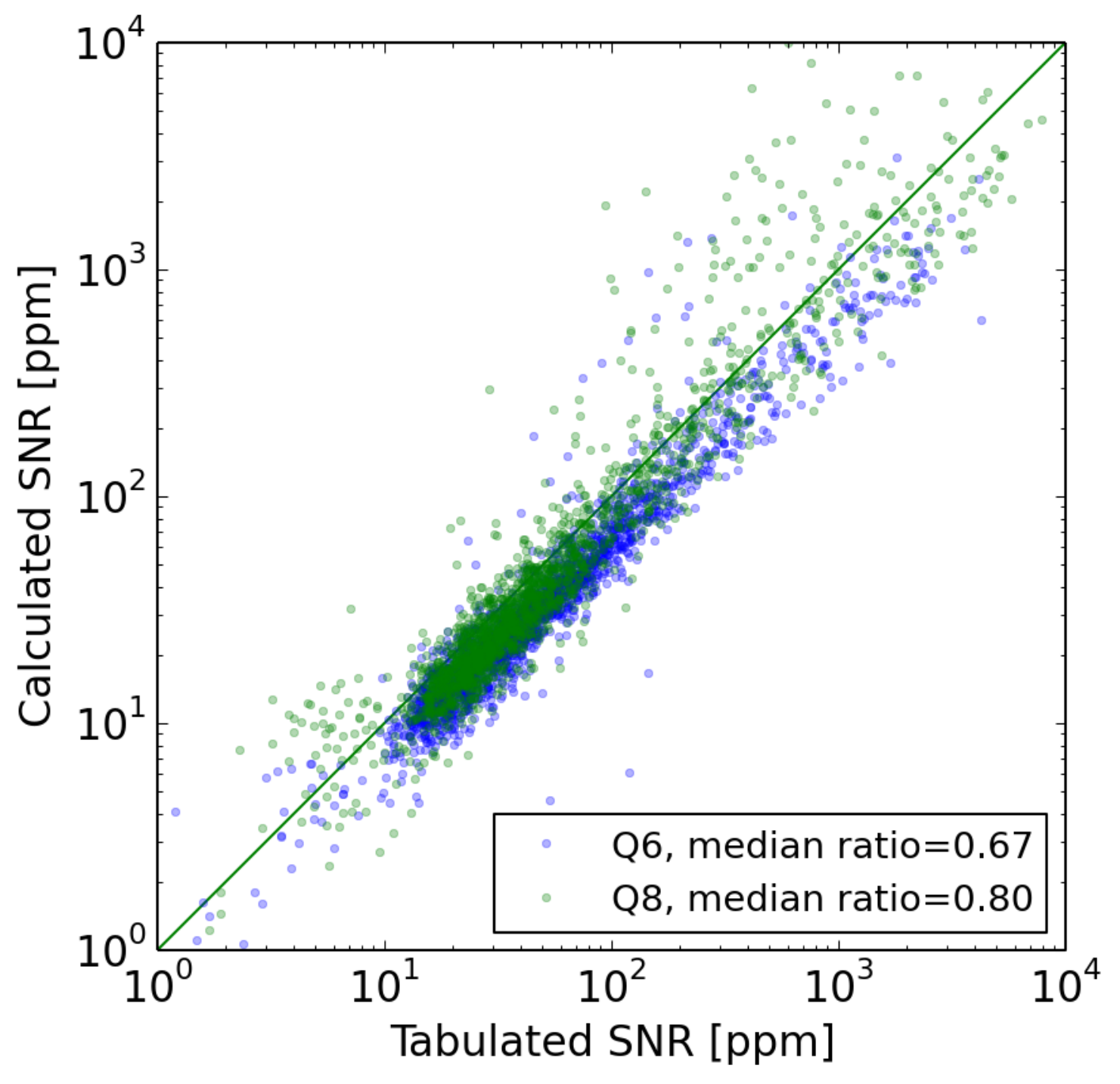}
		\caption{Calculated versus tabulated signal to noise for Poissonian noise as in \cite{2014ApJ...791...10M} (left), and interpolated noise based on equation \ref{eq:cdpp} and \ref{eq:snr} as in \cite{2013ApJ...767...95D} (right).
		The green line denotes a 1:1 slope. The scatter in the left panel is significantly larger, indicating Poissonian noise is not the best approximation for calculating the noise from a single transit. The right panel shows that the interpolated noise produces smaller scatter, but requires a correction factor to reproduce the tabulated signal-to-noise ratios.
		\label{fig:SNRcomp}}
	\end{figure}

	Figure \ref{fig:SNRcomp} shows the comparison between observed (\snrobs) and calculated (\snr) signal-to-noise ratios for all planetary candidates. The left panel shows the assumption made by \cite{2014ApJ...791...10M}, i.e. noise with a Poissonian time-behavior. In the right panel the signal-to-noise of a single transit is calculated by interpolating the CDPP to the transit duration, as in \cite{2013ApJ...767...95D}, while the total signal to noise is still scaled with the square root of the number of transits. The latter provides a smaller spread, but needs to be adjusted by a scaling factor of 1.33 to match the 1:1 slope. A signal-to-noise calculation where the total noise also scales with $n^{\cdppindex}$ performs significantly worse, and is not shown here. We conclude that interpolating the noise from the CDPP during the transit, and assuming Poissonian noise between transits results in the smallest scatter around the tabulated signal-to-noise values, but requires a correction factor of 1.33 to match the median signal-to-noise values.
	
\clearpage

\section{Tabulated occurrence rate}\label{app:occtable}
	Occurrence rates in the period-radius diagram are given in Table \ref{tab:occ} for the entire sample, and in Tables \ref{tab:occ0} to \ref{tab:occ3} for spectral type M to F.
	
	\begin{table}[ht]
		\title{Planet occurrence rates}
		\centering\tiny
		\begin{tabular}{l | llllllllllll}
\hline\hline
 & \multicolumn{12}{c}{Period [days]} \\ 
\hline
 & $0.4-$ & $0.68-$ & $1.2-$ & $2.0-$ & $3.4-$ & $5.8-$ & $10.0-$ & $17.0-$ & $29.0-$ & $50.0-$ & $85.0-$ & $150.0-$\\ 
 & \multicolumn{1}{r}{ 0.68 } & \multicolumn{1}{r}{ 1.2 } & \multicolumn{1}{r}{ 2.0 } & \multicolumn{1}{r}{ 3.4 } & \multicolumn{1}{r}{ 5.8 } & \multicolumn{1}{r}{ 10.0 } & \multicolumn{1}{r}{ 17.0 } & \multicolumn{1}{r}{ 29.0 } & \multicolumn{1}{r}{ 50.0 } & \multicolumn{1}{r}{ 85.0 } & \multicolumn{1}{r}{ 150.0 } & \multicolumn{1}{r}{ 250.0 }\\ 
\hline
$32.0-$ &  &  &  &  & $\, 0.0058$ &  &  & $\, 0.048$ & $\, 0.018$ & $\, 0.027$ &  & $\, 0.088$\\ 
\multicolumn{1}{r|}{ 45.0 }  & \multicolumn{1}{r}{ ${<}0.0012$ }  & \multicolumn{1}{r}{ ${<}0.0017$ }  & \multicolumn{1}{r}{ ${<}0.0025$ }  & \multicolumn{1}{r}{ ${<}0.0036$ }  & \multicolumn{1}{r}{ $\pm 0.0066$ }  & \multicolumn{1}{r}{ ${<}0.0073$ }  & \multicolumn{1}{r}{ ${<}0.010$ }  & \multicolumn{1}{r}{ $\pm 0.029$ }  & \multicolumn{1}{r}{ $\pm 0.028$ }  & \multicolumn{1}{r}{ $\pm 0.040$ }  & \multicolumn{1}{r}{ ${<}0.046$ }  & \multicolumn{1}{r}{ $\pm 0.103$ } \\[1ex] 
$23.0-$ &  &  & $\, 0.0048$ & $\, 0.037$ & $\, 0.023$ & $\, 0.022$ & $\, 0.035$ & $\, 0.082$ & $\, 0.12$ & $\, 0.067$ & $\, 0.10$ & $\, 0.14$\\ 
\multicolumn{1}{r|}{ 32.0 }  & \multicolumn{1}{r}{ ${<}0.0013$ }  & \multicolumn{1}{r}{ ${<}0.0019$ }  & \multicolumn{1}{r}{ $\pm 0.0044$ }  & \multicolumn{1}{r}{ $\pm 0.012$ }  & \multicolumn{1}{r}{ $\pm 0.012$ }  & \multicolumn{1}{r}{ $\pm 0.015$ }  & \multicolumn{1}{r}{ $\pm 0.022$ }  & \multicolumn{1}{r}{ $\pm 0.038$ }  & \multicolumn{1}{r}{ $\pm 0.06$ }  & \multicolumn{1}{r}{ $\pm 0.056$ }  & \multicolumn{1}{r}{ $\pm 0.08$ }  & \multicolumn{1}{r}{ $\pm 0.13$ } \\[1ex] 
$16.0-$ & $\, 0.0031$ & $\, 0.0045$ & $\, 0.011$ & $\, 0.050$ & $\, 0.090$ & $\, 0.041$ & $\, 0.13$ & $\, 0.097$ & $\, 0.18$ & $\, 0.076$ & $\, 0.11$ & $\, 0.17$\\ 
\multicolumn{1}{r|}{ 23.0 }  & \multicolumn{1}{r}{ $\pm 0.0023$ }  & \multicolumn{1}{r}{ $\pm 0.0033$ }  & \multicolumn{1}{r}{ $\pm 0.006$ }  & \multicolumn{1}{r}{ $\pm 0.015$ }  & \multicolumn{1}{r}{ $\pm 0.023$ }  & \multicolumn{1}{r}{ $\pm 0.020$ }  & \multicolumn{1}{r}{ $\pm 0.04$ }  & \multicolumn{1}{r}{ $\pm 0.044$ }  & \multicolumn{1}{r}{ $\pm 0.07$ }  & \multicolumn{1}{r}{ $\pm 0.059$ }  & \multicolumn{1}{r}{ $\pm 0.09$ }  & \multicolumn{1}{r}{ $\pm 0.14$ } \\[1ex] 
$11.0-$ & $\, 0.0017$ & $\, 0.0087$ & $\, 0.033$ & $\, 0.090$ & $\, 0.099$ & $\, 0.096$ & $\, 0.14$ & $\, 0.21$ & $\, 0.28$ & $\, 0.18$ & $\, 0.27$ & \\ 
\multicolumn{1}{r|}{ 16.0 }  & \multicolumn{1}{r}{ $\pm 0.0019$ }  & \multicolumn{1}{r}{ $\pm 0.0045$ }  & \multicolumn{1}{r}{ $\pm 0.010$ }  & \multicolumn{1}{r}{ $\pm 0.019$ }  & \multicolumn{1}{r}{ $\pm 0.026$ }  & \multicolumn{1}{r}{ $\pm 0.030$ }  & \multicolumn{1}{r}{ $\pm 0.04$ }  & \multicolumn{1}{r}{ $\pm 0.06$ }  & \multicolumn{1}{r}{ $\pm 0.09$ }  & \multicolumn{1}{r}{ $\pm 0.09$ }  & \multicolumn{1}{r}{ $\pm 0.13$ }  & \multicolumn{1}{r}{ ${<}0.087$ } \\[1ex] 
$8.0-$ & $\, 0.0048$ & $\, 0.011$ & $\, 0.028$ & $\, 0.029$ & $\, 0.066$ & $\, 0.13$ & $\, 0.15$ & $\, 0.27$ & $\, 0.33$ & $\, 0.44$ & $\, 0.61$ & $\, 0.83$\\ 
\multicolumn{1}{r|}{ 11.0 }  & \multicolumn{1}{r}{ $\pm 0.0029$ }  & \multicolumn{1}{r}{ $\pm 0.005$ }  & \multicolumn{1}{r}{ $\pm 0.009$ }  & \multicolumn{1}{r}{ $\pm 0.011$ }  & \multicolumn{1}{r}{ $\pm 0.021$ }  & \multicolumn{1}{r}{ $\pm 0.03$ }  & \multicolumn{1}{r}{ $\pm 0.05$ }  & \multicolumn{1}{r}{ $\pm 0.07$ }  & \multicolumn{1}{r}{ $\pm 0.10$ }  & \multicolumn{1}{r}{ $\pm 0.13$ }  & \multicolumn{1}{r}{ $\pm 0.19$ }  & \multicolumn{1}{r}{ $\pm 0.29$ } \\[1ex] 
$5.7-$ & $\, 0.0016$ & $\, 0.022$ & $\, 0.028$ & $\, 0.039$ & $\, 0.16$ & $\, 0.14$ & $\, 0.19$ & $\, 0.17$ & $\, 0.41$ & $\, 0.55$ & $\, 0.97$ & $\, 1.0$\\ 
\multicolumn{1}{r|}{ 8.0 }  & \multicolumn{1}{r}{ $\pm 0.0020$ }  & \multicolumn{1}{r}{ $\pm 0.007$ }  & \multicolumn{1}{r}{ $\pm 0.010$ }  & \multicolumn{1}{r}{ $\pm 0.013$ }  & \multicolumn{1}{r}{ $\pm 0.03$ }  & \multicolumn{1}{r}{ $\pm 0.04$ }  & \multicolumn{1}{r}{ $\pm 0.05$ }  & \multicolumn{1}{r}{ $\pm 0.06$ }  & \multicolumn{1}{r}{ $\pm 0.11$ }  & \multicolumn{1}{r}{ $\pm 0.15$ }  & \multicolumn{1}{r}{ $\pm 0.24$ }  & \multicolumn{1}{r}{ $\pm 0.3$ } \\[1ex] 
$4.0-$ & $\, 0.0096$ & $\, 0.016$ & $\, 0.047$ & $\, 0.081$ & $\, 0.15$ & $\, 0.27$ & $\, 0.38$ & $\, 0.46$ & $\, 0.49$ & $\, 0.78$ & $\, 0.73$ & $\, 0.81$\\ 
\multicolumn{1}{r|}{ 5.7 }  & \multicolumn{1}{r}{ $\pm 0.0041$ }  & \multicolumn{1}{r}{ $\pm 0.006$ }  & \multicolumn{1}{r}{ $\pm 0.012$ }  & \multicolumn{1}{r}{ $\pm 0.019$ }  & \multicolumn{1}{r}{ $\pm 0.03$ }  & \multicolumn{1}{r}{ $\pm 0.05$ }  & \multicolumn{1}{r}{ $\pm 0.07$ }  & \multicolumn{1}{r}{ $\pm 0.09$ }  & \multicolumn{1}{r}{ $\pm 0.12$ }  & \multicolumn{1}{r}{ $\pm 0.18$ }  & \multicolumn{1}{r}{ $\pm 0.22$ }  & \multicolumn{1}{r}{ $\pm 0.29$ } \\[1ex] 
$2.8-$ & $\, 0.0083$ & $\, 0.023$ & $\, 0.059$ & $\, 0.12$ & $\, 0.35$ & $\, 0.59$ & $\, 1.4$ & $\, 1.5$ & $\, 1.8$ & $\, 1.6$ & $\, 1.7$ & $\, 1.5$\\ 
\multicolumn{1}{r|}{ 4.0 }  & \multicolumn{1}{r}{ $\pm 0.0038$ }  & \multicolumn{1}{r}{ $\pm 0.007$ }  & \multicolumn{1}{r}{ $\pm 0.014$ }  & \multicolumn{1}{r}{ $\pm 0.02$ }  & \multicolumn{1}{r}{ $\pm 0.05$ }  & \multicolumn{1}{r}{ $\pm 0.08$ }  & \multicolumn{1}{r}{ $\pm 0.1$ }  & \multicolumn{1}{r}{ $\pm 0.2$ }  & \multicolumn{1}{r}{ $\pm 0.2$ }  & \multicolumn{1}{r}{ $\pm 0.3$ }  & \multicolumn{1}{r}{ $\pm 0.3$ }  & \multicolumn{1}{r}{ $\pm 0.4$ } \\[1ex] 
$2.0-$ & $\, 0.028$ & $\, 0.043$ & $\, 0.099$ & $\, 0.28$ & $\, 0.70$ & $\, 1.4$ & $\, 2.8$ & $\, 3.6$ & $\, 3.8$ & $\, 2.7$ & $\, 2.1$ & $\, 1.3$\\ 
\multicolumn{1}{r|}{ 2.8 }  & \multicolumn{1}{r}{ $\pm 0.007$ }  & \multicolumn{1}{r}{ $\pm 0.010$ }  & \multicolumn{1}{r}{ $\pm 0.018$ }  & \multicolumn{1}{r}{ $\pm 0.04$ }  & \multicolumn{1}{r}{ $\pm 0.07$ }  & \multicolumn{1}{r}{ $\pm 0.1$ }  & \multicolumn{1}{r}{ $\pm 0.2$ }  & \multicolumn{1}{r}{ $\pm 0.3$ }  & \multicolumn{1}{r}{ $\pm 0.3$ }  & \multicolumn{1}{r}{ $\pm 0.4$ }  & \multicolumn{1}{r}{ $\pm 0.4$ }  & \multicolumn{1}{r}{ $\pm 0.4$ } \\[1ex] 
$1.4-$ & $\, 0.039$ & $\, 0.12$ & $\, 0.20$ & $\, 0.52$ & $\, 1.0$ & $\, 1.7$ & $\, 2.4$ & $\, 2.1$ & $\, 1.8$ & $\, 1.4$ & $\, 1.4$ & $\, 2.1$\\ 
\multicolumn{1}{r|}{ 2.0 }  & \multicolumn{1}{r}{ $\pm 0.008$ }  & \multicolumn{1}{r}{ $\pm 0.02$ }  & \multicolumn{1}{r}{ $\pm 0.03$ }  & \multicolumn{1}{r}{ $\pm 0.05$ }  & \multicolumn{1}{r}{ $\pm 0.1$ }  & \multicolumn{1}{r}{ $\pm 0.1$ }  & \multicolumn{1}{r}{ $\pm 0.2$ }  & \multicolumn{1}{r}{ $\pm 0.2$ }  & \multicolumn{1}{r}{ $\pm 0.3$ }  & \multicolumn{1}{r}{ $\pm 0.3$ }  & \multicolumn{1}{r}{ $\pm 0.4$ }  & \multicolumn{1}{r}{ $\pm 0.7$ } \\[1ex] 
$1.0-$ & $\, 0.038$ & $\, 0.11$ & $\, 0.18$ & $\, 0.43$ & $\, 1.0$ & $\, 1.3$ & $\, 1.9$ & $\, 1.6$ & $\, 1.6$ & $\, 1.5$ & $\, 3.8$ & $\, 7.5$\\ 
\multicolumn{1}{r|}{ 1.4 }  & \multicolumn{1}{r}{ $\pm 0.008$ }  & \multicolumn{1}{r}{ $\pm 0.02$ }  & \multicolumn{1}{r}{ $\pm 0.03$ }  & \multicolumn{1}{r}{ $\pm 0.05$ }  & \multicolumn{1}{r}{ $\pm 0.1$ }  & \multicolumn{1}{r}{ $\pm 0.1$ }  & \multicolumn{1}{r}{ $\pm 0.2$ }  & \multicolumn{1}{r}{ $\pm 0.3$ }  & \multicolumn{1}{r}{ $\pm 0.4$ }  & \multicolumn{1}{r}{ $\pm 0.5$ }  & \multicolumn{1}{r}{ $\pm 1.3$ }  & \multicolumn{1}{r}{ $\pm 2.6$ } \\[1ex] 
$0.71-$ & $\, 0.019$ & $\, 0.064$ & $\, 0.12$ & $\, 0.34$ & $\, 0.69$ & $\, 1.6$ & $\, 1.6$ & $\, 1.5$ & $\, 0.25$ & $\, 6.4$ & $\, 9.7$ & \\ 
\multicolumn{1}{r|}{ 1.0 }  & \multicolumn{1}{r}{ $\pm 0.006$ }  & \multicolumn{1}{r}{ $\pm 0.014$ }  & \multicolumn{1}{r}{ $\pm 0.03$ }  & \multicolumn{1}{r}{ $\pm 0.06$ }  & \multicolumn{1}{r}{ $\pm 0.10$ }  & \multicolumn{1}{r}{ $\pm 0.2$ }  & \multicolumn{1}{r}{ $\pm 0.3$ }  & \multicolumn{1}{r}{ $\pm 0.4$ }  & \multicolumn{1}{r}{ $\pm 0.34$ }  & \multicolumn{1}{r}{ $\pm 2.9$ }  & \multicolumn{1}{r}{ $\pm 6.3$ }  & \multicolumn{1}{r}{ ${<}13.2$ } \\[1ex] 
$0.5-$ & $\, 0.019$ & $\, 0.040$ & $\, 0.095$ & $\, 0.33$ & $\, 1.6$ & $\, 1.5$ & $\, 1.5$ & $\, 6.4$ & $\, 19.9$ &  & $\, 230.3$ & $\, 72.3$\\ 
\multicolumn{1}{r|}{ 0.71 }  & \multicolumn{1}{r}{ $\pm 0.008$ }  & \multicolumn{1}{r}{ $\pm 0.017$ }  & \multicolumn{1}{r}{ $\pm 0.037$ }  & \multicolumn{1}{r}{ $\pm 0.10$ }  & \multicolumn{1}{r}{ $\pm 0.3$ }  & \multicolumn{1}{r}{ $\pm 0.4$ }  & \multicolumn{1}{r}{ $\pm 0.8$ }  & \multicolumn{1}{r}{ $\pm 3.1$ }  & \multicolumn{1}{r}{ $\pm 6.4$ }  & \multicolumn{1}{r}{ ${<}16.3$ }  & \multicolumn{1}{r}{ $\pm 69.0$ }  & \multicolumn{1}{r}{ $\pm 53.6$ } \\[1ex] 
$0.35-$ &  & $\, 0.029$ & $\, 0.061$ & $\, 0.11$ &  & $\, 0.75$ &  & $\, 80.9$ &  & $\, 119.2$ &  & \\ 
\multicolumn{1}{r|}{ 0.5 }  & \multicolumn{1}{r}{ ${<}0.018$ }  & \multicolumn{1}{r}{ $\pm 0.029$ }  & \multicolumn{1}{r}{ $\pm 0.074$ }  & \multicolumn{1}{r}{ $\pm 0.14$ }  & \multicolumn{1}{r}{ ${<}0.82$ }  & \multicolumn{1}{r}{ $\pm 0.79$ }  & \multicolumn{1}{r}{ ${<}6.5$ }  & \multicolumn{1}{r}{ $\pm 12.8$ }  & \multicolumn{1}{r}{ ${<}54.7$ }  & \multicolumn{1}{r}{ $\pm 51.8$ }  & \multicolumn{1}{r}{ ${<}389.7$ }  & \multicolumn{1}{r}{ ${<}867.9$ } \\ 
\hline\hline
\end{tabular}

		\caption{Occurrence rates per bin, for the entire \textit{Kepler} sample. Columns are orbital period $P$ in days, rows are planet radius $R_p$ in Earth radii}
		\label{tab:occ}
	\end{table} 
	
	\begin{table}[ht]
		\title{Planet occurrence rates}
		\centering\tiny
		\begin{tabular}{l | llllllllllll}
\hline\hline
 & \multicolumn{12}{c}{Period [days]} \\ 
\hline
 & $0.4-$ & $0.68-$ & $1.2-$ & $2.0-$ & $3.4-$ & $5.8-$ & $10.0-$ & $17.0-$ & $29.0-$ & $50.0-$ & $85.0-$ & $150.0-$\\ 
 & \multicolumn{1}{r}{ 0.68 } & \multicolumn{1}{r}{ 1.2 } & \multicolumn{1}{r}{ 2.0 } & \multicolumn{1}{r}{ 3.4 } & \multicolumn{1}{r}{ 5.8 } & \multicolumn{1}{r}{ 10.0 } & \multicolumn{1}{r}{ 17.0 } & \multicolumn{1}{r}{ 29.0 } & \multicolumn{1}{r}{ 50.0 } & \multicolumn{1}{r}{ 85.0 } & \multicolumn{1}{r}{ 150.0 } & \multicolumn{1}{r}{ 250.0 }\\ 
\hline
$32.0-$ &  &  &  &  &  &  &  &  &  &  &  & \\ 
\multicolumn{1}{r|}{ 45.0 }  & \multicolumn{1}{r}{ ${<}0.12$ }  & \multicolumn{1}{r}{ ${<}0.18$ }  & \multicolumn{1}{r}{ ${<}0.26$ }  & \multicolumn{1}{r}{ ${<}0.37$ }  & \multicolumn{1}{r}{ ${<}0.52$ }  & \multicolumn{1}{r}{ ${<}0.75$ }  & \multicolumn{1}{r}{ ${<}1.1$ }  & \multicolumn{1}{r}{ ${<}1.6$ }  & \multicolumn{1}{r}{ ${<}2.6$ }  & \multicolumn{1}{r}{ ${<}4.0$ }  & \multicolumn{1}{r}{ ${<}6.5$ }  & \multicolumn{1}{r}{ ${<}10.7$ } \\[1ex] 
$23.0-$ &  &  &  &  &  &  &  &  &  &  &  & \\ 
\multicolumn{1}{r|}{ 32.0 }  & \multicolumn{1}{r}{ ${<}0.15$ }  & \multicolumn{1}{r}{ ${<}0.21$ }  & \multicolumn{1}{r}{ ${<}0.30$ }  & \multicolumn{1}{r}{ ${<}0.43$ }  & \multicolumn{1}{r}{ ${<}0.62$ }  & \multicolumn{1}{r}{ ${<}0.89$ }  & \multicolumn{1}{r}{ ${<}1.3$ }  & \multicolumn{1}{r}{ ${<}1.8$ }  & \multicolumn{1}{r}{ ${<}3.0$ }  & \multicolumn{1}{r}{ ${<}4.8$ }  & \multicolumn{1}{r}{ ${<}7.6$ }  & \multicolumn{1}{r}{ ${<}12.5$ } \\[1ex] 
$16.0-$ &  &  &  &  &  &  &  &  &  &  &  & \\ 
\multicolumn{1}{r|}{ 23.0 }  & \multicolumn{1}{r}{ ${<}0.17$ }  & \multicolumn{1}{r}{ ${<}0.24$ }  & \multicolumn{1}{r}{ ${<}0.35$ }  & \multicolumn{1}{r}{ ${<}0.50$ }  & \multicolumn{1}{r}{ ${<}0.71$ }  & \multicolumn{1}{r}{ ${<}1.0$ }  & \multicolumn{1}{r}{ ${<}1.5$ }  & \multicolumn{1}{r}{ ${<}2.1$ }  & \multicolumn{1}{r}{ ${<}3.5$ }  & \multicolumn{1}{r}{ ${<}5.5$ }  & \multicolumn{1}{r}{ ${<}8.7$ }  & \multicolumn{1}{r}{ ${<}14.3$ } \\[1ex] 
$11.0-$ &  &  &  &  &  &  &  &  &  &  &  & \\ 
\multicolumn{1}{r|}{ 16.0 }  & \multicolumn{1}{r}{ ${<}0.19$ }  & \multicolumn{1}{r}{ ${<}0.27$ }  & \multicolumn{1}{r}{ ${<}0.39$ }  & \multicolumn{1}{r}{ ${<}0.56$ }  & \multicolumn{1}{r}{ ${<}0.80$ }  & \multicolumn{1}{r}{ ${<}1.1$ }  & \multicolumn{1}{r}{ ${<}1.6$ }  & \multicolumn{1}{r}{ ${<}2.4$ }  & \multicolumn{1}{r}{ ${<}3.9$ }  & \multicolumn{1}{r}{ ${<}6.1$ }  & \multicolumn{1}{r}{ ${<}9.7$ }  & \multicolumn{1}{r}{ ${<}15.9$ } \\[1ex] 
$8.0-$ &  &  &  &  &  &  &  &  &  &  &  & \\ 
\multicolumn{1}{r|}{ 11.0 }  & \multicolumn{1}{r}{ ${<}0.21$ }  & \multicolumn{1}{r}{ ${<}0.30$ }  & \multicolumn{1}{r}{ ${<}0.43$ }  & \multicolumn{1}{r}{ ${<}0.61$ }  & \multicolumn{1}{r}{ ${<}0.87$ }  & \multicolumn{1}{r}{ ${<}1.2$ }  & \multicolumn{1}{r}{ ${<}1.8$ }  & \multicolumn{1}{r}{ ${<}2.6$ }  & \multicolumn{1}{r}{ ${<}4.2$ }  & \multicolumn{1}{r}{ ${<}6.7$ }  & \multicolumn{1}{r}{ ${<}10.6$ }  & \multicolumn{1}{r}{ ${<}17.2$ } \\[1ex] 
$5.7-$ &  & $\, 0.36$ & $\, 0.42$ &  &  &  &  &  &  &  &  & \\ 
\multicolumn{1}{r|}{ 8.0 }  & \multicolumn{1}{r}{ ${<}0.22$ }  & \multicolumn{1}{r}{ $\pm 0.41$ }  & \multicolumn{1}{r}{ $\pm 0.59$ }  & \multicolumn{1}{r}{ ${<}0.65$ }  & \multicolumn{1}{r}{ ${<}0.93$ }  & \multicolumn{1}{r}{ ${<}1.3$ }  & \multicolumn{1}{r}{ ${<}1.9$ }  & \multicolumn{1}{r}{ ${<}2.8$ }  & \multicolumn{1}{r}{ ${<}4.5$ }  & \multicolumn{1}{r}{ ${<}7.2$ }  & \multicolumn{1}{r}{ ${<}11.3$ }  & \multicolumn{1}{r}{ ${<}18.3$ } \\[1ex] 
$4.0-$ &  &  &  &  &  &  &  &  &  &  &  & \\ 
\multicolumn{1}{r|}{ 5.7 }  & \multicolumn{1}{r}{ ${<}0.24$ }  & \multicolumn{1}{r}{ ${<}0.34$ }  & \multicolumn{1}{r}{ ${<}0.48$ }  & \multicolumn{1}{r}{ ${<}0.69$ }  & \multicolumn{1}{r}{ ${<}0.98$ }  & \multicolumn{1}{r}{ ${<}1.4$ }  & \multicolumn{1}{r}{ ${<}2.0$ }  & \multicolumn{1}{r}{ ${<}2.9$ }  & \multicolumn{1}{r}{ ${<}4.8$ }  & \multicolumn{1}{r}{ ${<}7.5$ }  & \multicolumn{1}{r}{ ${<}11.8$ }  & \multicolumn{1}{r}{ ${<}19.2$ } \\[1ex] 
$2.8-$ &  &  & $\, 0.52$ &  &  &  & $\, 2.0$ &  &  &  &  & \\ 
\multicolumn{1}{r|}{ 4.0 }  & \multicolumn{1}{r}{ ${<}0.24$ }  & \multicolumn{1}{r}{ ${<}0.35$ }  & \multicolumn{1}{r}{ $\pm 0.64$ }  & \multicolumn{1}{r}{ ${<}0.72$ }  & \multicolumn{1}{r}{ ${<}1.0$ }  & \multicolumn{1}{r}{ ${<}1.5$ }  & \multicolumn{1}{r}{ $\pm 2.7$ }  & \multicolumn{1}{r}{ ${<}3.1$ }  & \multicolumn{1}{r}{ ${<}5.0$ }  & \multicolumn{1}{r}{ ${<}7.8$ }  & \multicolumn{1}{r}{ ${<}12.3$ }  & \multicolumn{1}{r}{ ${<}19.9$ } \\[1ex] 
$2.0-$ &  &  &  & $\, 1.6$ & $\, 3.4$ & $\, 1.8$ & $\, 2.2$ & $\, 3.2$ &  & $\, 7.4$ &  & \\ 
\multicolumn{1}{r|}{ 2.8 }  & \multicolumn{1}{r}{ ${<}0.25$ }  & \multicolumn{1}{r}{ ${<}0.36$ }  & \multicolumn{1}{r}{ ${<}0.52$ }  & \multicolumn{1}{r}{ $\pm 1.2$ }  & \multicolumn{1}{r}{ $\pm 2.0$ }  & \multicolumn{1}{r}{ $\pm 2.0$ }  & \multicolumn{1}{r}{ $\pm 2.8$ }  & \multicolumn{1}{r}{ $\pm 4.2$ }  & \multicolumn{1}{r}{ ${<}5.2$ }  & \multicolumn{1}{r}{ $\pm 10.6$ }  & \multicolumn{1}{r}{ ${<}12.7$ }  & \multicolumn{1}{r}{ ${<}20.6$ } \\[1ex] 
$1.4-$ &  &  & $\, 0.59$ & $\, 0.87$ & $\, 2.7$ & $\, 4.0$ & $\, 2.6$ & $\, 9.6$ & $\, 5.4$ &  &  & \\ 
\multicolumn{1}{r|}{ 2.0 }  & \multicolumn{1}{r}{ ${<}0.26$ }  & \multicolumn{1}{r}{ ${<}0.37$ }  & \multicolumn{1}{r}{ $\pm 0.69$ }  & \multicolumn{1}{r}{ $\pm 1.02$ }  & \multicolumn{1}{r}{ $\pm 1.9$ }  & \multicolumn{1}{r}{ $\pm 2.7$ }  & \multicolumn{1}{r}{ $\pm 3.2$ }  & \multicolumn{1}{r}{ $\pm 6.8$ }  & \multicolumn{1}{r}{ $\pm 7.2$ }  & \multicolumn{1}{r}{ ${<}8.7$ }  & \multicolumn{1}{r}{ ${<}13.7$ }  & \multicolumn{1}{r}{ ${<}22.6$ } \\[1ex] 
$1.0-$ &  & $\, 0.80$ & $\, 0.60$ & $\, 2.7$ & $\, 1.2$ & $\, 7.1$ & $\, 8.5$ &  & $\, 13.8$ & $\, 9.9$ &  & \\ 
\multicolumn{1}{r|}{ 1.4 }  & \multicolumn{1}{r}{ ${<}0.27$ }  & \multicolumn{1}{r}{ $\pm 0.64$ }  & \multicolumn{1}{r}{ $\pm 0.72$ }  & \multicolumn{1}{r}{ $\pm 1.6$ }  & \multicolumn{1}{r}{ $\pm 1.5$ }  & \multicolumn{1}{r}{ $\pm 3.9$ }  & \multicolumn{1}{r}{ $\pm 5.2$ }  & \multicolumn{1}{r}{ ${<}4.3$ }  & \multicolumn{1}{r}{ $\pm 11.5$ }  & \multicolumn{1}{r}{ $\pm 13.1$ }  & \multicolumn{1}{r}{ ${<}19.3$ }  & \multicolumn{1}{r}{ ${<}36.8$ } \\[1ex] 
$0.71-$ & $\, 0.27$ &  & $\, 1.9$ & $\, 1.2$ &  & $\, 4.6$ &  &  &  & $\, 21.4$ &  & \\ 
\multicolumn{1}{r|}{ 1.0 }  & \multicolumn{1}{r}{ $\pm 0.38$ }  & \multicolumn{1}{r}{ ${<}0.44$ }  & \multicolumn{1}{r}{ $\pm 1.2$ }  & \multicolumn{1}{r}{ $\pm 1.4$ }  & \multicolumn{1}{r}{ ${<}1.5$ }  & \multicolumn{1}{r}{ $\pm 4.0$ }  & \multicolumn{1}{r}{ ${<}3.9$ }  & \multicolumn{1}{r}{ ${<}6.8$ }  & \multicolumn{1}{r}{ ${<}12.4$ }  & \multicolumn{1}{r}{ $\pm 27.9$ }  & \multicolumn{1}{r}{ ${<}50.6$ }  & \multicolumn{1}{r}{ ${<}110.7$ } \\[1ex] 
$0.5-$ & $\, 0.29$ & $\, 0.51$ & $\, 0.88$ & $\, 1.1$ & $\, 4.9$ & $\, 4.5$ &  &  &  &  &  & $\, 278.8$\\ 
\multicolumn{1}{r|}{ 0.71 }  & \multicolumn{1}{r}{ $\pm 0.43$ }  & \multicolumn{1}{r}{ $\pm 0.75$ }  & \multicolumn{1}{r}{ $\pm 0.99$ }  & \multicolumn{1}{r}{ $\pm 1.6$ }  & \multicolumn{1}{r}{ $\pm 4.8$ }  & \multicolumn{1}{r}{ $\pm 5.0$ }  & \multicolumn{1}{r}{ ${<}9.4$ }  & \multicolumn{1}{r}{ ${<}19.2$ }  & \multicolumn{1}{r}{ ${<}39.5$ }  & \multicolumn{1}{r}{ ${<}82.2$ }  & \multicolumn{1}{r}{ ${<}179.6$ }  & \multicolumn{1}{r}{ $\pm 271.0$ } \\[1ex] 
$0.35-$ &  &  &  &  &  &  &  &  &  &  &  & \\ 
\multicolumn{1}{r|}{ 0.5 }  & \multicolumn{1}{r}{ ${<}0.54$ }  & \multicolumn{1}{r}{ ${<}0.98$ }  & \multicolumn{1}{r}{ ${<}1.9$ }  & \multicolumn{1}{r}{ ${<}3.7$ }  & \multicolumn{1}{r}{ ${<}7.5$ }  & \multicolumn{1}{r}{ ${<}15.2$ }  & \multicolumn{1}{r}{ ${<}32.2$ }  & \multicolumn{1}{r}{ ${<}68.4$ }  & \multicolumn{1}{r}{ ${<}137.4$ }  & \multicolumn{1}{r}{ ${<}275.3$ }  & \multicolumn{1}{r}{ ${<}545.7$ }  & \multicolumn{1}{r}{ ${<}1001.2$ } \\ 
\hline\hline
\end{tabular}

		\caption{Same as table \ref{tab:occ}, but for M stars.}
		\label{tab:occ0}
	\end{table} 

	\begin{table}[ht]
		\title{Planet occurrence rates}
		\centering\tiny
		\begin{tabular}{l | llllllllllll}
\hline\hline
 & \multicolumn{12}{c}{Period [days]} \\ 
\hline
 & $0.4-$ & $0.68-$ & $1.2-$ & $2.0-$ & $3.4-$ & $5.8-$ & $10.0-$ & $17.0-$ & $29.0-$ & $50.0-$ & $85.0-$ & $150.0-$\\ 
 & \multicolumn{1}{r}{ 0.68 } & \multicolumn{1}{r}{ 1.2 } & \multicolumn{1}{r}{ 2.0 } & \multicolumn{1}{r}{ 3.4 } & \multicolumn{1}{r}{ 5.8 } & \multicolumn{1}{r}{ 10.0 } & \multicolumn{1}{r}{ 17.0 } & \multicolumn{1}{r}{ 29.0 } & \multicolumn{1}{r}{ 50.0 } & \multicolumn{1}{r}{ 85.0 } & \multicolumn{1}{r}{ 150.0 } & \multicolumn{1}{r}{ 250.0 }\\ 
\hline
$32.0-$ &  &  &  &  &  &  &  &  & $\, 0.14$ &  &  & \\ 
\multicolumn{1}{r|}{ 45.0 }  & \multicolumn{1}{r}{ ${<}0.0091$ }  & \multicolumn{1}{r}{ ${<}0.013$ }  & \multicolumn{1}{r}{ ${<}0.019$ }  & \multicolumn{1}{r}{ ${<}0.027$ }  & \multicolumn{1}{r}{ ${<}0.038$ }  & \multicolumn{1}{r}{ ${<}0.054$ }  & \multicolumn{1}{r}{ ${<}0.078$ }  & \multicolumn{1}{r}{ ${<}0.11$ }  & \multicolumn{1}{r}{ $\pm 0.21$ }  & \multicolumn{1}{r}{ ${<}0.24$ }  & \multicolumn{1}{r}{ ${<}0.36$ }  & \multicolumn{1}{r}{ ${<}0.57$ } \\[1ex] 
$23.0-$ &  &  &  &  &  &  &  &  & $\, 0.20$ &  &  & \\ 
\multicolumn{1}{r|}{ 32.0 }  & \multicolumn{1}{r}{ ${<}0.010$ }  & \multicolumn{1}{r}{ ${<}0.014$ }  & \multicolumn{1}{r}{ ${<}0.021$ }  & \multicolumn{1}{r}{ ${<}0.030$ }  & \multicolumn{1}{r}{ ${<}0.042$ }  & \multicolumn{1}{r}{ ${<}0.060$ }  & \multicolumn{1}{r}{ ${<}0.087$ }  & \multicolumn{1}{r}{ ${<}0.12$ }  & \multicolumn{1}{r}{ $\pm 0.24$ }  & \multicolumn{1}{r}{ ${<}0.27$ }  & \multicolumn{1}{r}{ ${<}0.40$ }  & \multicolumn{1}{r}{ ${<}0.63$ } \\[1ex] 
$16.0-$ &  &  &  &  &  & $\, 0.13$ &  & $\, 0.12$ &  &  &  & \\ 
\multicolumn{1}{r|}{ 23.0 }  & \multicolumn{1}{r}{ ${<}0.011$ }  & \multicolumn{1}{r}{ ${<}0.016$ }  & \multicolumn{1}{r}{ ${<}0.022$ }  & \multicolumn{1}{r}{ ${<}0.032$ }  & \multicolumn{1}{r}{ ${<}0.046$ }  & \multicolumn{1}{r}{ $\pm 0.11$ }  & \multicolumn{1}{r}{ ${<}0.094$ }  & \multicolumn{1}{r}{ $\pm 0.17$ }  & \multicolumn{1}{r}{ ${<}0.20$ }  & \multicolumn{1}{r}{ ${<}0.29$ }  & \multicolumn{1}{r}{ ${<}0.43$ }  & \multicolumn{1}{r}{ ${<}0.68$ } \\[1ex] 
$11.0-$ &  &  &  & $\, 0.068$ & $\, 0.091$ &  &  &  & $\, 0.38$ &  & $\, 0.48$ & \\ 
\multicolumn{1}{r|}{ 16.0 }  & \multicolumn{1}{r}{ ${<}0.012$ }  & \multicolumn{1}{r}{ ${<}0.017$ }  & \multicolumn{1}{r}{ ${<}0.024$ }  & \multicolumn{1}{r}{ $\pm 0.055$ }  & \multicolumn{1}{r}{ $\pm 0.079$ }  & \multicolumn{1}{r}{ ${<}0.069$ }  & \multicolumn{1}{r}{ ${<}0.099$ }  & \multicolumn{1}{r}{ ${<}0.14$ }  & \multicolumn{1}{r}{ $\pm 0.34$ }  & \multicolumn{1}{r}{ ${<}0.31$ }  & \multicolumn{1}{r}{ $\pm 0.59$ }  & \multicolumn{1}{r}{ ${<}0.72$ } \\[1ex] 
$8.0-$ &  &  & $\, 0.056$ & $\, 0.040$ &  &  & $\, 0.44$ & $\, 0.43$ & $\, 0.39$ & $\, 0.36$ & $\, 0.50$ & $\, 2.0$\\ 
\multicolumn{1}{r|}{ 11.0 }  & \multicolumn{1}{r}{ ${<}0.012$ }  & \multicolumn{1}{r}{ ${<}0.017$ }  & \multicolumn{1}{r}{ $\pm 0.040$ }  & \multicolumn{1}{r}{ $\pm 0.046$ }  & \multicolumn{1}{r}{ ${<}0.051$ }  & \multicolumn{1}{r}{ ${<}0.072$ }  & \multicolumn{1}{r}{ $\pm 0.22$ }  & \multicolumn{1}{r}{ $\pm 0.28$ }  & \multicolumn{1}{r}{ $\pm 0.35$ }  & \multicolumn{1}{r}{ $\pm 0.42$ }  & \multicolumn{1}{r}{ $\pm 0.62$ }  & \multicolumn{1}{r}{ $\pm 1.5$ } \\[1ex] 
$5.7-$ &  &  & $\, 0.053$ & $\, 0.040$ & $\, 0.060$ & $\, 0.14$ & $\, 0.45$ & $\, 0.31$ & $\, 0.45$ & $\, 0.36$ & $\, 0.52$ & $\, 0.83$\\ 
\multicolumn{1}{r|}{ 8.0 }  & \multicolumn{1}{r}{ ${<}0.013$ }  & \multicolumn{1}{r}{ ${<}0.018$ }  & \multicolumn{1}{r}{ $\pm 0.042$ }  & \multicolumn{1}{r}{ $\pm 0.047$ }  & \multicolumn{1}{r}{ $\pm 0.067$ }  & \multicolumn{1}{r}{ $\pm 0.12$ }  & \multicolumn{1}{r}{ $\pm 0.23$ }  & \multicolumn{1}{r}{ $\pm 0.25$ }  & \multicolumn{1}{r}{ $\pm 0.37$ }  & \multicolumn{1}{r}{ $\pm 0.43$ }  & \multicolumn{1}{r}{ $\pm 0.64$ }  & \multicolumn{1}{r}{ $\pm 1.05$ } \\[1ex] 
$4.0-$ & $\, 0.028$ &  & $\, 0.027$ & $\, 0.16$ & $\, 0.23$ & $\, 0.44$ & $\, 0.58$ & $\, 0.63$ & $\, 0.22$ & $\, 1.1$ & $\, 0.58$ & \\ 
\multicolumn{1}{r|}{ 5.7 }  & \multicolumn{1}{r}{ $\pm 0.021$ }  & \multicolumn{1}{r}{ ${<}0.018$ }  & \multicolumn{1}{r}{ $\pm 0.034$ }  & \multicolumn{1}{r}{ $\pm 0.08$ }  & \multicolumn{1}{r}{ $\pm 0.12$ }  & \multicolumn{1}{r}{ $\pm 0.20$ }  & \multicolumn{1}{r}{ $\pm 0.26$ }  & \multicolumn{1}{r}{ $\pm 0.34$ }  & \multicolumn{1}{r}{ $\pm 0.30$ }  & \multicolumn{1}{r}{ $\pm 0.7$ }  & \multicolumn{1}{r}{ $\pm 0.66$ }  & \multicolumn{1}{r}{ ${<}0.80$ } \\[1ex] 
$2.8-$ & $\, 0.028$ & $\, 0.035$ & $\, 0.083$ & $\, 0.11$ & $\, 0.70$ & $\, 0.62$ & $\, 1.4$ & $\, 1.5$ & $\, 2.1$ & $\, 1.7$ & $\, 1.4$ & $\, 2.0$\\ 
\multicolumn{1}{r|}{ 4.0 }  & \multicolumn{1}{r}{ $\pm 0.021$ }  & \multicolumn{1}{r}{ $\pm 0.030$ }  & \multicolumn{1}{r}{ $\pm 0.051$ }  & \multicolumn{1}{r}{ $\pm 0.07$ }  & \multicolumn{1}{r}{ $\pm 0.20$ }  & \multicolumn{1}{r}{ $\pm 0.23$ }  & \multicolumn{1}{r}{ $\pm 0.4$ }  & \multicolumn{1}{r}{ $\pm 0.5$ }  & \multicolumn{1}{r}{ $\pm 0.7$ }  & \multicolumn{1}{r}{ $\pm 0.8$ }  & \multicolumn{1}{r}{ $\pm 1.0$ }  & \multicolumn{1}{r}{ $\pm 1.5$ } \\[1ex] 
$2.0-$ & $\, 0.014$ & $\, 0.022$ & $\, 0.090$ & $\, 0.65$ & $\, 0.78$ & $\, 3.1$ & $\, 5.5$ & $\, 5.4$ & $\, 3.3$ & $\, 3.8$ & $\, 5.6$ & $\, 3.1$\\ 
\multicolumn{1}{r|}{ 2.8 }  & \multicolumn{1}{r}{ $\pm 0.017$ }  & \multicolumn{1}{r}{ $\pm 0.025$ }  & \multicolumn{1}{r}{ $\pm 0.053$ }  & \multicolumn{1}{r}{ $\pm 0.16$ }  & \multicolumn{1}{r}{ $\pm 0.21$ }  & \multicolumn{1}{r}{ $\pm 0.5$ }  & \multicolumn{1}{r}{ $\pm 0.8$ }  & \multicolumn{1}{r}{ $\pm 1.0$ }  & \multicolumn{1}{r}{ $\pm 0.9$ }  & \multicolumn{1}{r}{ $\pm 1.2$ }  & \multicolumn{1}{r}{ $\pm 1.8$ }  & \multicolumn{1}{r}{ $\pm 2.0$ } \\[1ex] 
$1.4-$ & $\, 0.072$ & $\, 0.26$ & $\, 0.32$ & $\, 0.95$ & $\, 1.8$ & $\, 3.3$ & $\, 4.0$ & $\, 4.1$ & $\, 4.0$ & $\, 2.8$ & $\, 2.3$ & $\, 2.3$\\ 
\multicolumn{1}{r|}{ 2.0 }  & \multicolumn{1}{r}{ $\pm 0.033$ }  & \multicolumn{1}{r}{ $\pm 0.07$ }  & \multicolumn{1}{r}{ $\pm 0.10$ }  & \multicolumn{1}{r}{ $\pm 0.20$ }  & \multicolumn{1}{r}{ $\pm 0.3$ }  & \multicolumn{1}{r}{ $\pm 0.5$ }  & \multicolumn{1}{r}{ $\pm 0.7$ }  & \multicolumn{1}{r}{ $\pm 0.9$ }  & \multicolumn{1}{r}{ $\pm 1.1$ }  & \multicolumn{1}{r}{ $\pm 1.2$ }  & \multicolumn{1}{r}{ $\pm 1.4$ }  & \multicolumn{1}{r}{ $\pm 2.9$ } \\[1ex] 
$1.0-$ & $\, 0.093$ & $\, 0.21$ & $\, 0.32$ & $\, 0.90$ & $\, 1.8$ & $\, 1.7$ & $\, 2.3$ & $\, 4.0$ & $\, 4.3$ & $\, 2.0$ & $\, 4.6$ & $\, 6.2$\\ 
\multicolumn{1}{r|}{ 1.4 }  & \multicolumn{1}{r}{ $\pm 0.037$ }  & \multicolumn{1}{r}{ $\pm 0.07$ }  & \multicolumn{1}{r}{ $\pm 0.10$ }  & \multicolumn{1}{r}{ $\pm 0.20$ }  & \multicolumn{1}{r}{ $\pm 0.4$ }  & \multicolumn{1}{r}{ $\pm 0.4$ }  & \multicolumn{1}{r}{ $\pm 0.7$ }  & \multicolumn{1}{r}{ $\pm 1.1$ }  & \multicolumn{1}{r}{ $\pm 1.6$ }  & \multicolumn{1}{r}{ $\pm 1.6$ }  & \multicolumn{1}{r}{ $\pm 3.8$ }  & \multicolumn{1}{r}{ $\pm 7.4$ } \\[1ex] 
$0.71-$ & $\, 0.050$ & $\, 0.14$ & $\, 0.18$ & $\, 0.49$ & $\, 1.5$ & $\, 2.9$ & $\, 1.7$ & $\, 1.9$ &  & $\, 7.4$ & $\, 14.7$ & \\ 
\multicolumn{1}{r|}{ 1.0 }  & \multicolumn{1}{r}{ $\pm 0.030$ }  & \multicolumn{1}{r}{ $\pm 0.06$ }  & \multicolumn{1}{r}{ $\pm 0.09$ }  & \multicolumn{1}{r}{ $\pm 0.18$ }  & \multicolumn{1}{r}{ $\pm 0.4$ }  & \multicolumn{1}{r}{ $\pm 0.8$ }  & \multicolumn{1}{r}{ $\pm 0.8$ }  & \multicolumn{1}{r}{ $\pm 2.3$ }  & \multicolumn{1}{r}{ ${<}3.0$ }  & \multicolumn{1}{r}{ $\pm 7.3$ }  & \multicolumn{1}{r}{ $\pm 18.8$ }  & \multicolumn{1}{r}{ ${<}43.6$ } \\[1ex] 
$0.5-$ & $\, 0.032$ & $\, 0.099$ & $\, 0.28$ & $\, 0.27$ & $\, 3.5$ & $\, 1.8$ & $\, 1.7$ & $\, 12.6$ &  &  & $\, 519.9$ & \\ 
\multicolumn{1}{r|}{ 0.71 }  & \multicolumn{1}{r}{ $\pm 0.037$ }  & \multicolumn{1}{r}{ $\pm 0.069$ }  & \multicolumn{1}{r}{ $\pm 0.19$ }  & \multicolumn{1}{r}{ $\pm 0.35$ }  & \multicolumn{1}{r}{ $\pm 1.0$ }  & \multicolumn{1}{r}{ $\pm 1.2$ }  & \multicolumn{1}{r}{ $\pm 2.3$ }  & \multicolumn{1}{r}{ $\pm 13.0$ }  & \multicolumn{1}{r}{ ${<}16.5$ }  & \multicolumn{1}{r}{ ${<}45.2$ }  & \multicolumn{1}{r}{ $\pm 457.2$ }  & \multicolumn{1}{r}{ ${<}436.6$ } \\[1ex] 
$0.35-$ &  &  & $\, 0.29$ &  &  &  &  & $\, 219.9$ &  &  &  & \\ 
\multicolumn{1}{r|}{ 0.5 }  & \multicolumn{1}{r}{ ${<}0.10$ }  & \multicolumn{1}{r}{ ${<}0.21$ }  & \multicolumn{1}{r}{ $\pm 0.37$ }  & \multicolumn{1}{r}{ ${<}1.0$ }  & \multicolumn{1}{r}{ ${<}2.4$ }  & \multicolumn{1}{r}{ ${<}6.1$ }  & \multicolumn{1}{r}{ ${<}16.0$ }  & \multicolumn{1}{r}{ $\pm 197.9$ }  & \multicolumn{1}{r}{ ${<}148.0$ }  & \multicolumn{1}{r}{ ${<}501.1$ }  & \multicolumn{1}{r}{ ${<}1832.4$ }  & \multicolumn{1}{r}{ ${<}7288.2$ } \\ 
\hline\hline
\end{tabular}

		\caption{Same as table \ref{tab:occ}, but for K stars.}
		\label{tab:occ1}
	\end{table} 

	\begin{table}[ht]
		\title{Planet occurrence rates}
		\centering\tiny
		\begin{tabular}{l | llllllllllll}
\hline\hline
 & \multicolumn{12}{c}{Period [days]} \\ 
\hline
 & $0.4-$ & $0.68-$ & $1.2-$ & $2.0-$ & $3.4-$ & $5.8-$ & $10.0-$ & $17.0-$ & $29.0-$ & $50.0-$ & $85.0-$ & $150.0-$\\ 
 & \multicolumn{1}{r}{ 0.68 } & \multicolumn{1}{r}{ 1.2 } & \multicolumn{1}{r}{ 2.0 } & \multicolumn{1}{r}{ 3.4 } & \multicolumn{1}{r}{ 5.8 } & \multicolumn{1}{r}{ 10.0 } & \multicolumn{1}{r}{ 17.0 } & \multicolumn{1}{r}{ 29.0 } & \multicolumn{1}{r}{ 50.0 } & \multicolumn{1}{r}{ 85.0 } & \multicolumn{1}{r}{ 150.0 } & \multicolumn{1}{r}{ 250.0 }\\ 
\hline
$32.0-$ &  &  &  &  & $\, 0.014$ &  &  & $\, 0.032$ &  &  &  & $\, 0.21$\\ 
\multicolumn{1}{r|}{ 45.0 }  & \multicolumn{1}{r}{ ${<}0.0029$ }  & \multicolumn{1}{r}{ ${<}0.0041$ }  & \multicolumn{1}{r}{ ${<}0.0058$ }  & \multicolumn{1}{r}{ ${<}0.0084$ }  & \multicolumn{1}{r}{ $\pm 0.015$ }  & \multicolumn{1}{r}{ ${<}0.017$ }  & \multicolumn{1}{r}{ ${<}0.024$ }  & \multicolumn{1}{r}{ $\pm 0.045$ }  & \multicolumn{1}{r}{ ${<}0.050$ }  & \multicolumn{1}{r}{ ${<}0.073$ }  & \multicolumn{1}{r}{ ${<}0.11$ }  & \multicolumn{1}{r}{ $\pm 0.24$ } \\[1ex] 
$23.0-$ &  &  &  & $\, 0.037$ & $\, 0.012$ & $\, 0.018$ & $\, 0.028$ & $\, 0.084$ & $\, 0.053$ &  & $\, 0.14$ & \\ 
\multicolumn{1}{r|}{ 32.0 }  & \multicolumn{1}{r}{ ${<}0.0031$ }  & \multicolumn{1}{r}{ ${<}0.0044$ }  & \multicolumn{1}{r}{ ${<}0.0064$ }  & \multicolumn{1}{r}{ $\pm 0.020$ }  & \multicolumn{1}{r}{ $\pm 0.017$ }  & \multicolumn{1}{r}{ $\pm 0.024$ }  & \multicolumn{1}{r}{ $\pm 0.034$ }  & \multicolumn{1}{r}{ $\pm 0.062$ }  & \multicolumn{1}{r}{ $\pm 0.071$ }  & \multicolumn{1}{r}{ ${<}0.079$ }  & \multicolumn{1}{r}{ $\pm 0.15$ }  & \multicolumn{1}{r}{ ${<}0.19$ } \\[1ex] 
$16.0-$ & $\, 0.0035$ & $\, 0.0051$ & $\, 0.0058$ & $\, 0.061$ & $\, 0.10$ & $\, 0.039$ & $\, 0.18$ & $\, 0.084$ & $\, 0.11$ & $\, 0.075$ &  & $\, 0.22$\\ 
\multicolumn{1}{r|}{ 23.0 }  & \multicolumn{1}{r}{ $\pm 0.0043$ }  & \multicolumn{1}{r}{ $\pm 0.0061$ }  & \multicolumn{1}{r}{ $\pm 0.0088$ }  & \multicolumn{1}{r}{ $\pm 0.025$ }  & \multicolumn{1}{r}{ $\pm 0.04$ }  & \multicolumn{1}{r}{ $\pm 0.032$ }  & \multicolumn{1}{r}{ $\pm 0.07$ }  & \multicolumn{1}{r}{ $\pm 0.066$ }  & \multicolumn{1}{r}{ $\pm 0.10$ }  & \multicolumn{1}{r}{ $\pm 0.109$ }  & \multicolumn{1}{r}{ ${<}0.13$ }  & \multicolumn{1}{r}{ $\pm 0.27$ } \\[1ex] 
$11.0-$ & $\, 0.0039$ & $\, 0.011$ & $\, 0.020$ & $\, 0.084$ & $\, 0.16$ & $\, 0.13$ & $\, 0.24$ & $\, 0.17$ & $\, 0.38$ & $\, 0.25$ & $\, 0.11$ & \\ 
\multicolumn{1}{r|}{ 16.0 }  & \multicolumn{1}{r}{ $\pm 0.0045$ }  & \multicolumn{1}{r}{ $\pm 0.008$ }  & \multicolumn{1}{r}{ $\pm 0.014$ }  & \multicolumn{1}{r}{ $\pm 0.030$ }  & \multicolumn{1}{r}{ $\pm 0.05$ }  & \multicolumn{1}{r}{ $\pm 0.05$ }  & \multicolumn{1}{r}{ $\pm 0.09$ }  & \multicolumn{1}{r}{ $\pm 0.09$ }  & \multicolumn{1}{r}{ $\pm 0.16$ }  & \multicolumn{1}{r}{ $\pm 0.17$ }  & \multicolumn{1}{r}{ $\pm 0.17$ }  & \multicolumn{1}{r}{ ${<}0.21$ } \\[1ex] 
$8.0-$ & $\, 0.011$ & $\, 0.021$ & $\, 0.041$ & $\, 0.022$ & $\, 0.092$ & $\, 0.15$ & $\, 0.090$ & $\, 0.21$ & $\, 0.40$ & $\, 0.48$ & $\, 0.60$ & $\, 0.84$\\ 
\multicolumn{1}{r|}{ 11.0 }  & \multicolumn{1}{r}{ $\pm 0.007$ }  & \multicolumn{1}{r}{ $\pm 0.011$ }  & \multicolumn{1}{r}{ $\pm 0.018$ }  & \multicolumn{1}{r}{ $\pm 0.017$ }  & \multicolumn{1}{r}{ $\pm 0.039$ }  & \multicolumn{1}{r}{ $\pm 0.06$ }  & \multicolumn{1}{r}{ $\pm 0.059$ }  & \multicolumn{1}{r}{ $\pm 0.11$ }  & \multicolumn{1}{r}{ $\pm 0.16$ }  & \multicolumn{1}{r}{ $\pm 0.22$ }  & \multicolumn{1}{r}{ $\pm 0.30$ }  & \multicolumn{1}{r}{ $\pm 0.46$ } \\[1ex] 
$5.7-$ &  & $\, 0.031$ & $\, 0.017$ & $\, 0.047$ & $\, 0.23$ & $\, 0.19$ & $\, 0.23$ & $\, 0.18$ & $\, 0.62$ & $\, 0.73$ & $\, 1.1$ & $\, 1.3$\\ 
\multicolumn{1}{r|}{ 8.0 }  & \multicolumn{1}{r}{ ${<}0.0037$ }  & \multicolumn{1}{r}{ $\pm 0.014$ }  & \multicolumn{1}{r}{ $\pm 0.012$ }  & \multicolumn{1}{r}{ $\pm 0.023$ }  & \multicolumn{1}{r}{ $\pm 0.06$ }  & \multicolumn{1}{r}{ $\pm 0.07$ }  & \multicolumn{1}{r}{ $\pm 0.09$ }  & \multicolumn{1}{r}{ $\pm 0.10$ }  & \multicolumn{1}{r}{ $\pm 0.21$ }  & \multicolumn{1}{r}{ $\pm 0.28$ }  & \multicolumn{1}{r}{ $\pm 0.4$ }  & \multicolumn{1}{r}{ $\pm 0.6$ } \\[1ex] 
$4.0-$ & $\, 0.011$ & $\, 0.021$ & $\, 0.058$ & $\, 0.11$ & $\, 0.15$ & $\, 0.33$ & $\, 0.43$ & $\, 0.61$ & $\, 0.67$ & $\, 0.77$ & $\, 1.0$ & $\, 1.4$\\ 
\multicolumn{1}{r|}{ 5.7 }  & \multicolumn{1}{r}{ $\pm 0.007$ }  & \multicolumn{1}{r}{ $\pm 0.012$ }  & \multicolumn{1}{r}{ $\pm 0.021$ }  & \multicolumn{1}{r}{ $\pm 0.04$ }  & \multicolumn{1}{r}{ $\pm 0.05$ }  & \multicolumn{1}{r}{ $\pm 0.09$ }  & \multicolumn{1}{r}{ $\pm 0.12$ }  & \multicolumn{1}{r}{ $\pm 0.17$ }  & \multicolumn{1}{r}{ $\pm 0.22$ }  & \multicolumn{1}{r}{ $\pm 0.28$ }  & \multicolumn{1}{r}{ $\pm 0.4$ }  & \multicolumn{1}{r}{ $\pm 0.6$ } \\[1ex] 
$2.8-$ & $\, 0.0071$ & $\, 0.040$ & $\, 0.053$ & $\, 0.10$ & $\, 0.39$ & $\, 0.78$ & $\, 1.9$ & $\, 2.2$ & $\, 2.2$ & $\, 2.0$ & $\, 2.1$ & $\, 1.8$\\ 
\multicolumn{1}{r|}{ 4.0 }  & \multicolumn{1}{r}{ $\pm 0.0062$ }  & \multicolumn{1}{r}{ $\pm 0.015$ }  & \multicolumn{1}{r}{ $\pm 0.022$ }  & \multicolumn{1}{r}{ $\pm 0.03$ }  & \multicolumn{1}{r}{ $\pm 0.08$ }  & \multicolumn{1}{r}{ $\pm 0.13$ }  & \multicolumn{1}{r}{ $\pm 0.3$ }  & \multicolumn{1}{r}{ $\pm 0.3$ }  & \multicolumn{1}{r}{ $\pm 0.4$ }  & \multicolumn{1}{r}{ $\pm 0.5$ }  & \multicolumn{1}{r}{ $\pm 0.6$ }  & \multicolumn{1}{r}{ $\pm 0.7$ } \\[1ex] 
$2.0-$ & $\, 0.036$ & $\, 0.044$ & $\, 0.094$ & $\, 0.30$ & $\, 0.86$ & $\, 1.5$ & $\, 3.3$ & $\, 3.9$ & $\, 5.0$ & $\, 2.4$ & $\, 2.3$ & $\, 1.7$\\ 
\multicolumn{1}{r|}{ 2.8 }  & \multicolumn{1}{r}{ $\pm 0.012$ }  & \multicolumn{1}{r}{ $\pm 0.016$ }  & \multicolumn{1}{r}{ $\pm 0.027$ }  & \multicolumn{1}{r}{ $\pm 0.06$ }  & \multicolumn{1}{r}{ $\pm 0.12$ }  & \multicolumn{1}{r}{ $\pm 0.2$ }  & \multicolumn{1}{r}{ $\pm 0.3$ }  & \multicolumn{1}{r}{ $\pm 0.4$ }  & \multicolumn{1}{r}{ $\pm 0.6$ }  & \multicolumn{1}{r}{ $\pm 0.5$ }  & \multicolumn{1}{r}{ $\pm 0.6$ }  & \multicolumn{1}{r}{ $\pm 0.7$ } \\[1ex] 
$1.4-$ & $\, 0.041$ & $\, 0.11$ & $\, 0.21$ & $\, 0.59$ & $\, 1.2$ & $\, 1.6$ & $\, 2.3$ & $\, 1.3$ & $\, 1.8$ & $\, 1.3$ & $\, 1.6$ & $\, 3.0$\\ 
\multicolumn{1}{r|}{ 2.0 }  & \multicolumn{1}{r}{ $\pm 0.013$ }  & \multicolumn{1}{r}{ $\pm 0.03$ }  & \multicolumn{1}{r}{ $\pm 0.04$ }  & \multicolumn{1}{r}{ $\pm 0.08$ }  & \multicolumn{1}{r}{ $\pm 0.1$ }  & \multicolumn{1}{r}{ $\pm 0.2$ }  & \multicolumn{1}{r}{ $\pm 0.3$ }  & \multicolumn{1}{r}{ $\pm 0.3$ }  & \multicolumn{1}{r}{ $\pm 0.4$ }  & \multicolumn{1}{r}{ $\pm 0.5$ }  & \multicolumn{1}{r}{ $\pm 0.6$ }  & \multicolumn{1}{r}{ $\pm 1.3$ } \\[1ex] 
$1.0-$ & $\, 0.046$ & $\, 0.14$ & $\, 0.20$ & $\, 0.47$ & $\, 1.1$ & $\, 1.4$ & $\, 1.8$ & $\, 1.9$ & $\, 0.82$ & $\, 1.8$ & $\, 2.8$ & $\, 13.5$\\ 
\multicolumn{1}{r|}{ 1.4 }  & \multicolumn{1}{r}{ $\pm 0.013$ }  & \multicolumn{1}{r}{ $\pm 0.03$ }  & \multicolumn{1}{r}{ $\pm 0.04$ }  & \multicolumn{1}{r}{ $\pm 0.08$ }  & \multicolumn{1}{r}{ $\pm 0.2$ }  & \multicolumn{1}{r}{ $\pm 0.2$ }  & \multicolumn{1}{r}{ $\pm 0.3$ }  & \multicolumn{1}{r}{ $\pm 0.4$ }  & \multicolumn{1}{r}{ $\pm 0.40$ }  & \multicolumn{1}{r}{ $\pm 0.9$ }  & \multicolumn{1}{r}{ $\pm 2.4$ }  & \multicolumn{1}{r}{ $\pm 5.6$ } \\[1ex] 
$0.71-$ & $\, 0.016$ & $\, 0.060$ & $\, 0.060$ & $\, 0.41$ & $\, 0.77$ & $\, 1.8$ & $\, 2.8$ & $\, 2.2$ & $\, 0.56$ & $\, 6.8$ &  & \\ 
\multicolumn{1}{r|}{ 1.0 }  & \multicolumn{1}{r}{ $\pm 0.010$ }  & \multicolumn{1}{r}{ $\pm 0.022$ }  & \multicolumn{1}{r}{ $\pm 0.029$ }  & \multicolumn{1}{r}{ $\pm 0.10$ }  & \multicolumn{1}{r}{ $\pm 0.17$ }  & \multicolumn{1}{r}{ $\pm 0.4$ }  & \multicolumn{1}{r}{ $\pm 0.7$ }  & \multicolumn{1}{r}{ $\pm 0.9$ }  & \multicolumn{1}{r}{ $\pm 0.82$ }  & \multicolumn{1}{r}{ $\pm 8.3$ }  & \multicolumn{1}{r}{ ${<}9.4$ }  & \multicolumn{1}{r}{ ${<}27.6$ } \\[1ex] 
$0.5-$ & $\, 0.014$ & $\, 0.014$ & $\, 0.084$ & $\, 0.61$ & $\, 1.2$ & $\, 1.5$ & $\, 2.3$ &  & $\, 43.3$ &  &  & \\ 
\multicolumn{1}{r|}{ 0.71 }  & \multicolumn{1}{r}{ $\pm 0.016$ }  & \multicolumn{1}{r}{ $\pm 0.019$ }  & \multicolumn{1}{r}{ $\pm 0.065$ }  & \multicolumn{1}{r}{ $\pm 0.21$ }  & \multicolumn{1}{r}{ $\pm 0.4$ }  & \multicolumn{1}{r}{ $\pm 0.9$ }  & \multicolumn{1}{r}{ $\pm 1.6$ }  & \multicolumn{1}{r}{ ${<}4.3$ }  & \multicolumn{1}{r}{ $\pm 18.7$ }  & \multicolumn{1}{r}{ ${<}41.1$ }  & \multicolumn{1}{r}{ ${<}139.7$ }  & \multicolumn{1}{r}{ ${<}544.2$ } \\[1ex] 
$0.35-$ &  & $\, 0.070$ &  & $\, 0.25$ &  &  &  &  &  & $\, 506.0$ &  & \\ 
\multicolumn{1}{r|}{ 0.5 }  & \multicolumn{1}{r}{ ${<}0.044$ }  & \multicolumn{1}{r}{ $\pm 0.076$ }  & \multicolumn{1}{r}{ ${<}0.25$ }  & \multicolumn{1}{r}{ $\pm 0.35$ }  & \multicolumn{1}{r}{ ${<}1.9$ }  & \multicolumn{1}{r}{ ${<}5.6$ }  & \multicolumn{1}{r}{ ${<}18.3$ }  & \multicolumn{1}{r}{ ${<}62.9$ }  & \multicolumn{1}{r}{ ${<}243.0$ }  & \multicolumn{1}{r}{ $\pm 319.5$ }  & \multicolumn{1}{r}{ ${<}5335.0$ }  & \multicolumn{1}{r}{ ${<}1000.0$ } \\ 
\hline\hline
\end{tabular}

		\caption{Same as table \ref{tab:occ}, but for G stars.}
		\label{tab:occ2}
	\end{table} 
	
	\begin{table}[ht]
		\title{Planet occurrence rates}
		\centering\tiny
		\begin{tabular}{l | llllllllllll}
\hline\hline
 & \multicolumn{12}{c}{Period [days]} \\ 
\hline
 & $0.4-$ & $0.68-$ & $1.2-$ & $2.0-$ & $3.4-$ & $5.8-$ & $10.0-$ & $17.0-$ & $29.0-$ & $50.0-$ & $85.0-$ & $150.0-$\\ 
 & \multicolumn{1}{r}{ 0.68 } & \multicolumn{1}{r}{ 1.2 } & \multicolumn{1}{r}{ 2.0 } & \multicolumn{1}{r}{ 3.4 } & \multicolumn{1}{r}{ 5.8 } & \multicolumn{1}{r}{ 10.0 } & \multicolumn{1}{r}{ 17.0 } & \multicolumn{1}{r}{ 29.0 } & \multicolumn{1}{r}{ 50.0 } & \multicolumn{1}{r}{ 85.0 } & \multicolumn{1}{r}{ 150.0 } & \multicolumn{1}{r}{ 250.0 }\\ 
\hline
$32.0-$ &  &  &  &  &  &  &  & $\, 0.083$ &  & $\, 0.065$ &  & \\ 
\multicolumn{1}{r|}{ 45.0 }  & \multicolumn{1}{r}{ ${<}0.0030$ }  & \multicolumn{1}{r}{ ${<}0.0042$ }  & \multicolumn{1}{r}{ ${<}0.0060$ }  & \multicolumn{1}{r}{ ${<}0.0086$ }  & \multicolumn{1}{r}{ ${<}0.012$ }  & \multicolumn{1}{r}{ ${<}0.018$ }  & \multicolumn{1}{r}{ ${<}0.025$ }  & \multicolumn{1}{r}{ $\pm 0.059$ }  & \multicolumn{1}{r}{ ${<}0.052$ }  & \multicolumn{1}{r}{ $\pm 0.097$ }  & \multicolumn{1}{r}{ ${<}0.11$ }  & \multicolumn{1}{r}{ ${<}0.17$ } \\[1ex] 
$23.0-$ &  &  & $\, 0.011$ & $\, 0.038$ & $\, 0.042$ & $\, 0.034$ & $\, 0.054$ & $\, 0.11$ & $\, 0.17$ & $\, 0.16$ & $\, 0.10$ & $\, 0.34$\\ 
\multicolumn{1}{r|}{ 32.0 }  & \multicolumn{1}{r}{ ${<}0.0032$ }  & \multicolumn{1}{r}{ ${<}0.0045$ }  & \multicolumn{1}{r}{ $\pm 0.011$ }  & \multicolumn{1}{r}{ $\pm 0.020$ }  & \multicolumn{1}{r}{ $\pm 0.025$ }  & \multicolumn{1}{r}{ $\pm 0.031$ }  & \multicolumn{1}{r}{ $\pm 0.044$ }  & \multicolumn{1}{r}{ $\pm 0.07$ }  & \multicolumn{1}{r}{ $\pm 0.11$ }  & \multicolumn{1}{r}{ $\pm 0.13$ }  & \multicolumn{1}{r}{ $\pm 0.15$ }  & \multicolumn{1}{r}{ $\pm 0.30$ } \\[1ex] 
$16.0-$ & $\, 0.0037$ & $\, 0.0055$ & $\, 0.013$ & $\, 0.057$ & $\, 0.11$ & $\, 0.020$ & $\, 0.092$ & $\, 0.11$ & $\, 0.31$ & $\, 0.10$ & $\, 0.14$ & $\, 0.18$\\ 
\multicolumn{1}{r|}{ 23.0 }  & \multicolumn{1}{r}{ $\pm 0.0043$ }  & \multicolumn{1}{r}{ $\pm 0.0062$ }  & \multicolumn{1}{r}{ $\pm 0.011$ }  & \multicolumn{1}{r}{ $\pm 0.025$ }  & \multicolumn{1}{r}{ $\pm 0.04$ }  & \multicolumn{1}{r}{ $\pm 0.026$ }  & \multicolumn{1}{r}{ $\pm 0.055$ }  & \multicolumn{1}{r}{ $\pm 0.08$ }  & \multicolumn{1}{r}{ $\pm 0.14$ }  & \multicolumn{1}{r}{ $\pm 0.11$ }  & \multicolumn{1}{r}{ $\pm 0.16$ }  & \multicolumn{1}{r}{ $\pm 0.25$ } \\[1ex] 
$11.0-$ &  & $\, 0.0096$ & $\, 0.059$ & $\, 0.11$ & $\, 0.052$ & $\, 0.10$ & $\, 0.094$ & $\, 0.32$ & $\, 0.18$ & $\, 0.17$ & $\, 0.28$ & \\ 
\multicolumn{1}{r|}{ 16.0 }  & \multicolumn{1}{r}{ ${<}0.0035$ }  & \multicolumn{1}{r}{ $\pm 0.0081$ }  & \multicolumn{1}{r}{ $\pm 0.021$ }  & \multicolumn{1}{r}{ $\pm 0.03$ }  & \multicolumn{1}{r}{ $\pm 0.031$ }  & \multicolumn{1}{r}{ $\pm 0.05$ }  & \multicolumn{1}{r}{ $\pm 0.057$ }  & \multicolumn{1}{r}{ $\pm 0.13$ }  & \multicolumn{1}{r}{ $\pm 0.12$ }  & \multicolumn{1}{r}{ $\pm 0.14$ }  & \multicolumn{1}{r}{ $\pm 0.22$ }  & \multicolumn{1}{r}{ ${<}0.20$ } \\[1ex] 
$8.0-$ &  & $\, 0.0056$ & $\, 0.0076$ & $\, 0.022$ & $\, 0.064$ & $\, 0.15$ & $\, 0.13$ & $\, 0.22$ & $\, 0.19$ & $\, 0.44$ & $\, 0.69$ & $\, 0.53$\\ 
\multicolumn{1}{r|}{ 11.0 }  & \multicolumn{1}{r}{ ${<}0.0036$ }  & \multicolumn{1}{r}{ $\pm 0.0066$ }  & \multicolumn{1}{r}{ $\pm 0.0094$ }  & \multicolumn{1}{r}{ $\pm 0.017$ }  & \multicolumn{1}{r}{ $\pm 0.032$ }  & \multicolumn{1}{r}{ $\pm 0.06$ }  & \multicolumn{1}{r}{ $\pm 0.07$ }  & \multicolumn{1}{r}{ $\pm 0.10$ }  & \multicolumn{1}{r}{ $\pm 0.12$ }  & \multicolumn{1}{r}{ $\pm 0.22$ }  & \multicolumn{1}{r}{ $\pm 0.32$ }  & \multicolumn{1}{r}{ $\pm 0.38$ } \\[1ex] 
$5.7-$ & $\, 0.0037$ & $\, 0.017$ & $\, 0.021$ & $\, 0.034$ & $\, 0.13$ & $\, 0.11$ & $\, 0.089$ & $\, 0.13$ & $\, 0.22$ & $\, 0.45$ & $\, 1.0$ & $\, 0.61$\\ 
\multicolumn{1}{r|}{ 8.0 }  & \multicolumn{1}{r}{ $\pm 0.0047$ }  & \multicolumn{1}{r}{ $\pm 0.010$ }  & \multicolumn{1}{r}{ $\pm 0.014$ }  & \multicolumn{1}{r}{ $\pm 0.020$ }  & \multicolumn{1}{r}{ $\pm 0.04$ }  & \multicolumn{1}{r}{ $\pm 0.05$ }  & \multicolumn{1}{r}{ $\pm 0.059$ }  & \multicolumn{1}{r}{ $\pm 0.09$ }  & \multicolumn{1}{r}{ $\pm 0.12$ }  & \multicolumn{1}{r}{ $\pm 0.22$ }  & \multicolumn{1}{r}{ $\pm 0.4$ }  & \multicolumn{1}{r}{ $\pm 0.41$ } \\[1ex] 
$4.0-$ & $\, 0.0034$ & $\, 0.016$ & $\, 0.046$ & $\, 0.022$ & $\, 0.13$ & $\, 0.15$ & $\, 0.27$ & $\, 0.29$ & $\, 0.42$ & $\, 0.62$ & $\, 0.53$ & $\, 0.55$\\ 
\multicolumn{1}{r|}{ 5.7 }  & \multicolumn{1}{r}{ $\pm 0.0048$ }  & \multicolumn{1}{r}{ $\pm 0.010$ }  & \multicolumn{1}{r}{ $\pm 0.019$ }  & \multicolumn{1}{r}{ $\pm 0.018$ }  & \multicolumn{1}{r}{ $\pm 0.05$ }  & \multicolumn{1}{r}{ $\pm 0.06$ }  & \multicolumn{1}{r}{ $\pm 0.10$ }  & \multicolumn{1}{r}{ $\pm 0.12$ }  & \multicolumn{1}{r}{ $\pm 0.17$ }  & \multicolumn{1}{r}{ $\pm 0.25$ }  & \multicolumn{1}{r}{ $\pm 0.30$ }  & \multicolumn{1}{r}{ $\pm 0.41$ } \\[1ex] 
$2.8-$ & $\, 0.0043$ & $\, 0.0056$ & $\, 0.045$ & $\, 0.15$ & $\, 0.23$ & $\, 0.41$ & $\, 0.80$ & $\, 0.79$ & $\, 1.5$ & $\, 1.2$ & $\, 1.4$ & $\, 0.99$\\ 
\multicolumn{1}{r|}{ 4.0 }  & \multicolumn{1}{r}{ $\pm 0.0048$ }  & \multicolumn{1}{r}{ $\pm 0.0069$ }  & \multicolumn{1}{r}{ $\pm 0.020$ }  & \multicolumn{1}{r}{ $\pm 0.04$ }  & \multicolumn{1}{r}{ $\pm 0.06$ }  & \multicolumn{1}{r}{ $\pm 0.10$ }  & \multicolumn{1}{r}{ $\pm 0.16$ }  & \multicolumn{1}{r}{ $\pm 0.20$ }  & \multicolumn{1}{r}{ $\pm 0.3$ }  & \multicolumn{1}{r}{ $\pm 0.4$ }  & \multicolumn{1}{r}{ $\pm 0.5$ }  & \multicolumn{1}{r}{ $\pm 0.52$ } \\[1ex] 
$2.0-$ & $\, 0.025$ & $\, 0.049$ & $\, 0.11$ & $\, 0.15$ & $\, 0.50$ & $\, 1.0$ & $\, 1.7$ & $\, 3.0$ & $\, 2.9$ & $\, 2.7$ & $\, 0.93$ & $\, 0.48$\\ 
\multicolumn{1}{r|}{ 2.8 }  & \multicolumn{1}{r}{ $\pm 0.010$ }  & \multicolumn{1}{r}{ $\pm 0.017$ }  & \multicolumn{1}{r}{ $\pm 0.03$ }  & \multicolumn{1}{r}{ $\pm 0.04$ }  & \multicolumn{1}{r}{ $\pm 0.09$ }  & \multicolumn{1}{r}{ $\pm 0.2$ }  & \multicolumn{1}{r}{ $\pm 0.2$ }  & \multicolumn{1}{r}{ $\pm 0.4$ }  & \multicolumn{1}{r}{ $\pm 0.5$ }  & \multicolumn{1}{r}{ $\pm 0.5$ }  & \multicolumn{1}{r}{ $\pm 0.42$ }  & \multicolumn{1}{r}{ $\pm 0.43$ } \\[1ex] 
$1.4-$ & $\, 0.029$ & $\, 0.098$ & $\, 0.16$ & $\, 0.34$ & $\, 0.63$ & $\, 1.3$ & $\, 2.0$ & $\, 2.2$ & $\, 1.0$ & $\, 1.0$ & $\, 1.0$ & $\, 0.63$\\ 
\multicolumn{1}{r|}{ 2.0 }  & \multicolumn{1}{r}{ $\pm 0.011$ }  & \multicolumn{1}{r}{ $\pm 0.023$ }  & \multicolumn{1}{r}{ $\pm 0.04$ }  & \multicolumn{1}{r}{ $\pm 0.06$ }  & \multicolumn{1}{r}{ $\pm 0.10$ }  & \multicolumn{1}{r}{ $\pm 0.2$ }  & \multicolumn{1}{r}{ $\pm 0.3$ }  & \multicolumn{1}{r}{ $\pm 0.4$ }  & \multicolumn{1}{r}{ $\pm 0.3$ }  & \multicolumn{1}{r}{ $\pm 0.5$ }  & \multicolumn{1}{r}{ $\pm 0.6$ }  & \multicolumn{1}{r}{ $\pm 0.90$ } \\[1ex] 
$1.0-$ & $\, 0.016$ & $\, 0.054$ & $\, 0.11$ & $\, 0.22$ & $\, 0.77$ & $\, 1.1$ & $\, 1.8$ & $\, 0.61$ & $\, 0.81$ & $\, 0.50$ & $\, 5.2$ & \\ 
\multicolumn{1}{r|}{ 1.4 }  & \multicolumn{1}{r}{ $\pm 0.009$ }  & \multicolumn{1}{r}{ $\pm 0.018$ }  & \multicolumn{1}{r}{ $\pm 0.03$ }  & \multicolumn{1}{r}{ $\pm 0.06$ }  & \multicolumn{1}{r}{ $\pm 0.13$ }  & \multicolumn{1}{r}{ $\pm 0.2$ }  & \multicolumn{1}{r}{ $\pm 0.3$ }  & \multicolumn{1}{r}{ $\pm 0.27$ }  & \multicolumn{1}{r}{ $\pm 0.45$ }  & \multicolumn{1}{r}{ $\pm 0.70$ }  & \multicolumn{1}{r}{ $\pm 2.7$ }  & \multicolumn{1}{r}{ ${<}4.7$ } \\[1ex] 
$0.71-$ & $\, 0.0099$ & $\, 0.048$ & $\, 0.13$ & $\, 0.21$ & $\, 0.36$ & $\, 0.66$ & $\, 0.35$ & $\, 0.26$ &  &  & $\, 46.4$ & \\ 
\multicolumn{1}{r|}{ 1.0 }  & \multicolumn{1}{r}{ $\pm 0.0076$ }  & \multicolumn{1}{r}{ $\pm 0.020$ }  & \multicolumn{1}{r}{ $\pm 0.04$ }  & \multicolumn{1}{r}{ $\pm 0.08$ }  & \multicolumn{1}{r}{ $\pm 0.12$ }  & \multicolumn{1}{r}{ $\pm 0.21$ }  & \multicolumn{1}{r}{ $\pm 0.28$ }  & \multicolumn{1}{r}{ $\pm 0.35$ }  & \multicolumn{1}{r}{ ${<}2.3$ }  & \multicolumn{1}{r}{ ${<}8.1$ }  & \multicolumn{1}{r}{ $\pm 55.2$ }  & \multicolumn{1}{r}{ ${<}144.8$ } \\[1ex] 
$0.5-$ & $\, 0.015$ & $\, 0.022$ &  &  & $\, 0.20$ & $\, 0.79$ &  & $\, 16.2$ & $\, 31.1$ &  & $\, 2844.7$ & \\ 
\multicolumn{1}{r|}{ 0.71 }  & \multicolumn{1}{r}{ $\pm 0.011$ }  & \multicolumn{1}{r}{ $\pm 0.025$ }  & \multicolumn{1}{r}{ ${<}0.047$ }  & \multicolumn{1}{r}{ ${<}0.12$ }  & \multicolumn{1}{r}{ $\pm 0.18$ }  & \multicolumn{1}{r}{ $\pm 0.61$ }  & \multicolumn{1}{r}{ ${<}4.0$ }  & \multicolumn{1}{r}{ $\pm 18.2$ }  & \multicolumn{1}{r}{ $\pm 39.3$ }  & \multicolumn{1}{r}{ ${<}384.3$ }  & \multicolumn{1}{r}{ $\pm 1808.8$ }  & \multicolumn{1}{r}{ ${<}1000.0$ } \\[1ex] 
$0.35-$ &  &  &  &  &  & $\, 5.5$ &  & $\, 124.5$ &  &  &  & \\ 
\multicolumn{1}{r|}{ 0.5 }  & \multicolumn{1}{r}{ ${<}0.049$ }  & \multicolumn{1}{r}{ ${<}0.14$ }  & \multicolumn{1}{r}{ ${<}0.48$ }  & \multicolumn{1}{r}{ ${<}1.8$ }  & \multicolumn{1}{r}{ ${<}7.6$ }  & \multicolumn{1}{r}{ $\pm 7.0$ }  & \multicolumn{1}{r}{ ${<}180.9$ }  & \multicolumn{1}{r}{ $\pm 157.2$ }  & \multicolumn{1}{r}{ ${<}1000.0$ }  & \multicolumn{1}{r}{ ${<}1000.0$ }  & \multicolumn{1}{r}{ ${<}1000.0$ }  & \multicolumn{1}{r}{ ${<}1000.0$ } \\ 
\hline\hline
\end{tabular}

		\caption{Same as table \ref{tab:occ}, but for F stars.}
		\label{tab:occ3}
	\end{table}

\end{document}